\documentclass[twocolumn]{aastex631}
\usepackage{appendix}

\begin{document}

\title{The JWST View of Cygnus A: Jet-Driven Coronal Outflow with a Twist}

\author[0000-0002-0786-7307]{Patrick M. Ogle}
\affiliation{Space Telescope Science Institute, 3700 San Martin Drive, Baltimore, MD 21218, USA}
\email{pogle@stsci.edu}

\author{B. Sebastian}
\affiliation{Space Telescope Science Institute, 3700 San Martin Drive, Baltimore, MD 21218, USA}

\author{A. Aravindan}
\affiliation{Department of Physics and Astronomy, University of California, Riverside, CA 92521, USA}

\author{M. McDonald}
\affiliation{Department of Physics and Astronomy, University of California, Riverside, CA 92521, USA}

\author{G. Canalizo}
\affiliation{Department of Physics and Astronomy, University of California, Riverside, CA 92521, USA}

\author{M. L.\ N.\ Ashby}
\affiliation{Center for Astrophysics \textbar\ Harvard \& Smithsonian, 
Cambridge, MA 02138, USA}

\author{M. Azadi}
\affiliation{Center for Astrophysics \textbar\ Harvard \& Smithsonian, 
Cambridge, MA 02138, USA}

\author{R. Antonucci}
\affiliation{Physics Department, University of California, Santa Barbara, CA, 93106-9530, USA}

\author{P. Barthel}
\affiliation{Kapteyn Astronomical Institute, University of Groningen, P.O. Box 800, 9700 AV Groningen, The Netherlands}

\author{S. Baum}
\affiliation{Department of Physics and Astronomy, University of Manitoba, Winnipeg, MB R3T 2N2, Canada}
\affiliation{Center for Space Plasma \& Aeronomic Research, 
University of Alabama in Huntsville
Huntsville, AL 35899, USA}

\author{M. Birkinshaw}
\affiliation{HH Wills Physics Laboratory, University of Bristol, Tyndall Avenue, Bristol BS8 1TL, UK}

\author{C. Carilli}
\affiliation{National Radio Astronomy Observatory, P.O. Box O, Socorro, NM 87801, USA}

\author{M. Chiaberge}
\affiliation{Space Telescope Science Institute, 3700 San Martin Drive, Baltimore, MD 21218, USA}

\author{C. Duggal}
\affiliation{Department of Physics and Astronomy, University of Manitoba, Winnipeg, MB R3T 2N2, Canada}

\author{K. Gebhardt} 
\affiliation{Department of Astronomy, The University of Texas at Austin, 2515 Speedway Boulevard, Austin, TX 78712, USA}

\author{S. Hyman}
\affiliation{Steward Observatory, University of Arizona, Tucson, AZ 85721, USA}

\author{J. Kuraszkiewicz}
\affiliation{Center for Astrophysics \textbar\ Harvard \& Smithsonian, 
Cambridge, MA 02138, USA}

\author{E. Lopez-Rodriguez}
\affiliation{Department of Physics \& Astronomy, University of South Carolina, Columbia, SC 29208, USA}
\affiliation{Kavli Institute for Particle Astrophysics \& Cosmology (KIPAC), Stanford University, Stanford, CA 94305, USA}

\author[0000-0001-7421-2944]{A. M. Medling}
\affiliation{Ritter Astrophysical Research Center and Department of Physics \& Astronomy, University of Toledo, Toledo, OH 43606, USA}

\author{G. Miley}
\affiliation{Leiden Observatory, University of Leiden, PO Box 9513, Leiden, 2300 RA, The Netherlands}

\author{O. Omoruyi}
\affiliation{Center for Astrophysics \textbar\ Harvard \& Smithsonian, 
Cambridge, MA 02138, USA}

\author{C. O'Dea}
\affiliation{Department of Physics and Astronomy, University of Manitoba, Winnipeg, MB R3T 2N2, Canada}
\affiliation{Center for Space Plasma \& Aeronomic Research, 
University of Alabama in Huntsville
Huntsville, AL 35899, USA}

\author{D. Perley}
\affiliation{Astrophysics Research Institute, Liverpool John Moores University, IC2, Liverpool Science Park, 146 Brownlow Hill, Liverpool L3 5RF, UK}

\author{R. A. Perley}
\affiliation{National Radio Astronomy Observatory, P.O. Box O, Socorro, NM 87801, USA}

\author[0000-0002-3099-1664]{E. Perlman}
\affiliation{Aerospace, Physics and Space Sciences Department, Florida Institute of Technology, 150 W. University Boulevard, Melbourne, FL 32901, USA}

\author{V. Reynaldi}
\affiliation{Instituto de Astrofisica de la Plata (CONICET--UNLP), Paseo del Bosque s/n, 1900, La Plata, Argentina }

\author{M. Singha}
\affiliation{Astrophysics Science Division, NASA Goddard Space Flight Center, Greenbelt, MD 20771, USA}

\author{W. Sparks}
\affiliation{SETI Institute, 339 N Bernado Ave, Mountain View, CA 94043, USA}
\affiliation{Space Telescope Science Institute, 3700 San Martin Drive, Baltimore, MD 21218, USA}

\author{G. Tremblay}
\affiliation{Center for Astrophysics \textbar\ Harvard \& Smithsonian, 
Cambridge, MA 02138, USA}

\author{B. Wilkes}
\affiliation{Center for Astrophysics \textbar\ Harvard \& Smithsonian, 
Cambridge, MA 02138, USA}
\affiliation{HH Wills Physics Laboratory, University of Bristol, Tyndall Avenue, Bristol BS8 1TL, UK}

\author[0000-0002-9895-5758]{S.~P.~Willner}
\affiliation{Center for Astrophysics \textbar\ Harvard \& Smithsonian, 
Cambridge, MA 02138, USA}

\author{D. M. Worrall}
\affiliation{HH Wills Physics Laboratory, University of Bristol, Tyndall Avenue, Bristol BS8 1TL, UK}

\begin{abstract}
We present first results from James Webb Space Telescope (JWST) Near-Infrared Spectrograph (NIRSpec), Mid-Infrared Instrument (MIRI), and Keck Cosmic Webb Imager (KCWI) integral field spectroscopy of the powerful but highly obscured host-galaxy of the jetted radio source Cygnus A.  We detect 169 infrared emission lines at $1.7$--$27 ~\mu$m and explore the  kinematics and physical properties of the extended narrow-line region (NLR) in unprecedented detail.  The density-stratified NLR appears to be shaped by the initial blow-out and ongoing interaction of the radio jet with the interstellar medium, creating a multi-phase bicone with a layered structure composed of molecular and ionized gas. The NLR spectrum, with strong coronal emission at kpc-scale, is well-modeled by AGN photoionization. We find evidence that the NLR is rotating around the radio axis, perhaps mediated by magnetic fields and driven by angular momentum transfer from the radio jet. The overall velocity field of the NLR is well described by 250 km s$^{-1}$ outflow along biconical spiral flow lines, combining both rotation and outflow signatures. There is particularly bright [Fe {\sc ii}] $\lambda 1.644 ~\mu$m emission from a dense, high-velocity dispersion, photoionized clump of clouds found near the projected radio axis. Outflows of 600--2000 km s$^{-1}$ are found in bullets and streamers of ionized gas that may be ablated by the radio jet from these clouds, driving a local outflow rate of $40 M_\odot$ yr$^{-1}$.

\end{abstract}


\section{Introduction} \label{sec:intro}

Radio-jet feedback may be the key to suppressing star formation in high-mass elliptical galaxies \citep{Morganti2005,2006MNRAS.365...11C,Cattaneo2009,Fabian2012,Heckman2014,Lanz2016,Girdhar2022,Kondapally2023,Heckman2023}. 
It may do so both by ejecting gas from the galaxy and by heating the circumgalactic medium and thereby stalling gas accretion \citep{Crain2015,Somerville2015,Forster2019,Gaibler2012,Dugan2017,Guillard2012,McNamara2012,Holt2008}. However, the mass outflow rates observed in the ionized NLRs of nearby radio-loud active galactic nuclei (AGNs) are relatively modest, suggesting either that ionized gas represents a small fraction of the mass outflow or that outflows are less powerful at late cosmic epochs \citep{2007NewAR..51..153T}. The manner in which the radio jet couples to the host-galaxy interstellar medium (ISM) in the case of ejective feedback is an area of active observational and theoretical study \citep{2018MNRAS.479.5544M,Mandal2021,Mandal2024,Meenakshi2022,Nesvadba2021}.  JWST IFU spectroscopy has begun to reveal new details about how radio jet feedback works in both the local and high-redshift universe \citep{2022A&A...665L..11P, 2024A&A...683A.169W, 2024arXiv240111612R,2024A&A...689A.314L,2024ApJ...974..127C,2024arXiv240603218D}. 
 In particular, it is becoming apparent that the impact of the radio jet on the ISM depends on a number of factors, including jet power, age, and geometry with respect to the host ISM. It is therefore crucial to observe jets with a range of power and age in different environments in order to fully understand their impact on the ISM and massive galaxy evolution.

Cygnus A ($z = 0.0557$, $1\arcsec = 1.09$~kpc) is by far the most powerful radio source in the nearby universe, powered by a collimated, relativistic jet launched from the active galactic nucleus \citep{1954ApJ...119..206B,1984ApJ...285L..35P,1991AJ....102.1691C,1996MNRAS.283L..45B,1996A&ARv...7....1C}. The AGN is highly obscured by dust at visible wavelengths and has an absorption column density of $N_{\rm{H}} = 1.7\times10^{23}$ cm$^{-2}$ to X-rays \citep{1991ApJ...372L..67D,1994ApJ...431L...1U,2002ApJ...564..176Y,2015ApJ...808..154R}. A hidden Type 1 quasar is revealed by spectropolarimetry, which shows emission lines with full width at half maximum (FWHM) up to $26,000$ km s$^{-1}$ in polarized flux \citep{1997ApJ...482L..37O, 1994Natur.371..313A}. Radiation absorbed by the pc-scale dusty torus surrounding the AGN is re-radiated at mid-infrared wavelengths, with a bolometric luminosity of $L_\mathrm{IR} = 3.0\times 10^{45}$ erg s$^{-1}$ \citep{2012ApJ...747...46P}\footnote{Herschel FIR observations admit the possibility of a contribution from star formation at a rate of $26 M_\odot$ yr$^{-1}$, which could lower the inferred AGN luminosity to $8\times 10^{44}$ erg s$^{-1}$ \citep{2012ApJ...757L..26B}. However, we find no evidence for emission at $<27$ $\mu$m from this component in the JWST data. The FIR emission observed by Herschel may have an extranuclear starburst origin or alternatively arise from dust heated by a combination of the AGN and the old stellar population in the galaxy bulge.}. The 10--100 $\mu$m polarimetric observations show a highly ($\sim$10\%) polarized  nucleus with an inferred B-field parallel to the axis of the obscuring torus, as traced by magnetically aligned dust grains, demonstrating the influence of the magnetic field on the ISM of a jetted radio source \citep{2018ApJ...861L..23L, 2018MNRAS.478.2350L,Lopez-Rodriguez2023}. 

A prominent ionization cone is seen in polarized optical continuum and emission lines \citep{1997ApJ...482L..37O, 1998MNRAS.301..131J}. The structure of this biconical NLR is even more apparent at near infrared (NIR) wavelengths, which are less affected by dust obscuration in the host-galaxy \citep{1999ApJ...512L..91T, 2003ApJ...597..823C}. Highly asymmetric polarized flux at 2~$\mu$m could be indicative of beamed NIR continuum \citep{Tadhunter2000}. However, the polarized flux may alternatively follow the asymmetry in the spatial distribution of the dusty scattering medium. The NLR of Cygnus A contains a mix of low and high-ionization photoionized components \citep{1994MNRAS.268..989T, 1975ApJ...197..535O,1994AJ....108..414S}.  \cite{1994MNRAS.268..989T} suggested that the extreme [N\,{\sc{ii}}] line strength may require an N/O abundance $4\times$ solar. However, the [N\,{\sc{ii}}] emission appears to be enhanced in a disk-like structure perpendicular to the ionization cone \citep{1994AJ....108..414S}, and its strength may potentially be explained by shock ionization. Rotation dominates the overall kinematics of ionized gas, H$_2$, and CO \citep{2003MNRAS.342..861T, 2021MNRAS.506.2950R, 2022ApJ...937..106C}. Additionally, line-splitting and high velocity dispersion indicate the presence of a high-velocity ionized outflow in the NLR \citep{1994AJ....108..414S, 1991MNRAS.251P..46T, 2021MNRAS.506.2950R}.  

In this contribution we present JWST NIRSpec IFU and MIRI MRS observations of the central 2 kpc of the Cygnus A host-galaxy (\S 2). The spectra reveal a rich collection of ionized and molecular gas emission lines (\S 3). We explore the spatial distribution of ionized and molecular gas, demonstrating a stratified biconical NLR that appears to be shaped both by the radio jet and ionizing radiation from the AGN (\S 4). In \S 5, we map the extinction towards the ionized gas emission lines from the NLR.  We describe the NLR kinematics in \S 6 and confirm that the multitude of ionized gas lines from the biconical NLR are excited by photoionization in \S7. We model the velocity field by spiral outflow in \S 7. We  conclude  in \S 8 by discussing the implications of our results for radio jet feedback on the multiphase ISM in Cygnus A.

\section{Observations}

We present new observations of Cyg A taken with JWST \citep{2023PASP..135f8001G}, utilizing the NIRSpec IFU \citep{2022A&A...661A..80J,2023PASP..135c8001B} and MIRI MRS \citep{2015PASP..127..646W,2023A&A...675A.111A}. We also present new observations obtained with the Keck Cosmic Web Imager \citep[KCWI;][]{2018ApJ...864...93M}.

\subsection{NIRSpec IFU}

Cygnus A was observed with the NIRSpec IFU  on 2023 August 16  as part of JWST PID 4065 (PI Ogle). Spectra were obtained with the G235H/F170P and G395H/F290LP filter/grating combinations at a spectral resoluton $R \sim 2700$, covering the wavelength range 1.66--5.27 $\mu$m with small gaps in wavelength coverage at the chip gaps. Data were taken at four dither points, yielding a $3\farcs6 \times 3\farcs4 $ arcsec$^2$ field of view (FOV). Dithering better samples the telescope PSF, which is undersampled by the $0\farcs1$ IFU pixels at $<3.2$ $\mu$m.  Exposures were taken using the NRSIRS2 readout pattern to enable subtraction of the variable detector bias level and thereby reduce 1/f noise. The total exposure time was 5894s in each grating. The IFU data were retrieved from MAST and reprocessed through all stages (with the 1281 pmap set of calibration reference files) to take advantage of improvements in the JWST Build 11.0 pipeline. In particular, a bug fixed in this pipeline version yielded improved rejection of bright cosmic rays for the IRS2 readout mode.  We used a clipped median filter with a 3x3x3 kernel to further remove outliers from the Level 3 datacube. While such a median filter will spatially and spectrally smooth the data, the use of a conservative flux clip limit means that only isolated high or low voxels were replaced with the median of the surrounding kernel.  

Emission line analysis was performed at the native instrumental spatial and spectral resolution for individual emission line flux maps and kinematics. The spatial resolution of the IFU does not change significantly with wavelength below  $3.2$ $\mu$m because the IFU is spatially undersampled. Therefore, the emission line ratios used for our extinction analysis have the same effective spatial resolution. Our photoionization modeling is performed on spectra extracted in large apertures and is therefore not impacted by variations in spatial resolution with wavelength. In our kinematics analyses, we subtract the instrumental FWHM in quadrature to obtain velocity dispersion. We used the {\sc jdaviz} \citep{jdaviz2024} Moment Map tool to generate total flux, velocity, and velocity dispersion maps, utilizing a local linear fit around each line for continuum subtraction. 

NIRSpec spectra of the nucleus were extracted in an  $0\farcs7 \times 0\farcs7$ ($0.76 \times 0.76$ kpc$^{2}$) square aperture centered on the nucleus (Fig. 2). We additionally extracted spectra from a $0\farcs3 \times 0\farcs3$ (0.32$\times$0.32~kpc) square aperture at the NW reference (NWR) region and an $0\farcs8 \times 1\farcs0$ (0.87$\times$1.09~kpc$^2$) elliptical aperture encompassing the Fe {\sc ii} Clump. The continuum spectra from the two NIRSpec gratings match quite well without any renormalization. There is a small ($\sim 10\%
$) mismatch between the NIRspec and MIRI spectra of the nucleus at 5 $\mu$m, perhaps related to the different pixel sizes of the two instruments, which does not significantly impact any of our results.  We used the Penalized Pixel-Fitting method ({\tt pPXF})\citep{2023MNRAS.526.3273C,2017MNRAS.466..798C,2011MNRAS.413..813C} to fit the galaxy continuum and decompose the NIRSpec emission line spectrum into multiple velocity components.

\subsection{MIRI MRS}

 MIRI MRS observations of Cyg A were obtained on 2023 October 30, also as part of JWST PID 4065.
 We defer a full presentation of the MIRI MRS results, but do make use of the MRS spectra to illustrate a few key results. Data were taken in all MRS channels and sub-bands, covering the continuous wavelength range 4.9--27.9 $\mu$m. The observations were made using a 4-point dither, optimized for an extended source.  We retrieved the standard Level 3 data products from MAST, as generated by the JWST Build 10.0.1 pipeline using the 1141 pmap. MIRI MRS spectra were extracted in fractional pixel apertures matching the NIRSpec apertures, however, pixel interpolation errors may cause the small mismatch noted in \S2.1.

\subsection{KCWI}

Cyg A was observed with KCWI for Keck program 2024B-U230 (PI: G. Canalizo) on 2024 August 4 and 2024 August 9. KCWI is an integral field spectrograph that covers the optical and near-infrared spectrum. The observations utilized the blue grating (BL), covering 3500–5500 \AA, and the red grating (RL), spanning 5600–10810 \AA, with the $1 \times 1$ binning on the small slicer setup. This configuration gives a spectral resolution of $\sim 80$ km s$^{-1}$ FWHM at 4550 \AA~ and an 8" $\times$  20" FOV. The net exposure time was $14 \times 1000$ s for the blue data cube containing the [O {\sc iii}] 5007 \AA ~line presented here. The observations were obtained under clear sky conditions and a seeing of $\sim 0\farcs8$. The data were reduced using version 1.1.0 of the KCWI Data Reduction Pipeline \citep[DRP;][]{Neill2023}, following the standard procedures listed in the documentation, including wavelength and flux calibration. To construct a full mosaic of the galaxy, we dithered between exposures each night. A foreground star common between both nights was used to align and stack the individual data cubes using custom Python scripts. We then employed routines from the IFSRED library \citep{2014ascl.soft09004R} to resample the data cube, converting the originally rectangular $0\farcs35 \times 0\farcs15$ spaxels to $0\farcs15 \times 0\farcs15$ using IFSR-KCWIRESAMPLE. This yielded undistorted, square pixels matching the detector spatial sampling along IFU slices, but the actual spatial resolution of the data is seeing-limited.

\subsection{VLA}

We used an existing 11 GHz map of the entire source and a new 33 GHz VLA map of the kpc-scale radio jet for comparison with the NLR. The source was observed by the JVLA \citep{Perley2011} in 2014 and 2015 as part of a multi-configuration, multi-frequency study of the polarimetric properties of Cygnus A and its interaction with the cluster medium.  Details of the calibration and imaging, and results from this comprehensive study, are presented by \cite{2020ApJ...903...36S}.  The  11 GHz image shown was produced by combining data from all four VLA configurations, after self-calibration was performed to remove instrumental and atmospheric gain fluctuations. The 33 GHz map was generated from observations of Cygnus A with the VLA from 2017 through 2023.   The goal of this ongoing study is to track the evolution of the A2 transient, following its 2016 discovery \citep{2017ApJ...841..117P, 2019ApJ...874L..32C}.  The displayed image was made from a combination of all A-configuration observations taken in this program, following standard self-calibration techniques. 

\section{Multiwavelength Perspective}

\begin{figure*}[ht]
  \includegraphics[trim=0.0cm 0.0cm 0.0cm 0.0cm, clip, width=\linewidth]{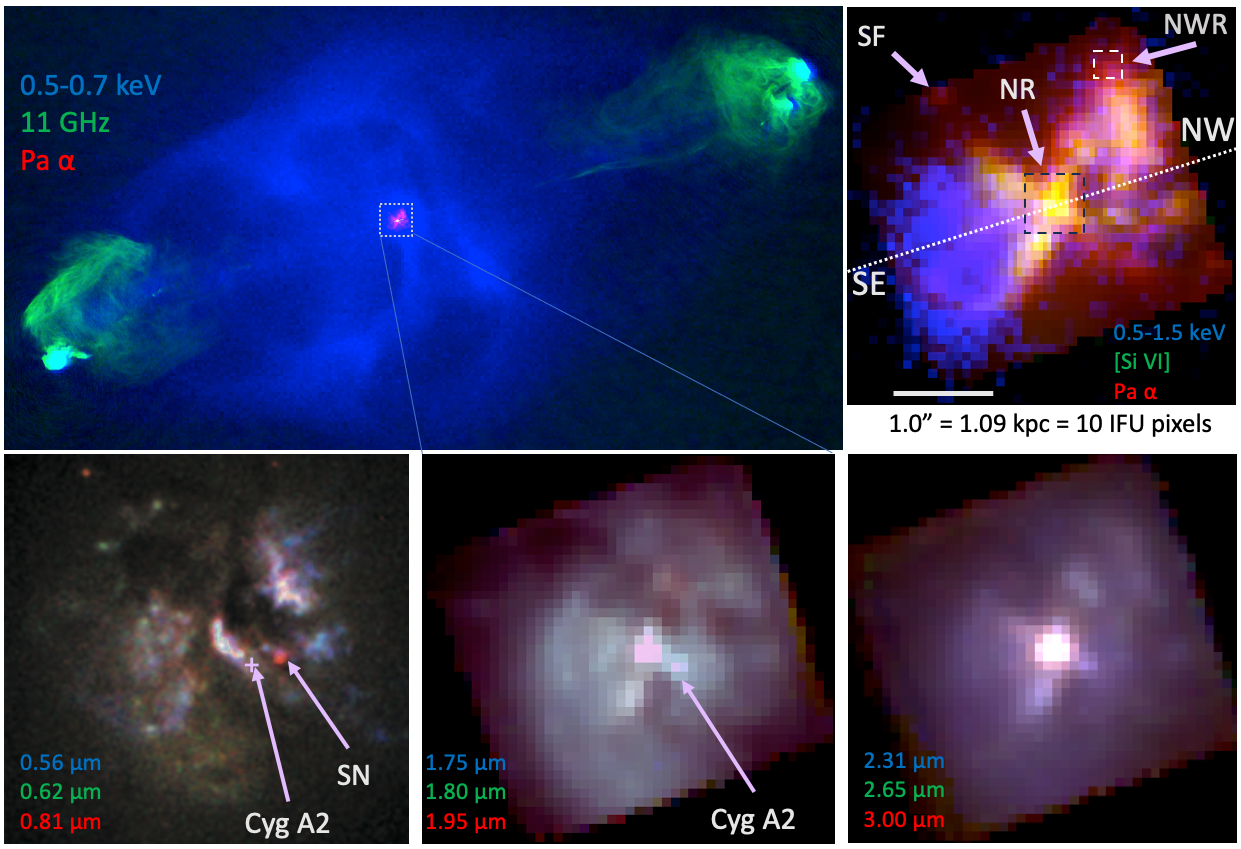}
 \caption{Cygnus A NLR in context.  Top left: The radio jet, with its hot spots and enormous radio lobes, drives a shock into the intracluster medium, mapped at 11 GHz by the VLA \citep{2020ApJ...903...36S} and imaged in 0.5--0.7 keV X-rays by {\it Chandra} \citep{2018ApJ...855...71S}. Smaller panels: the JWST NIRSpec IFU and HST observations zoom in a $3\farcs6 \times 3\farcs4 $ region centered on the galaxy nucleus. Top right: Comparison of Pa $\alpha$, [Si {\sc vi}] 1.963 $\mu$m from NIRSpec, and 0.5--1.5 keV X-rays from {\it Chandra}. Various features are labeled, including the SE and NW lobes of the bicone, the nuclear extraction region (NR), the northwest reference extraction region (NWR), and SF, the only identified star-forming region. X-ray photoionized gas appears to fill the NLR bicone, but soft X-rays are obscured by dense patches of gas in the NW cone and by a curved filament in the SE cone. 
 Bottom row: series of continuum images from HST (left) and NIRSpec IFU (center, right), featuring patchy dust extinction, stellar continuum, and scattered light. Dust scattering of the hidden QSO optical continuum gives rise to the blue patches of emission in the HST image. Scattered light originating from the hot inner edge of the dusty torus traces the edges of the hollow NLR bicone at 1.8--3.0 $\mu$m.  The Cyg A2 secondary nucleus appears strongest in starlight at $\sim 1.7\mu$m, but has no obvious counterpart in the emission line maps. Cyg A2 should not be confused with the apparent (red) supernova that only appears in the HST $0.81\mu$m image. \textit{NB: All images in this paper are presented in standard orientation, with N up, E to the left.}}
 \label{fig:collage}
\end{figure*}

\begin{figure*}[ht]
  \includegraphics[trim=0.0cm 0.0cm 0.0cm 0.0cm, clip, width=\linewidth]{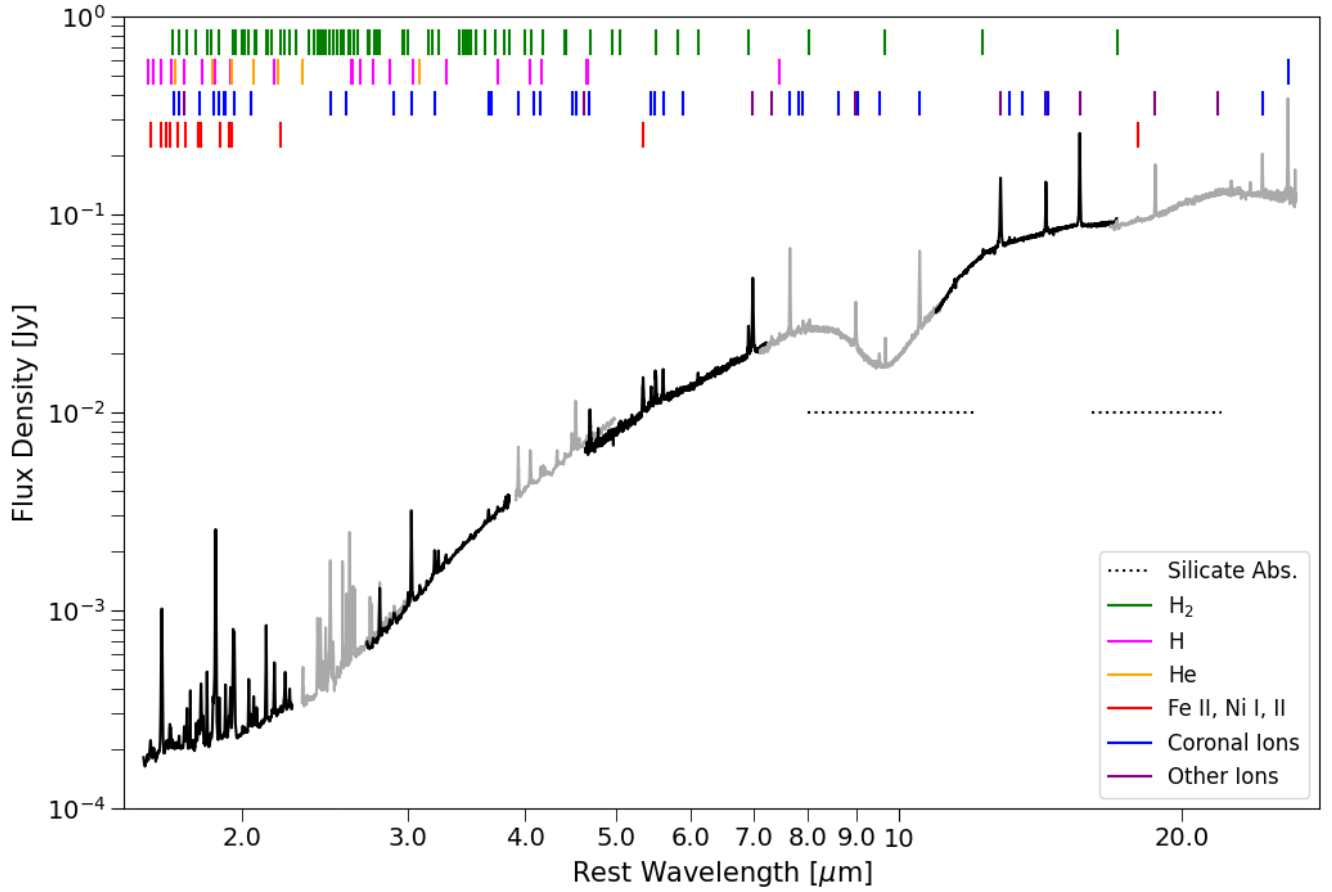}
 \caption{NIRSpec IFU and MIRI MRS spectra of the nucleus, extracted within an $0\farcs7 \times 0\farcs7$ square aperture. The continuum at $>2 ~\mu$m is hot dust emission from the inner AGN torus, with silicate absorption from cooler dust in the outer torus or the shielded faces of dense clumps. Emission lines from warm H$_2$ are indicated by green tick marks and ions by other colored marks as indicated in the legend.}
 \label{fig:integrated_spectrum}
\end{figure*}

\begin{figure*}[ht]
  \includegraphics[trim=0.0cm 0.0cm 0.0cm 0.0cm, clip, width=0.95\linewidth]{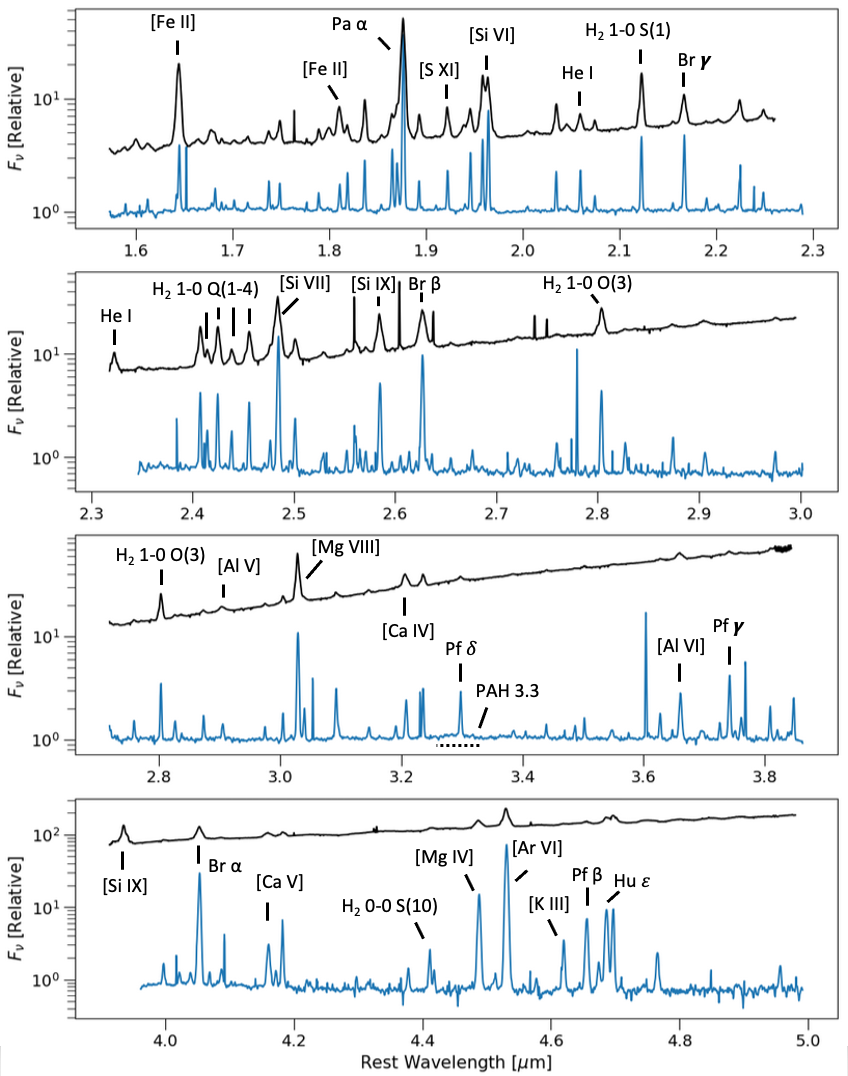}
 \caption{NIRSpec IFU spectra of the nucleus (NR, black) and the NW reference region (NWR, blue). Emission lines of interest are marked. Spectra are scaled to show them together. The plethora of rovibrational H$_2$ lines are important diagnostics of shocks or turbulence in warm molecular gas. The Paschen, Pfund, and Brackett series of hydrogen have characteristic ratios that are useful tracers of extinction. The [Fe {\sc ii}] 1.644 $~\mu$m line is produced in high density clouds, while coronal lines of S, Si, Al, Mg, Ar, and Ca trace photoionization by FUV and X-ray  photons in relatively low density regions. The weak PAH $3.3 ~\mu$m feature may trace either low-level star formation or the diffuse UV photon field from the galaxy bulge. The emission line widths are much broader in the nucleus than in the extended NLR, tracing the gravitational potential of the black hole and any non-virial radial motions.}
 \label{fig:nirspec_extraction}
\end{figure*}

\begin{figure*}[ht]
 \includegraphics[trim=0.0cm 0.0cm 0.0cm 0.0cm, clip, width=\linewidth]{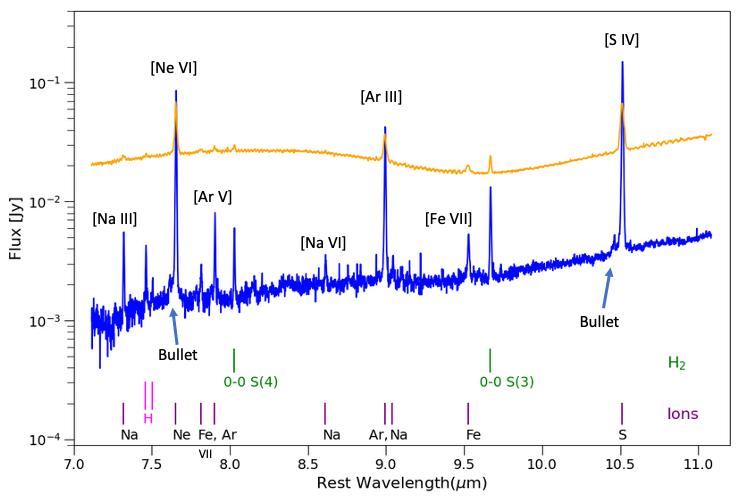}
 \caption{A segment of the MIRI MRS spectra of the Fe {\sc ii} Clump (blue) and nucleus (orange), displaying numerous coronal and intermediate ionization  emission lines.  The silicate absorption trough is considerably weaker in the scattered nuclear continuum spectrum from the Fe {\sc ii} Clump, which has a more polar, less-obscured view of the nucleus than we do--a striking confirmation of radio-loud AGN unification. The Cyg A Bullet coronal outflow makes an appearance blue-ward of the [Ne {\sc vi}] and [S {\sc iv}] lines in the Fe {\sc ii} Clump spectrum.}
 \label{fig:miri_extraction}
\end{figure*}

In order to put the JWST results in full perspective, we show them in relation to a rich suite of available  multiwavelength observations, including VLA radio, \textit{Chandra} X-ray, and Hubble optical imaging (Fig.~\ref{fig:collage}). The radio jets span a diameter of 64 kpc, ending in radio hotspots where they are slowed to subrelativistic speeds and their power is dissipated \citep{1996A&ARv...7....1C}. The enormous energy released creates large cavities in the intracluster medium and drives the shocks seen as bright edges in the \textit{Chandra} image at 0.5--07 keV \citep{2018ApJ...855...71S}.
Zooming in on the $3\farcs6 \times 3\farcs4$  ($3.9\times 3.7$ kpc) central region of the brightest cluster galaxy (BCG) of Cygnus A, a distinct biconical NLR is seen at optical through MIR wavelengths. X-ray ($0.5$--$1.5$ keV) coronal line emission \citep{2002ApJ...564..176Y} shows up preferentially inside the SE bicone, where there is a lower column density of gas, illustrating a marked asymmetry between the two sides of the bicone. The NLR bicone and the host-galaxy are heavily obscured by patchy dust clouds in the HST images, with bright patches of the optical NLR shining through in both direct and scattered light. The obscuration becomes less but still significant in the JWST NIRSpec continuum and emission line images extracted from the IFU data. 
The unresolved, heavily  obscured active nucleus lies near the center of the bicone. The optical continuum from the AGN and NIR continuum from hot dust at the inner edge of the obscuring torus are only seen indirectly via scattered light. The nucleus itself is only directly visible in hard X-rays \citep{2015ApJ...808..154R}. 

We present the combined NIRSpec and MRS $1.6$--$27 ~\mu$m spectrum of the nucleus in Fig.~\ref{fig:integrated_spectrum}. The continuum at $>2 ~\mu$m is likely dominated by emission from dust in the the AGN torus with a wide range of temperatures \citep{2014MNRAS.439.1648A, 2019ApJ...884..171H, 2024MNRAS.527.2371B}. Silicate absorption is seen at $8.3$--$12 ~\mu$m and weakly at $18$--$22 ~\mu$m. At $<2 ~\mu$m, there is a significant continuum contribution from starlight.  We detect 94 emission lines from ions (Table~\ref{Table1}) and 75 emission lines from H$_2$ (reported elsewhere). The ionization potentials of the ions range from 13.6 eV for  H {\sc i} to 351 eV for Si\,{\sc ix}.  The spectrum of the nucleus has much broader emission lines than the extended NLR (Figures~\ref{fig:nirspec_extraction}, \ref{fig:miri_extraction}).
It is also striking that the off-nuclear spectrum shows much weaker silicate absorption than the nucleus, consistent with a higher inclination, less obscured view of the dusty torus, as seen in scattered light and consistent with both the observed and predicted inclination-dependent silicate absorption depths of quasars and radio galaxies  \citep{2010ApJ...717..766L, 2008ApJ...685..147N}.

There is a weak 3.3 $\mu$m PAH feature in the NIRSpec spectrum of the NWR, but PAH features are not readily detected in the MRS spectrum (Figures~\ref{fig:nirspec_extraction}, \ref{fig:miri_extraction}). Only one location shows clear evidence of star formation: an unresolved knot at the NE edge of our map, seen most clearly in the Pa $\alpha$ line (Fig.~\ref{fig:collage}). The spectrum at that location only shows emission from H, He, and [Fe {\sc{ii}}], with no high-ionization lines detected. While it is possible that star formation occurs elsewhere in the region covered by our map, it would be difficult to distinguish in projection to the NLR bicone.

The secondary radio nucleus Cyg A2 \citep{2017ApJ...841..117P} does not stand out in any of the emission line maps (Figures~\ref{fig:collage} \& \ref{fig:nirspec_collage}), nor in MIR continuum. However, it does appear in continuum emission at $<3$ $\mu$m, consistent with its identification as a secondary stellar nucleus in ground-based and HST images. While a radio outburst at this location may indicate a secondary AGN, there is no indication in the JWST data of significant thermal AGN activity.

\section{NLR Morphology}

\begin{figure*}[ht]
  \includegraphics[trim=0.0cm 0.0cm 0.0cm 0.0cm, clip, width=\linewidth]{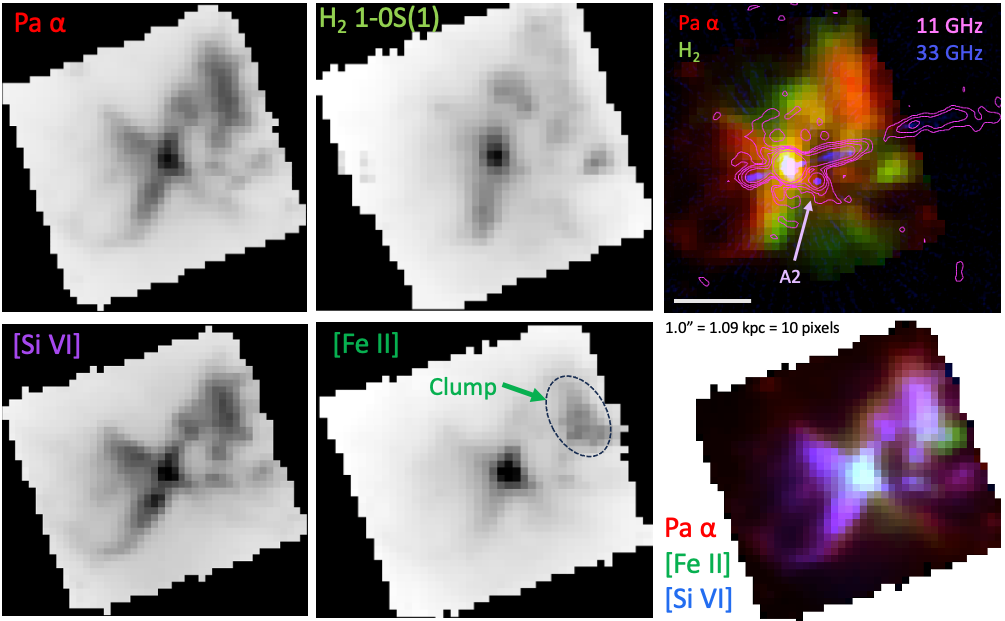}
 \caption{NIRSpec IFU continuum-subtracted line flux (moment 0 maps) of Pa $\alpha$, H$_2$ 1-0 S(1) 2.122$~\mu$m, [Si {\sc vi}] 1.963$~\mu$m, and [Fe {\sc ii}] 1.644$~\mu$m. The VLA radio maps at 33 GHz (blue) and 11 GHz (contours) feature the radio jet, nucleus, and the Cyg A2 secondary nucleus \citep{2017ApJ...841..117P, 2019ApJ...874L..32C}. There is no distinguishable narrow-line emission from Cyg A2. The H$_2$ rovibrational line emission appears to follow the NLR bicone, but with broader opening angle. High-ionization regions traced by [Si {\sc vi}] are distributed in a narrower bicone than low-ionization regions traced by Pa $\alpha$. Strong [Fe {\sc ii}] emission traces dense photoionized gas in the nucleus and Fe {\sc ii} Clump.}
 \label{fig:nirspec_collage}
\end{figure*}

The axis of symmetry of the biconical NLR is aligned with the radio jet  (Figures~\ref{fig:collage} \& \ref{fig:nirspec_collage}). Distinct limb brightening indicates that the bicone is hollow or much less dense along its axis.  Extinction obscures large swathes of the galaxy and NLR at visible wavelengths, becoming less severe but still significant in the NIR. The morphology of the NLR is echoed in scattered light at 1.8 -- 3 $\mu$m and beyond. Direct hot dust emission from the nucleus becomes increasingly apparent at  $>3$ $\mu$m.
There is a significant asymmetry between the SE and NW cones, marked by the presence of much more molecular gas and dust in the NW cone. The NW cone is clumpier, while the SE cone appears to be evacuated, with line emission found primarily along the edges of the cone. There is a remarkable difference in the spatial distribution of the ionized line emission and the rovibrational H$_2$ line emission. The general impression is that the bicone has a wider opening angle in H$_2$ than in ionized gas. The H$_2$ rovibrational lines may be powered by a number of mechanisms, including shocks and turbulence driven by the radio jet or outflows, X-ray heating, cosmic rays, or pumping by the NIR continuum. Any of these might be expected to have a broader reach than UV photoionization, which appears to be restricted to a well-defined cone by the dusty torus. Detailed analysis of the H$_2$ excitation mechanism and temperature distribution is outside of the scope of this paper.

There is a clear variation in structure with ionization parameter (Fig. \ref{fig:nirspec_collage}). Emission in the [Si {\sc vi}]\, 1.963\,$~\mu$m coronal line has a narrower half-opening angle of $46\arcdeg$ compared to Pa $\alpha$, which has a half-opening angle of $54\arcdeg$. FUV emission from the nucleus is likely restricted to a narrower cone than the UV emission, and the density may be higher along the edges of the biconical NLR because its center has been partially cleared by the radio jet. X-ray emission from extremely highly photoionized gas, including [Ne {\sc ix}] and [Si {\sc xi}] is observed in both lobes of the bicone with {\it Chandra} \citep{2002ApJ...564..176Y}. Dense ionized gas clumps in the NW cone and also a distinct ionized gas filament in the SE cone pass in front of and absorb the X-ray emission (Fig.~\ref{fig:collage}), indicating that the X-ray emission comes from the cone interior. The [Fe {\sc ii}] 1.644 $~\mu$m line is enhanced in a cloud complex in the NW cone of the NLR, which we refer to as the Fe {\sc ii} Clump. While this line is potentially a shock indicator \citep{Rodriguez-ardila2004,Mouri2000}, the enhanced [Fe {\sc ii}] emission is most likely produced by dense, photoionized gas, as we demonstrate below.

\begin{figure*}[ht]
 \includegraphics[trim=0.0cm 0.0cm 0.0cm 0.0cm, clip, width=\linewidth]{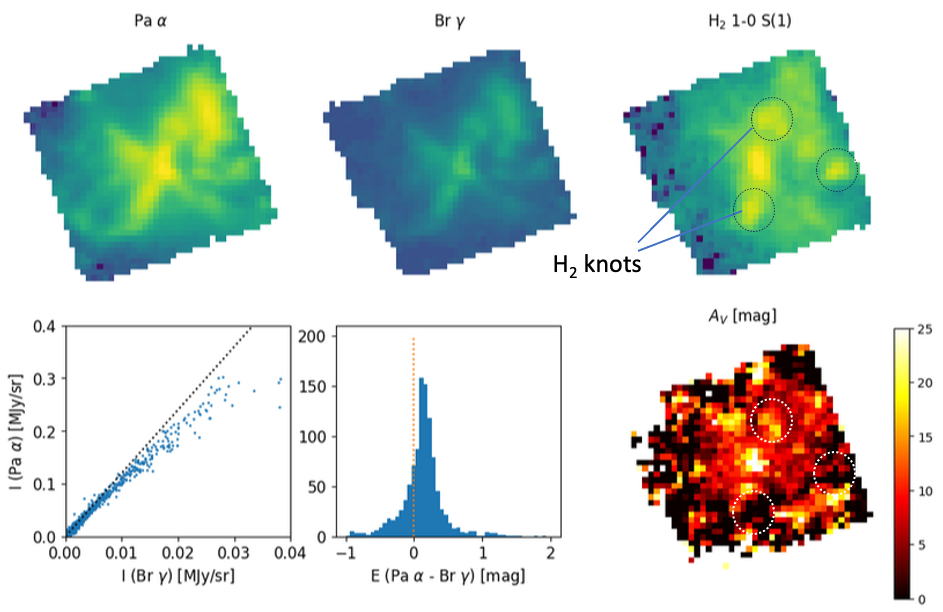}
 \caption{Extinction at Pa $\alpha$ relative to Br $\gamma$. Top row: Emission line (moment 0) intensity maps.
 Bottom left: Pa $\alpha$ vs. Br $\gamma$ for each spaxel compared to the Case B ratio of 12.4 (dotted line). Bottom center: Relative extinction assuming Case B. Bottom right: Extinction map. Some patches of higher extinction correspond to knots in the H$_2$ 1-0 S(1) intensity map and the nucleus. Other H$_2$ knots that do not obscure Pa $\alpha$ significantly may be located behind the NLR.}
 \label{fig:extinction}
\end{figure*}

\section{Extinction}

Fig.~\ref{fig:extinction} shows the relative extinction $E$(Pa $\alpha$ - Br $\gamma$) of the NLR, mapped from the ratio of the Pa $\alpha$ and Br $\gamma$ emission lines. We assumed an intrinsic line ratio of 12.4 for Case B recombination \citep{1987MNRAS.224..801H} at a characteristic nebular temperature of $1\times 10^4$ K. 
The Galactic extinction curve of \cite{2016ApJ...821...78S} implies $A_V = 45.7 \times E$(Pa $\alpha$ - Br $\gamma$). We find a median relative extinction of $E$(Pa $\alpha$ - Br $\gamma$) $= 0.13$ mag in the NLR, corresponding to a median $V$-band extinction of $A_V = 6.01$ mag (mean $A_V = 5.42$ mag). This level of extinction results in $<13\%$ loss for emission lines with rest wavelengths larger than Pa $\alpha$ and therefore does not significantly affect our overall analysis of NLR emission line ratios. The extinction includes the foreground Galactic extinction $A_V = 1.50 \pm 0.05$ mag, estimated from the spectrum of a nearby interstellar extinction probe star \citep{1997ApJ...482L..37O}, which is greater than the value of $A_V = 1.03$ mag given by the NED extinction calculator \citep{2011ApJ...737..103S}.
The differential extinction increases up to 0.5 mag  ($A_V = 25$) at the locations of some H$_2$ knots. The H$_2$ knots associated with elevated extinction must therefore be dusty and located in front of the bulk of the NLR. In comparison, \cite{2021MNRAS.506.2950R} found a mean NLR extinction $A_V = 12.5\pm 6.3$ mag and extinction values up to $A_V = 20$ along the NE and NW edges of the ionization cone. While the range of extinction values is similar, our extinction map looks qualitatively different from theirs, perhaps because they use Gauss--Hermite moments to measure emission line flux, while we integrated the line flux. Unlike our map, there is no peak in extinction at the nucleus. In a $0\farcs3 \times 0\farcs3$ box surrounding the nucleus, we find $E$(Pa $\alpha$ - Br $\gamma$) $= 0.57$ mag ($A_V =26.2$ mag), compared to the value in their map of $A_V \sim 14$ mag at the nucleus.  Past measurements of nuclear extinction yield a wide range of values ($A_V = 1.2$--$9.6$ mag) depending on line wavelength \cite{2000MNRAS.318.1232W}. The visible measurements yield selectively lower extinction than our NIR measurements because they are flux-averaged and therefore probe line emission from regions of lower extinction. Our JWST measurement also naturally yields a higher extinction value than ground based measurements because of improved spatial resolution.

\section{Ionized Gas Kinematics}

\begin{figure*}[ht]
 \includegraphics[trim=0.0cm 0.0cm 0.0cm 0.0cm, clip, width=\linewidth]{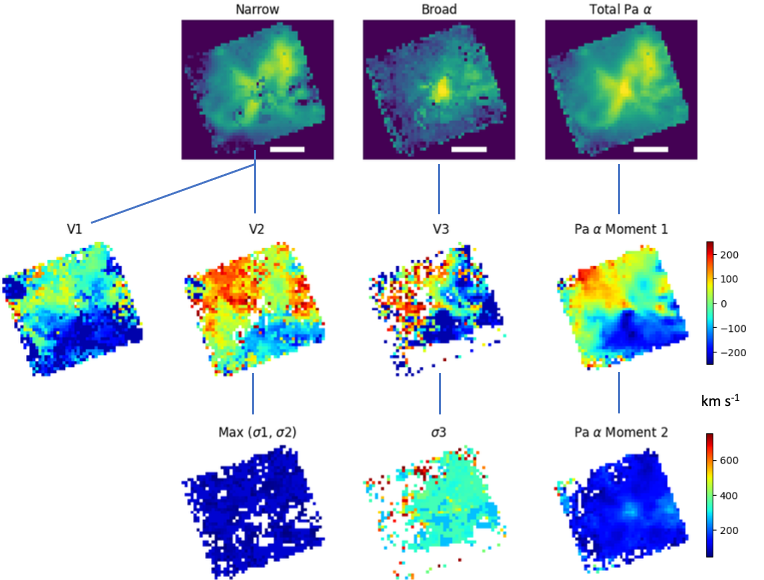}
 \caption{Gas velocity component map for emission lines in the 1.7--2.3 $\mu$m region of the NIRSpec G235H spectrum. The velocity map for each component (V1, V2, V3) is compared to the corresponding Pa $\alpha$ moment maps. Blue and red narrow velocity components V1 and V2 follow an overall rotational pattern aligned with the radio jet. They have widths of $\sigma = 70$--$100$ km s$^{-1}$, while broad velocity component V3 has $\sigma \sim 350$ km s$^{-1}$. The 3-component velocity map characterizes the multiple velocity components in the extended NLR, while the moment maps present a flux-weighted average of these components. }
 \label{fig:velocity_maps}
\end{figure*}

\cite{2003MNRAS.342..861T} measured kinematics from Keck long-slit observations of Pa$\alpha$ and H$_2$ across the nucleus at PA $= 0\arcdeg$, yielding a systemic redshift of 0.0559. Their aperture was not exactly perpendicular to the radio axis and this redshift estimate may be affected by asymmetric outflow. Alternatively, \cite{2022ApJ...937..106C} adopted the H {\sc i} 21 cm absorption redshift $z = 0.05634$ as zero velocity. Rather than using either of these values, we measured gas kinematics relative to the mean redshift (0.0557 $\pm$ 0.0001) of CO absorption bands arising from old stellar populations in the NIRSpec FOV, different by  -60 km s$^{-1}$ from the value of \cite{2003MNRAS.342..861T}. This difference is significant given the $\pm 10$ km s$^{-1}$ accuracy of the JWST NIRSpec wavelength calibration and the $\pm 8$ km s$^{-1}$ accuracy of the \cite{2003MNRAS.342..861T} observations.

The emission lines were masked and the NIRSpec spectral cubes were fit in {\tt pPXF} with the extended MILES (EMILES) galaxy templates \citep{2016MNRAS.463.3409V} plus a polynomial AGN continuum component to establish the continuum level. We then fit 3 Gaussian emission line components (V1, V2, V3)  to each emission line, with constraints on velocity and velocity dispersion. Blue and red narrow velocity components (V1 and V2) with $\sigma \sim 80$ km s$^{-1}$ are needed to accommodate line splitting between the near and far sides of the NLR and a third, broad component (V3) with $\sigma \sim 350$ km s$^{-1}$ to fit the large line widths in the NLR interior. The three velocity components are constrained to be within $\pm 250$ km s$^{-1}$ of systemic velocity in order to minimize cross-contamination among neighboring emission lines. Unmodeled, higher velocity components are present, but represent a small fraction of the total line flux. The velocity components were simultaneously fit to create emission line maps for every emission line in the $1.6$--$2.3 ~\mu$m region of the spectrum. Example spectral fits and residuals are shown in the Appendix. The resulting velocity and velocity dispersion maps are presented in Fig.~\ref{fig:velocity_maps} along with the narrow and broad component flux maps (F1+F2, F3) of Pa $\alpha$.  

\begin{figure*}[ht]
 \includegraphics[trim=0.0cm 0.0cm 0.0cm 0.0cm, clip, width=0.8\linewidth]{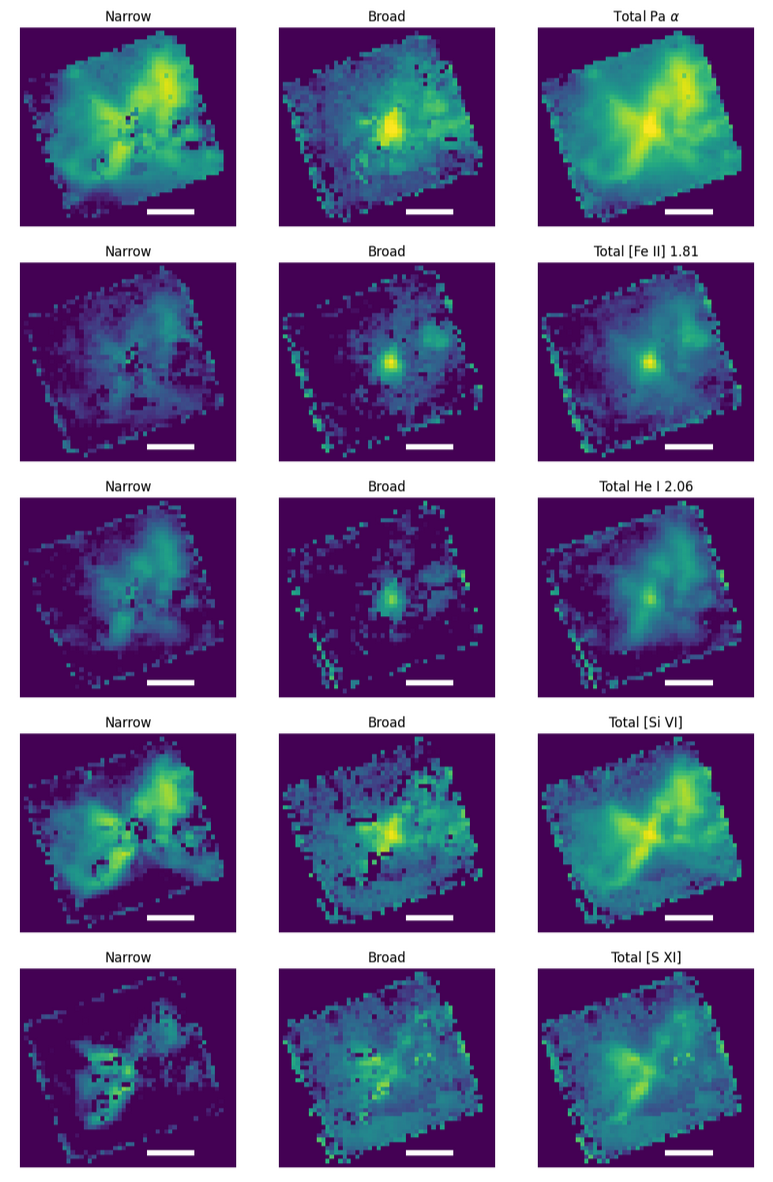}
 \caption{Emission line flux maps for narrow (V1 + V2) and broad (V3) velocity components, ordered by increasing ionization potential. Scale bar is $1\farcs0 = 1.09$ kpc}
 \label{fig:emission_line_flux_maps}
\end{figure*}

\begin{figure*}[ht]
 \includegraphics[trim=0.0cm 0.0cm 0.0cm 0.0cm, clip, width=\linewidth]{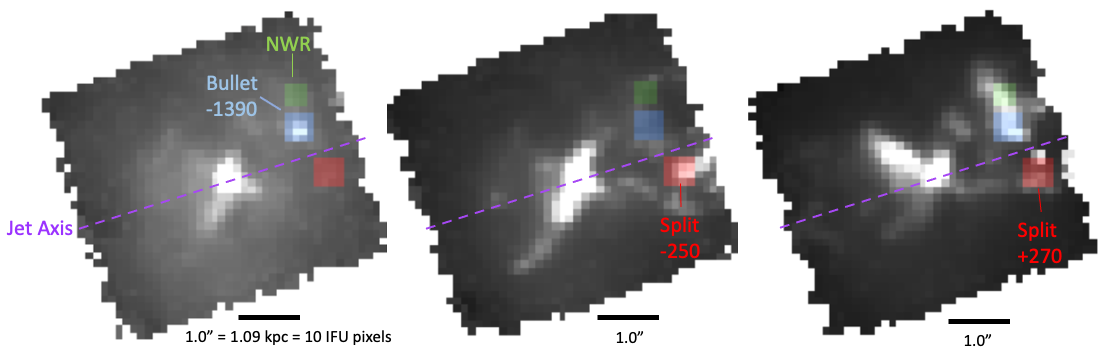}
 \includegraphics[trim=0.0cm 0.0cm 0.0cm 0.0cm, clip, width=\linewidth]{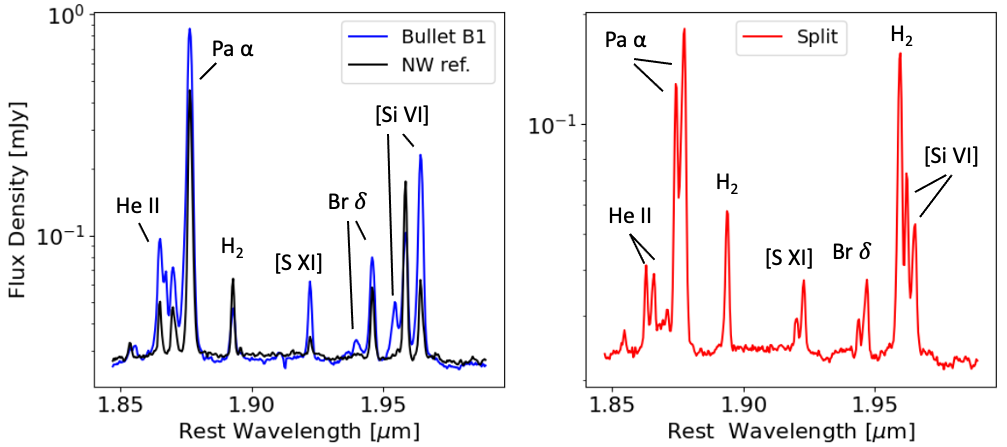}
 \caption{High velocity outflows. Top: [Si {\sc vi}] channel maps at Bullet B1 peak velocity ($v = -1390$ km s$^{-1}$), blue peak of W Split (-250 km s$^{-1}$), and red peak of W Split ($+270$ km s$^{-1}$). Extraction regions for Bullet B1 (blue), NW reference spectrum (green), and W Split (red) are indicated. Bottom Left: Bullet B1 is seen in the very broad, blue-shifted velocity component in Br $\delta$ and [Si {\sc vi}] 1.963 $\mu$m. It is not seen in the H$_2$ lines nor the very high ionization [S {\sc xi}] line. Pa$\alpha$ emission from B1 is blended with He {\sc ii} emission.  Bottom right: Example of line splitting from outflow in the NW cone at the W Split extraction region. Ionized lines are split by outflow while H$_2$ lines are not.}
 \label{fig:high_velocity_outflows}
\end{figure*}

\begin{figure*}
  \includegraphics[width=\linewidth]{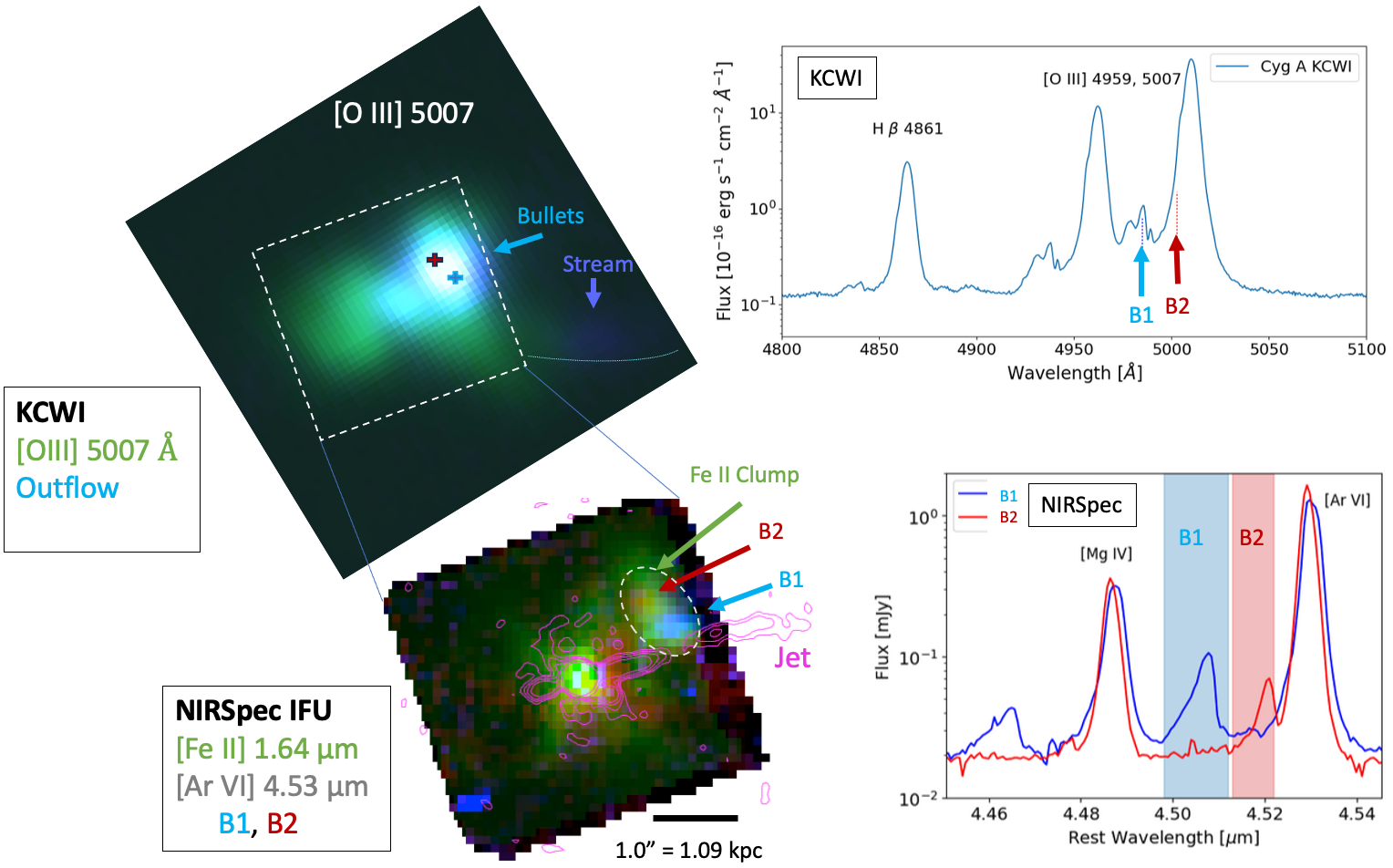}
 \caption{Comparison of KCWI and NIRSpec IFU data. Top left: [O {\sc iii}] 5007 \text{\AA} line maps in two velocity ranges (blue: $-1700 < v_r < -1200$ km s$^{-1}$, green: $-700< v_r < 500$ km s$^{-1}$), with much harder stretch on the blue, high velocity outflow component. Top right: KCWI optical spectrum integrated over the Fe {\sc ii} Clump region. H$\beta$ and [O {\sc iii}] 4959, 5007 \text{\AA} lines show triple-peaked velocity structure in all three lines from bullet components B1 and B2 plus diffuse high-velocity outflow extended across the Fe {\sc ii} Clump. Bottom left: [Ar {\sc vi}] 4.530 $\mu$m emission from high velocity bullets B1 and B2 appears to be associated with the Fe {\sc ii} Clump in the NW cone.   Bottom right: $0\farcs3 \times 0\farcs3$ spectral extractions centered on bullets B1 and B2 show distinct satellite peaks from localized high velocity outflows. The highest velocities, up to 2000 km s$^{-1}$, are found in bullet B1, which lies close to the projected radio axis.} 
 \label{fig:KCWI_NIRSPEC_comparison}
\end{figure*}

 The overall kinematics of the NLR display a combination of rotation and outflow (Fig.~\ref{fig:velocity_maps}). The range of velocity dispersion in the narrow components is found to be $70$--$100$ km s$^{-1}$, while the broad component has a velocity dispersion of $300$--$500$ km s$^{-1}$.  These two spatially and kinematically distinct velocity systems in the biconical NLR may correspond to low dispersion gas along the outer surface of the bicone and outflow along the inner surface driven by the AGN or radio jet. While it has been assumed in the past that the overall rotation pattern reflects the rotation of a gas disk in the galaxy potential \citep{2003MNRAS.342..861T, 2021MNRAS.506.2950R}, this hypothesis is incongruent with the biconical structure of the NLR. We suggest that the close alignment of the rotation axis with the projected radio jet axis supports our new idea that the biconical NLR is rotating around its own axis. It seems unlikely that the X-shaped emission we see comes from the intersection of the AGN ionization cone and a disk with nearly perpendicular yet kinematically aligned axes. Purely gravitational orbits of gas in the two structures would intersect each other, a dynamically untenable situation. It would also be difficult to keep the cone evacuated with disk orbits passing through it and the jet would likely be deflected out of the plane of the disk. 
 
The combined narrow (V1 + V2)  and broad (V3) velocity components are found in differing amounts and varying morphology, depending on ionization level (Fig.~\ref{fig:emission_line_flux_maps}). For example, [Fe {\sc ii}] emission ($\chi = 16.2$ eV ionization potential) from the broad velocity component is concentrated in the nucleus and the Fe {\sc ii} Clump near the projected radio axis, but there is only weak [Fe {\sc ii}] emission from the narrow  components along the edges of the NLR.  The relatively low ionization He {\sc i} 2.06 $\mu$m line ($\chi = 24.6$ eV) follows roughly the same morpho-kinematics described above for Pa $\alpha$ ($\chi = 13.6$ eV), equally distributed between the narrow components at the edges of the bicone and broad component in the interior of the bicone. The high-ionization [Si VI] 1.963 $\mu$m ($\chi = 205$ eV) and [S {\sc xi}] 1.920 $\mu$m lines ($\chi = 505$ eV) are edge-brightened, following distinct bicones with narrower opening angles than Pa$\alpha$.

Several lines of sight through the NLR show multiple kinematic components, further demonstrating the complexity of the velocity field. In particular, line splitting is seen near the NW bicone axis, with a velocity difference of 520~km~s$^{-1}$ (W Split; Fig.~\ref{fig:high_velocity_outflows}) indicating fast outflow. The H$_2$ emission lines from this region do not show line splitting, and their velocity falls roughly halfway between the red- and blue-shifted ionized gas emission velocity components. Line splitting is also seen in other locations in the ionization cone. A likely explanation is that we see outflow components from both the near and far sides of the ionization cone projected along the same line of sight.   The spotty distribution of velocity splitting along the bicone axis, together with the appearance of elongated structures in the channel maps (Fig.~\ref{fig:high_velocity_outflows}) suggests that outflowing material is filamentary and occupies a small fraction of the NLR bicone.

High velocity outflows appear in various locations in the NLR. The most extraordinary examples, which we call the Cyg A Bullets (B1 \& B2), appear as emission line knots associated with the Fe {\sc ii} Clump (Figs.~\ref{fig:high_velocity_outflows} \& ~\ref{fig:KCWI_NIRSPEC_comparison}). The higher velocity  bullet (B1) appears as satellite velocity peak at -1380 km s$^{-1}$, with a tail that extends to $\sim -2000$ km s$^{-1}$. The lower velocity bullet (B2) peaks at -630 km s$^{-1}$. The bullets appear in H, He, intermediate ionization, and coronal emission lines in both NIRSpec and MIRI wavebands (Table 1, Fig.~\ref{fig:nirspec_extraction}, and  Fig.~\ref{fig:miri_extraction}).  However, they are not seen in H$_2$ molecular gas, nor the highest ionization lines, including [S {\sc xi}], [Si {\sc ix}], and [Al {\sc ix}].  Bullet B1 was first noticed in long-slit optical spectra of the [O {\sc iii}] 5007 \text{\AA} line \citep{1991MNRAS.251P..46T}, while B2 is a new find. Our NIRspec observations resolve the bullets from the Fe {\sc ii} clump for the first time. B1 and B2 are found at distances of 1.3 kpc and 1.2 kpc from the AGN, respectively. B1 appears to be adjacent to a gap or lower surface brightness segment of the radio jet, raising the possibility that the jet has been disrupted at this location by its interaction with the Fe {\sc ii} Clump. In addition to driving outflows, such an interaction might lead to the entrainment of gas by the radio jet \citep{Rosen1999,Wang2011,Rossi2020}. 

Our KCWI observations reveal details of the optical line emission from the Cyg A Bullets (Fig.~\ref{fig:KCWI_NIRSPEC_comparison}). We identify their kinematic signature  in many optical lines, including the [O {\sc iii}] 5007, 4959 \text{\AA} doublet and H$\beta$. Additional high velocity outflow is seen along the full extent of the Fe {\sc ii} Clump in our KCWI [O {\sc iii}] map, confirming that the entire clump is an active site for launching outflows. (We also note a faint streamer of high velocity outflow to the west of the region covered by the JWST observations.) The Cyg A Bullets may be related to the high-velocity associated absorbers found in the spectra of some radio-loud quasars \citep{1986ApJ...307..504F,1993ApJ...414..506A,2000ApJ...536..101H}. Such absorption features are often detached, at negative velocities, and have a high-velocity tail, but they are not as broad as the troughs found in radio-quiet BALQSOs. An even faster, surrounding wind may ablate the bullets, creating a high velocity cometary tail of gas. In fact, we do find evidence that Bullet B1 is extended in an E-W direction, with the highest velocity tail of gas extended to the west.

In summary, the extended NLR in Cyg A appears to be both rotating around the radio jet axis and outflowing from the AGN. Multiple velocity components are present across the rotating NLR bicone. Low velocity dispersion is found preferentially along the bicone edges and high dispersion closer to the bicone axis. Very high velocity outflows are associated with the Fe {\sc ii} Clump, which appears to be subject to both jet-ISM interaction and intense radiation, launching bullets of intermediate ionization and coronal gas that attain velocities as great as of $2000$ km s$^{-1}$. 

\begin{figure*}[ht]
  \includegraphics[trim=0.0cm 0.0cm 0.0cm 0.0cm, clip, width=\linewidth]{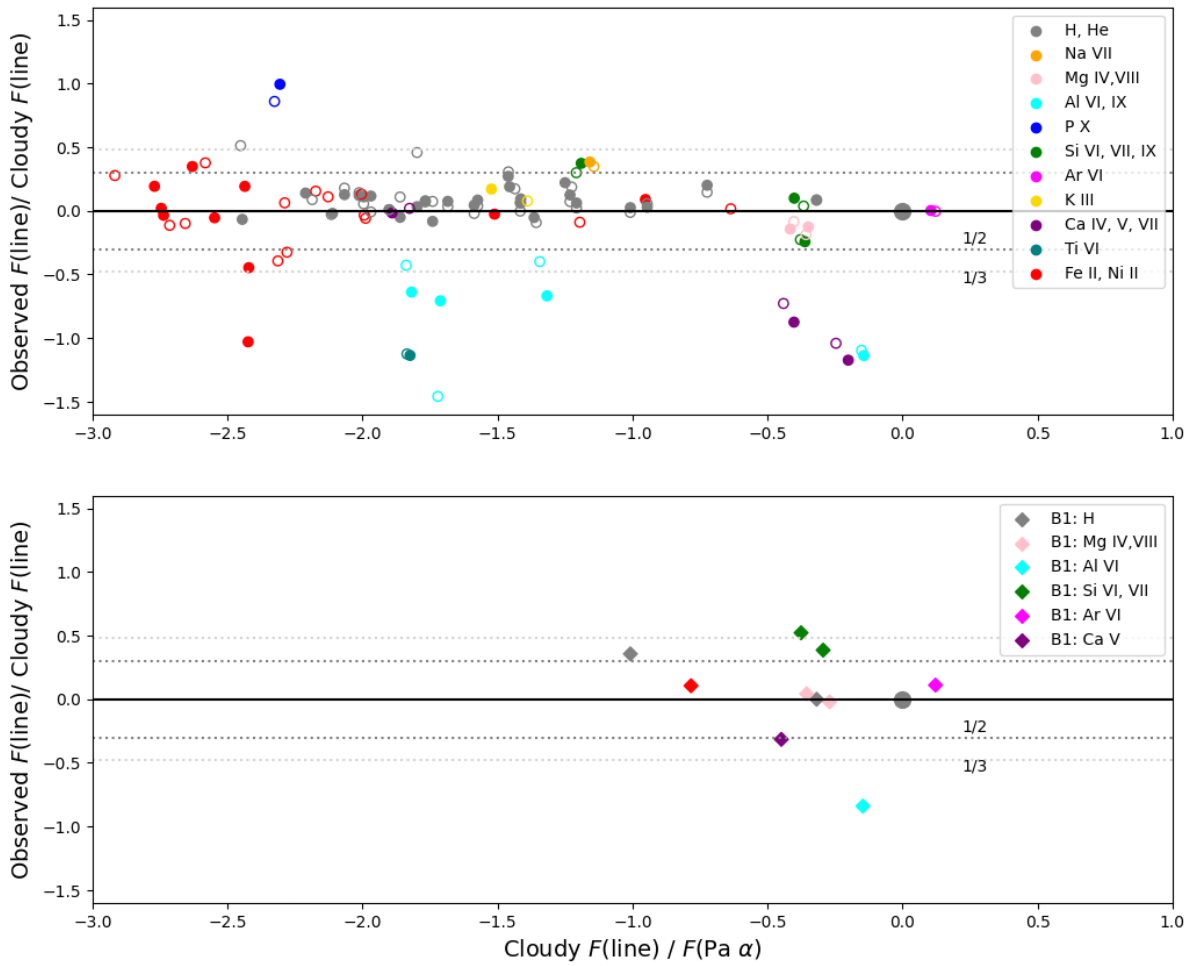}
 \caption{Observed line flux relative to Pa $\alpha$ compared to predictions of multi-component Cloudy AGN photoionization models (Table~\ref{Table1}). Top: NWR (solid circles) and Fe {\sc ii} Clump (empty circles). The models reproduce most H, He, [Fe {\sc ii}], and coronal line ratios to within a factor of 2 or better (inner set of dotted lines). Bright Al and Ca coronal lines appear to be overpredicted by the model. Bottom: Similar comparison to photoionization model for Cyg A Bullet B1.} 
 \label{fig:cloudy_model}
\end{figure*}

\section{Discussion}

\subsection{NLR Photoionization Models}

We used Cloudy C23.01 to model the NWR, Fe {\sc ii} Clump, and Cyg A Bullet B1 regions of the NLR (Tables 1, 2 \& Fig.~\ref{fig:cloudy_model}). A \cite{1987ApJ...323..456M} AGN SED with input bolometric luminosity of $\log L [\mathrm{erg ~s}^{-1}] = 45.5$ is incident on a spherical shell at the projected radius of each region. We employed Solar elemental abundances and no depletion onto dust, under the preliminary assumption that dust is destroyed in the highly ionized NLR clouds that contribute to the observed spectra. Five density components were required to model each region ($\log n_\mathrm{H}[\mathrm{cm}^{-3}] = 0.0, 1.0, 1.5, 2.0$, and 3.0). Most of the Pa $\alpha$ flux comes from the $\log n_\mathrm{H} = 1.5$ component; most of the  Fe {\sc ii} line flux from the $\log n_\mathrm{H} = 3$ component; and most of the coronal line emission from the two low density components. The relative normalizations for each model component were initially optimized to minimize the overall chi-squared in Fig. 11, giving equal weight to all well-detected emission lines.  After finding the global minimum, care was taken to separately minimize chi-squared for each ion.  Our Cloudy photoionization models match the observed H, He, [Fe {\sc ii}], intermediate ionization, and many of the coronal line fluxes in the NIRSpec spectra quite well (Fig.~\ref{fig:cloudy_model}). While the model for Bullet B1 is less well constrained by fewer detected lines, its ionization distribution is very similar to the Fe {\sc ii} clump. Several lines, including  [Al {\sc vi, ix}], [Ca {\sc iv, v}], and [Ti {\sc vi}]  are grossly overpredicted in both the NWR and Fe {\sc ii} Clump.  The over-prediction of these lines may indicate an under-abundance of the corresponding ions in the Cyg A NLR, perhaps due to residual dust depletion. 

Gas masses for each model component (Table 2) were estimated from the observed, spatially-resolved Pa $\alpha$ flux and the model hydrogen column density and model line luminosity for each region, following Equation 2 of \cite{2022ApJ...930...14R}. The total mass of ionized gas in the Fe {\sc ii} clump is $1.1\times 10^8 M_\odot$ and its outflow rate is $40 M_\odot$ yr$^{-1}$ for an estimated outflow velocity of 150 km s$^{-1}$ (see \S7.3). While the mass in the CygA Bullet B1 is only $4\times10^6 M_\odot$, its outflow rate is an outsized $13 M_\odot$ yr$^{-1}$ over its resolved length of 0.4 kpc, by virtue of its very high velocity. Almost all (97\%) of the mass in the Fe {\sc ii} clump resides in the low to intermediate density model components. The masses of the high density components are much lower, supporting a picture where small, dense knots are embedded in less-dense, more highly ionized wind. Compared to the NWR, the Fe {\sc ii} Clump has 5 times more mass in its highest density component relative to its total mass. This relatively small mass of additional dense photoionized gas appears to account for the enhanced [Fe {\sc ii}] emission from the Fe {\sc ii} Clump, without any need for shock heating. Strong [Fe {\sc ii}] emission is also seen in Seyfert galaxy outflows such as the one in NGC1068, and may be attributed to photoionization in dust-depleted dense gas at the boundary layer of molecular clouds \citep{1994ApJ...421...92B}.  The large range in gas pressure indicated by our multicomponent photoionization model (Table 2) rules out a two-phase isobaric NLR \citep{1981ApJ...249..422K}. Radiation pressure-dominated models may provide a solution, creating a density gradient on the inward faces of clouds \citep{2002ApJ...572..753D, 2019MNRAS.485..416B}. Complex density fields are also produced by jet-driven winds in multiphase jet-ISM hydrodynamical simulations \citep{2018MNRAS.479.5544M}. 

\subsection{Acceleration of Fast NLR Outflows}

\begin{figure*}
  \includegraphics[trim=0.0cm 0.0cm 0.0cm 0.0cm, clip, width=0.9\linewidth] {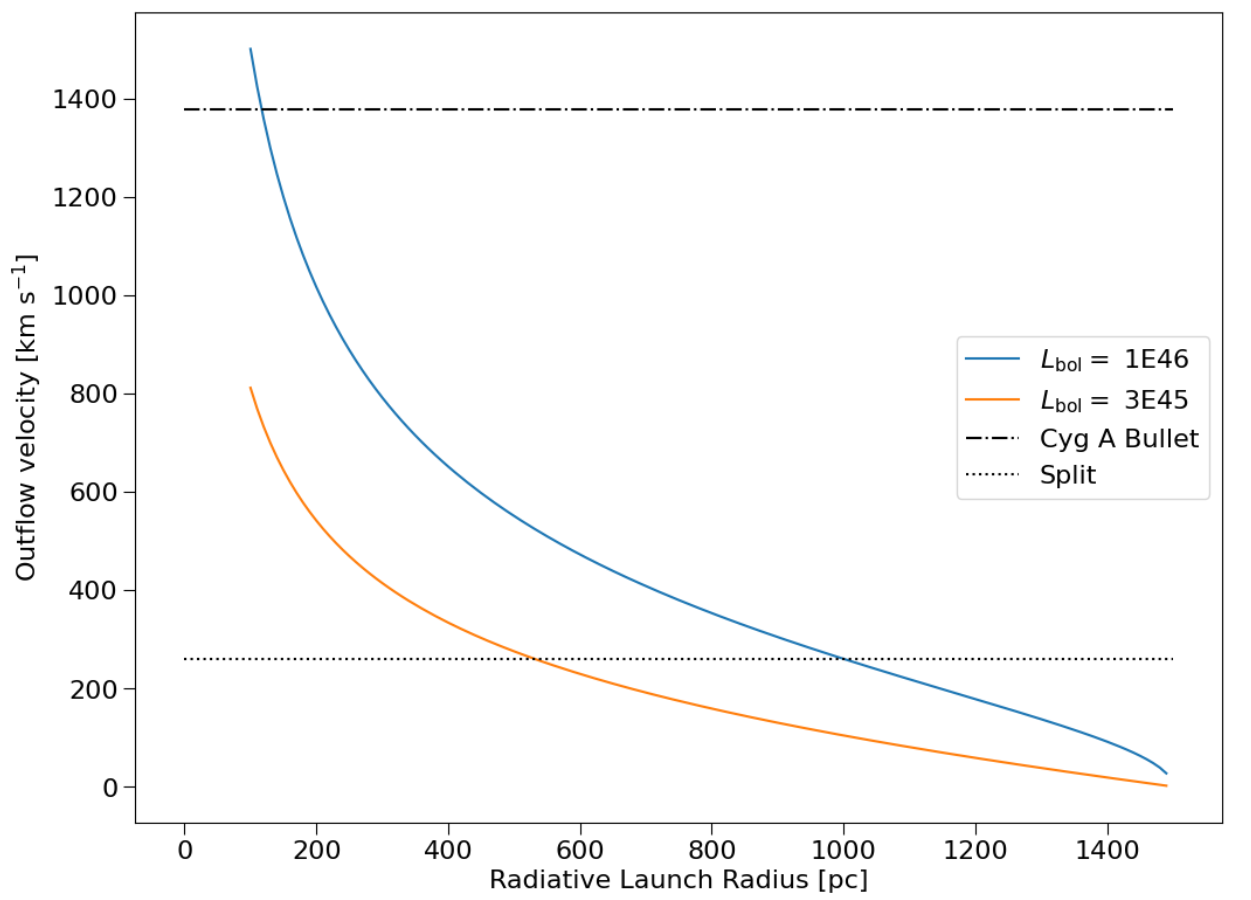}
   \caption{Outflow velocity of an NLR cloud observed at 1.5 kpc that was launched by radiation pressure at a smaller radius and slowed by gravity. The outflow velocity attained by a cloud launched at radius $R_\mathrm{launch}$ was computed assuming a constant density galaxy stellar core with mass of $2\times 10^{10} M_\odot$ within 2 kpc. The upper and lower curves allow for a factor of 3 variability in AGN bolometric luminosity. The Cyg A Bullets could attain the observed velocity with radiative driving only if launched very close to the nucleus at a radius of $< 120$ pc (comparable to the size of a NIRSpec IFU pixel), which seems unlikely. The radial velocity of the W Split feature and other, lower velocity outflows could plausibly be consistent with radiative driving.}
   \label{fig:outflow_velvsradius}
\end{figure*}

Outflows are commonly found in the extended NLRs of nearby Seyfert galaxies, radio galaxies, and distant quasars \citep{2022MNRAS.510..639B, 2024ApJ...965..103B, 2024ApJ...960..126V}.  Observations of Seyferts often show a misalignment of kinematical axes between the inner region of the NLR where outflows are prevalent, and outer regions that are in the ionization cone but follow the rotation of the galaxy disk \citep{2024arXiv240909771Z}. The kinematics of the outflow-dominated regions typically appear to be consistent with radial outflow. One exception is the pc-scale rotating molecular outflow reported in ESO 320-G030 and attributed to a magnetically driven wind \citep{2024A&A...684L..11G}. However, Cygnus A is unique so far in having an ionized NLR that rotates around its axis at kpc scale. It may be that only powerful radio jets can cause the extended NLR to spin about the radio axis. The well-defined, edge-brightened, apparently biconical structure of the Cyg A NLR may be a result of the effectiveness of its powerful radio jet in clearing out the center of the NLR. As discussed above, we consider it unlikely that the NLR lies in a flat disk illuminated by the AGN ionization cone, because of its distinct geometry and kinematics.

Radiative acceleration and jet-ISM interactions are the two mechanisms recognized to accelerate ionized gas and produce the outflows observed in Seyferts and radio galaxies. \cite{2023ApJ...943...98M} successfully model the outflows in nearby Seyferts with a radiative acceleration plus gravitational deceleration model, matching the observed maximum velocities as a function of radius. The enclosed mass of the Cyg A host galaxy inside 2 kpc, as estimated from the CO 2-1 velocity field, is $\sim 2\times 10^{10} M_\odot$ \citep{2022ApJ...937..106C}. Assuming a force-multiplier value of $M=500$ and AGN bolometric luminosity of $3\times 10^{45}$ erg s$^{-1}$ \citep{2012ApJ...747...46P}, we find that outflows launched in the kpc-scale Cyg A NLR do experience a net radiative acceleration outward. We plot the radial outflow velocity that could be attained by radiation pressure vs. launch radius $R_\mathrm{launch}$ for two different AGN luminosities, the observed luminosity and one that is brighter by a factor of 3.3, to allow for the possibility that the AGN may have been significantly brighter in the past (Fig.~\ref{fig:outflow_velvsradius}). Within this luminosity range, it is possible to radiatively launch an outflow at a distance as far as $0.5$--$1$ kpc from the nucleus and attain a radial velocity of $250$ km s$^{-1}$, as observed in the Cyg A NLR at the W split, 1.5 kpc from the nucleus (Fig.~\ref{fig:high_velocity_outflows}). Such radiation-driven outflows could potentially be launched across most of the region observed by NIRSpec. However, the 1400 km s$^{-1}$ velocity of the Cyg A Bullet (B1) could only be achieved by radiation pressure if it were launched at $R_\mathrm{launch} < 120$ pc. If that were the case, the bullet would have had to travel the distance of 1.3 kpc from the AGN ballistically, without running into anything, all the while staying confined inside a small knot. Furthermore, the maximum velocity of 2000 km s$^{-1}$ observed in the blue tail of the B1 line profile would have to be launched at an even smaller radius if driven by radiation pressure. Moreover, the association of both bullets B1 and B2 with the high-velocity dispersion Fe {\sc ii} Clump suggests they may have been launched from that location.

\begin{figure*}
  \includegraphics[trim=0.0cm 0.0cm 0.0cm 0.0cm, clip, width=\linewidth] {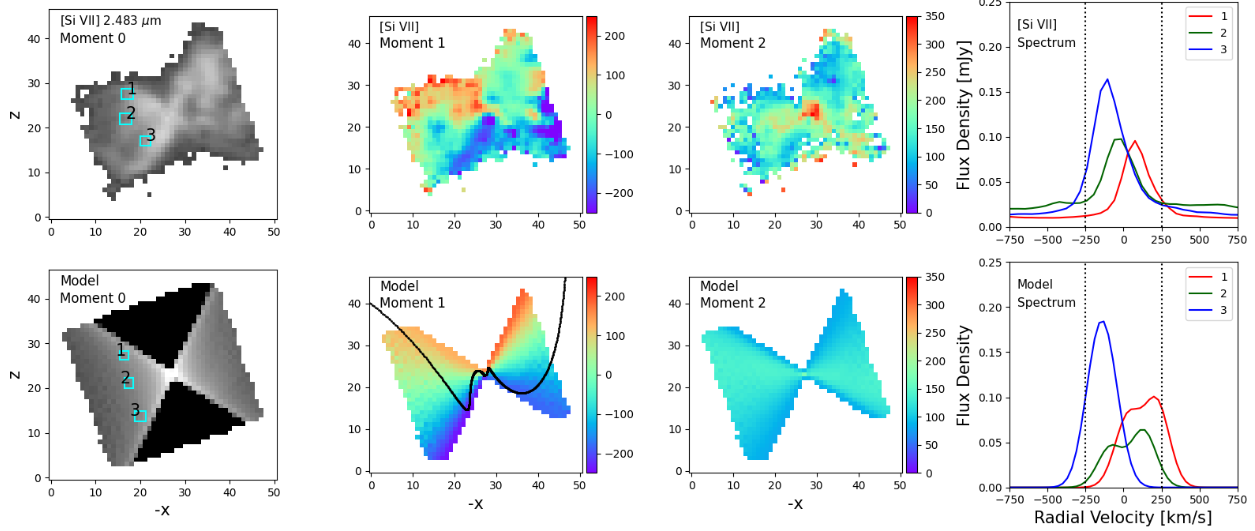}

 \caption{[Si {\sc vii}] 2.483 $\mu$m moment maps and spectral extractions compared to spiral outflow model. The moment 1 and moment 2 maps trace velocity and velocity dispersion in km s$^{-1}$. A model with constant velocity (250 km s$^{-1}$) outflow tangent to logarithmic conical spiral flow lines and an intrinsic velocity dispersion of 80 km s$^{-1}$ yields a similar velocity field to the observations. A representative spiral flow line is overplotted on the model moment 1 map. Line profiles are extracted from three different regions of the observations and the corresponding locations in the simulation.  The combination of rotation and radial outflow leads to an asymmetric velocity pattern, where the projected velocity is highest in the NW limb (receding) and SE limb (approaching). The velocity dispersion is lowest along the edges of the rotating cone where the rotation signal is coherent, and highest near the axis, where the radial outflow component causes line splitting. Detailed structure in the observed moment maps and spectra that deviates from the model appears to be caused in part by bright clouds that skew the velocity and lower the velocity dispersion in discrete patches. The nucleus also displays much higher velocity dispersion than the model.} 
 \label{fig:helical_modeling}
\end{figure*}

 The presence of $600$--$2000$ km s$^{-1}$ outflows at $>1$ kpc indicates that radiation pressure is not enough and additional forces are at work, most likely related to the radio jet. High velocity, jet-driven outflows are seen in other powerful radio galaxies by JWST, including 3C 326, 3C 293, and IC 5063 \citep{2024A&A...689A.314L, 2024ApJ...974..127C, 2024arXiv240603218D}. Hydrodynamical simulations of jet-ISM interactions can indeed produce such high velocity outflows \citep{1997ApJ...491L..73S, 2018MNRAS.479.5544M}.  In these simulations, clouds within the radio jet cocoon are subject to compression, ablation, and acceleration by the jet, resulting in enhanced NLR emission and jet-driven outflows close to the radio jet axis. While most of the jet power is currently being deposited in the intercluster medium at the Cyg A radio hot spots, the presence of jet-driven outflows and rotation in the NLR at kpc scale indicates that the jet still has a significant impact on the interstellar medium. As shown below, only a small fraction of the jet power is needed to drive the observed outflows. This raises the possibility that jet feedback on the ISM can persist as long as a jet is active, long after it breaks out of the host galaxy ISM. 
 
 \subsection{NLR Spiral Outflow Kinematic Model}

In order to better understand the observed complex kinematics described above, consisting of both rotation and outflow, we model the Cyg A NLR by a thin, hollow bicone with spiral outflow (Fig.~\ref{fig:helical_modeling}). While a variety of different functional forms might be used to represent a conical spiral, a logarithmic conical spiral naturally yields a continuous,  self-similar curve, with constant-velocity $v$ over any range of radii. {\it This mathematical model is intended to describe the observed velocity field, but is not intended to be a physical model of the outflow.} As discussed below, the model raises some unresolved questions regarding the acceleration and angular momentum of the NLR. Regardless, it conveniently combines both rotation around the cone axis and radial outflow from the center, with the relative contribution of each, together with the cone opening angle, determining the slope of the spiral. Intrinsic velocity dispersion ($\sigma_v$) is included as an extra parameter to match the observed line widths at the edge of the bicone. Logarithmic conical spiral flow lines are characterized by the cone half-opening angle $\Theta$ and constant velocity components $(v_R,v_{\phi})$:

\begin{equation} 
R = R_0 e^{k (\phi - \phi_0)}, z = R cos \Theta, k = (v_R /v_\phi)  sin \Theta
\end{equation}

with initial coordinates $(\phi,R,z) = (\phi_0,R_0,z_0)$ at time $t = 0$.

The cone half-opening angle  is estimated from the projected half-opening angle of the NLR, as measured from, e.g., the  [Si {\sc vii}]  emission line map  (see Fig. 13: $\Theta = 46\arcdeg$). The inclination of the cone is determined by assuming that the bicone is aligned with the radio jet. VLBI observations of Doppler boosting and apparent motion of radio jet components indicate that the pc-scale jet is inclined by $45\arcdeg < i \lesssim 74.5\arcdeg$ to the line of sight \citep{1995PNAS...9211371B, 2016A&A...585A..33B, 2016A&A...588L...9B}, with the WNW side approaching. The direction of rotation is selected to match the observed velocity field, with the angular momentum vector pointing ESE along the radio axis. We sample the model at 600 discrete points along each of 200 spiral flow lines in order to map the velocity field over the entire surface of the bicone. We populate a mock datacube matching the NIRSpec IFU spaxels by adding a Gaussian emission line component at the projected radial velocity from each model sample that falls within it. We then extract emission line moment maps and spectra from both the observations and the mock data cube, in the same $2\times 2$ pixel apertures, using {\sc jdaviz}.  

We compare our spiral outflow model to the observed moment maps of the [Si {\sc vii}] 2.483 $\mu$m line (Fig.~\ref{fig:helical_modeling}), selected for analysis because it is an isolated high-ionization line, relatively free of blending with the many H$_2$ lines in the spectrum. The observed velocity field is best matched by $(v_R,v_{\phi}) = (150, 200$) km s$^{-1}$ along the spiral flow lines and $\sigma_v = 80$ km s$^{-1}$.  Assuming that all of the mass in the Fe {\sc ii} Clump is participating in the spiral outflow, we estimate a mass outflow rate of $40 M_\odot$ yr$^{-1}$ the NW cone. There are no comparably bright clouds in the NLR at this distance from the AGN, so this gives a fair estimate of the total mass outflow rate at a radius of 1.2 kpc.

The model flux map shows edge-brightened limbs, similar to the observations. The model velocity map matches the observed velocity map in several respects, with maximal projected velocity along the NE and SW limbs of the bicone. The symmetry of the velocity field is broken by the rotation of the cone, such that the NW limb displays the lowest projected velocity.  As a consequence of the radial velocity component of the spiral outflow model, line splitting occurs and the velocity dispersion increases by $50$--$150$ km s$^{-1}$ SE and NW of the bicone axis. The model line profiles extracted away from the limbs of the cone therefore show a double peak, split between radial outflow from the front and back sides of the cone. This characteristic signature of radial outflow is seen in several locations of the observed data cube, and in the asymmetric line profiles and shoulders in Fig.~\ref{fig:helical_modeling}.

While a spiral outflow model does match several aspects of the data, there are also large deviations. There are scattered patches of emission observed near the axis of the cone that are significantly blue-shifted or redshifted, possibly associated with localized outflow components accelerated by the jet. Excess velocity dispersion is particularly apparent in channel-like regions of low flux where fainter, broader emission line components are not outshined by brighter, lower velocity dispersion components. The observed velocity dispersion in these patches ranges from  $150$--$350$ km s$^{-1}$, significantly larger than the velocity dispersion encompassed by the model. At the W Split extraction region in particular, separate blue and red streamers are seen in the channel maps at $-270, +250$ km s$^{-1}$ (Fig.~\ref{fig:high_velocity_outflows}). Such spots of high velocity, high velocity dispersion and larger-than-expected line splitting near the projected axis are clear signatures of additional, fast radial outflow components not included in the model. Even higher velocity ($600$--$2000$ km s$^{-1}$) outflows such as the Cyg A Bullets are found at low flux levels and therefore do not show up in the integrated moment maps.

\subsection{Possible Origins of NLR Rotation}

Spiral biconical outflow challenges our understanding of NLR dynamics. We consider two possibilities here: (1) The rotation of the NLR derives directly from the AGN or jet, or (2) that it is residual rotation from a disk or inflowing molecular gas that was disrupted by the AGN or jet.

\textit{(1) AGN or jet-driven NLR rotation}: The energy requirement of jet-driven spiral outflow is far from prohibitive. The power carried by the spiral outflow is only  $P_\mathrm{out} = M (v^2 + 3 \sigma_v^2) /t_\mathrm{out}  = 3.3 \times 10^{41} M_8/R$ erg s$^{-1}$ for a an NLR mass of $ M_8 = M/10^8 M_\odot$, over an outflow timescale of $ t_\mathrm{out} = R/v_r = 10 $ Myr. This would require only a very small fraction ($\sim1.0 \times 10^{-4}$) of the jet or AGN power to drive the outflow.  A magnetized disk or torus can potentially launch a wind along its rotation axis via magnetic torques \citep{1982MNRAS.199..883B, 1994ApJ...434..446K}. While this might explain high velocity outflows near the nucleus, it is unclear how such a mechanism would operate 1 kpc from the AGN. The rotation of the NLR about the radio axis may instead indicate that the radio jet-ISM interaction is transfering angular momentum to the ISM along its full length.

Clouds launched close to the AGN that followed the entire length of a spiral flow line from the nucleus at $< 0.2$ kpc to the edge of our map at 1.5 kpc would be expected to decelerate with a $1/r$ dependence if conserving angular momentum ($L = M_c r v_\phi = \mathrm{constant}$.) Instead, a constant velocity model appears to better describe the overall velocity field. This implies that the angular momentum increases with radius, assuming the clouds are not losing significant mass. The angular momentum of the biconical NLR inside a radius of 1.5 kpc is $L_\mathrm{NLR} = 2 M R v_\phi/3 = 1.2 \times 10^{70} M_8$ g cm$^2$ s$^{-1}$, for a rotation velocity of $v_\phi = 200$ km s$^{-1}$. Angular momentum may be transferred from the SMBH or accretion disk in the nucleus, the torus, or the molecular disk further out.  A maximally spinning supermassive black hole of mass $M_\mathrm{BH} = 2.5 \times 10^9 M_\odot$ would have angular momentum $J = G M_\mathrm{BH}^2 / c  = 5.5 \times 10^{67}$ g cm$^2$ s$^{-1}$ \citep{2003MNRAS.342..861T}, only $\sim0.5\%$ of the estimated NLR angular momentum. However, it is possible that sustained SMBH accretion might transfer enough angular momentum over time.

The observed centripetal acceleration is $v_\phi^2/R \sin \Theta = 1.8 \times 10^{-7}$  cm s$^{-2}$, or $\sim 4$ times the gravitational acceleration of $4.9\times 10^{-8}$ cm s$^{-2}$ from the enclosed mass of $3.5 \times 10^9 M_\odot$ at $R= 1$ kpc. There must be an additional force directed towards the axis of the cone, such as gas pressure gradient or electromagnetism, in order to sustain the centripetal acceleration of a spiral outflow. As shown in \S 6, it is clear that local fast outflows are accelerated far from the AGN and close to the radio jet axis, perhaps indicating that the radio jet may also be responsible for driving the overall spiral outflow.  

(2) \textit{Residual rotation from infalling gas or existing gas disk}: Alternatively, the rotation of the NLR could derive from the preexisting angular momentum of gas that fell into or continues to fall into the center of the galaxy, forming a disk. In particular the angular momentum might derive from the $2$ kpc radius CO disk/ring \citep{1991AJ....102.1691C}, which has a comparable size, velocity, and specific angular momentum to the NLR. The primary difficulty with this scenario is the large scale-height of the NLR, which is 1.3 kpc if measured from nucleus to bicone tip, or 0.9 kpc if we only include the X-shaped part of the NLR. That would mean the putative disk must be puffed up, with a $\sim 2:1$ ratio of radius to height. Interestingly, a puffed-up disk was envisioned by \cite{2018MNRAS.479.5544M} as the starting conditions for their jet-disk interaction simulations. In their work, this provides the necessary path-length for a jet launched perpendicular to the disk to interact significantly rather than escaping immediately.  In principle, such a puffed-up disk could either be in the process of gravitational collapse and settling or it could have been puffed up by the action of the AGN and radio jet. In either case, the disk must display sub-Keplerian rotation and be supported primarily by velocity dispersion.  However, the bicone rotation velocity of 200 km s$^{-1}$ is 1.7 times the Keplerian (circular) velocity of $v_c = (GM/R)^{1/2} = 120 R^{-1/2}$ km s$^{-1}$ at 1 kpc. Furthermore, the gravitational force is radial and cannot sustain this rotation velocity around the radio axis. As an illustration, a cloud launched radially from the disk plane at $R=0.1$ kpc from the $2.5\times 10^9 M_\odot$ SMBH, with an initial circular velocity of $330$  km s$^{-1}$, that ends up 1 kpc from the nucleus along the bicone surface will have a circular velocity of only one tenth its original circular velocity ($33$  km s$^{-1}$), in order to conserve angular momentum along the cone axis. Similarly, a cloud launched farther out, at 0.5 kpc, will have an initial velocity of $88$  km s$^{-1}$ and its velocity will drop to $44$  km s$^{-1}$ at 1 kpc, much less than the observed NLR rotation. Because of this, we prefer the idea that the radio jet may be torquing the NLR at kpc-scale distance.

\section{Conclusions}

JWST NIRSpec IFU and MIRI MRS spectra of Cyg A reveal new details about the structure, kinematics and ionization state of the narrow-line region. We suggest that its multilayer,  hollowed-out biconical structure is the result of the jet clearing a cavity in the ISM. A hot cocoon inflated by the jet naturally creates a bicone when the jet head breaks out of the galaxy. The kpc-scale segment of the jet is still interacting with the host ISM as it continues to ablate and clear away clouds that linger near the jet.  The ionization structure of the NLR is also consistent with this scenario, where the inner surface of the bicone is illuminated by the radiation pattern delineated by the dusty torus. A photoionization model with a range of densities can nicely explain the emission line strengths of the large number of ions observed in the NIRSpec spectra. Extended coronal line emission is found from low density $n_e \sim 1$--$10$ cm$^{-3}$ ionized gas throughout the bicone. Strong [Fe {\sc ii}] emission from high density ($n_e \sim 10^3$ cm$^{-3}$), photoionized clouds is found in a massive Fe {\sc ii} Clump in the NW cone. Radiative acceleration may explain the relatively low-velocity outflow of 150 km s$^{-1}$ seen in much of the Cyg A NLR. However, residual interaction of the jet  with remaining clumps of dense gas in the NLR leads to localized fast outflows. In particular, we find outflows of 600--2000 km s$^{-1}$ from fast Bullets that appear to originate from the Fe {\sc ii} Clump. The high mass outflow rate of $40 M_\odot$ yr$^{-1}$ from the Fe {\sc ii} Clump and significant aditional outflow of $13 M_\odot$ yr$^{-1}$ from Cyg A Bullet B1 demonstrates that jet feedback continues to strongly impact the ISM long after the jet has broken out of the host galaxy, and may eventually serve to eject much of the remaining gas.

We find evidence that the NLR is rotating about the radio jet axis. While this rotation was previously observed, it was attributed to the overall rotation of the molecular gas disk in the host-galaxy gravitational potential. However, we suggest that such an interpretation is geometrically and dynamically inconsistent with the clear, edge-brightened biconical geometry of the NLR. There are multiple indications that both rotational and radial outflow components are needed to explain the NLR kinematics. Rotation is required to explain the strong blue-shift and redshift seen at the limbs of the bicone, while radial outflow is required to explain the line-splitting and higher velocity dispersion in the interior of the bicone. The symmetry of the velocity pattern is also broken in a characteristic way, with different projected velocity at each of the four limbs of the bicone. We present a biconical spiral outflow model which captures many of the above kinematic features. We suggest that the rotation of the NLR is driven by angular momentum transfer from the radio jet, in order to maintain the observed rotation velocity over a large range of radius. Localized high velocity, jet-driven outflows like the Cyg A Bullets support such a picture, and may be the agents that transfer energy and angular momentum from the jet to the NLR.

\appendix

\begin{figure*}
  \includegraphics[trim=0.0cm 0.0cm 0.0cm 0.0cm, clip, width=0.9\linewidth]{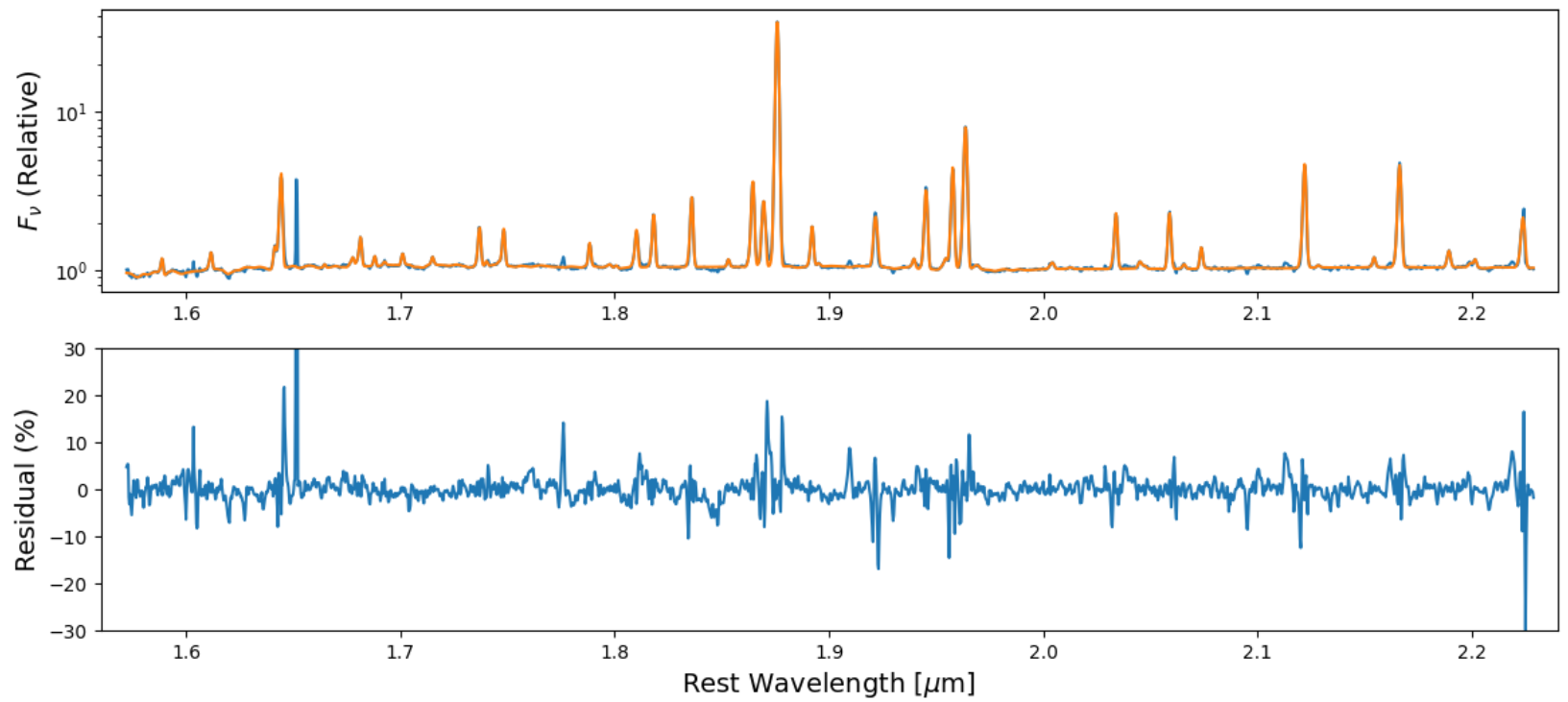}
  \includegraphics[trim=0.0cm 0.0cm 0.0cm 0.0cm, clip, width=0.9\linewidth]{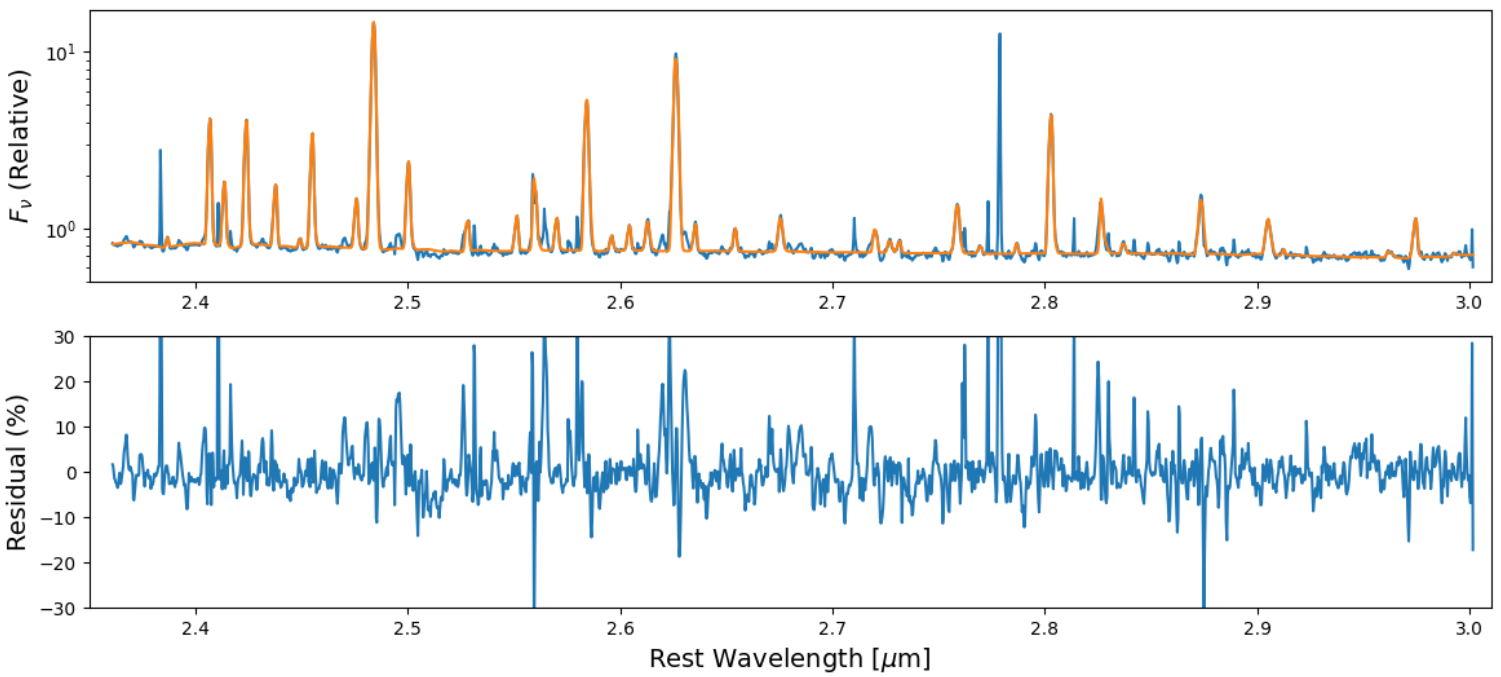}
  \includegraphics[trim=0.0cm 0.0cm 0.0cm 0.0cm, clip, width=0.9\linewidth]{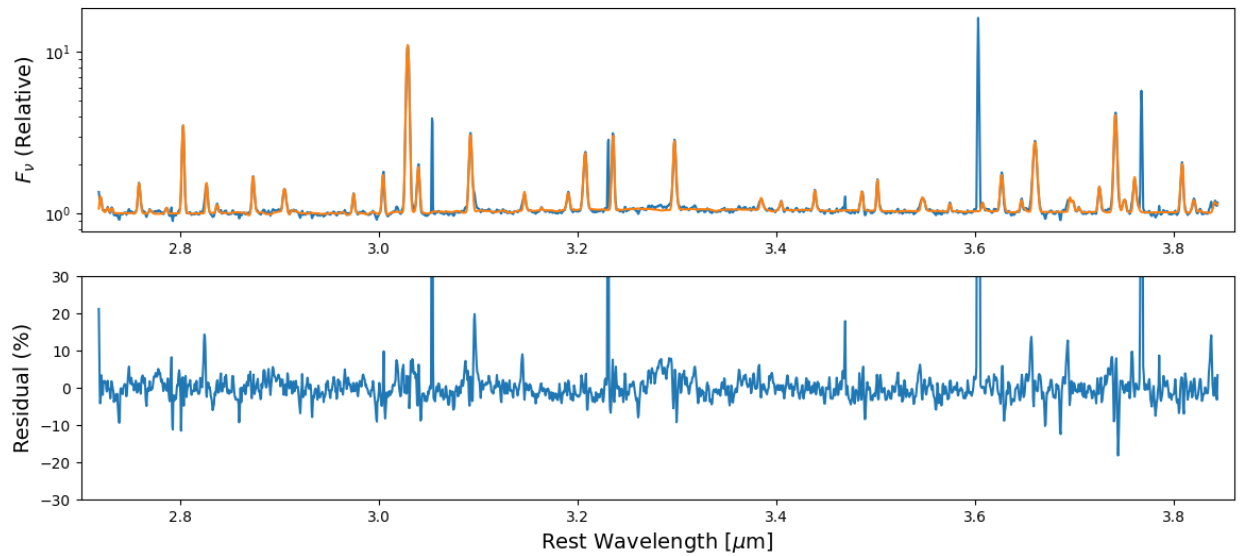}
\caption{pPXF 3-component fits to NIRSpec spectra of NW reference (NWR) region. Notable residuals include foreground Galactic Pa$\alpha$, the unmodeled 3.3 $\mu$m PAH feature, and a number of weak, unidentified lines not included in the model. Some narrow residual spikes are from bad data.}
\label{fig:ppxf-fits}
\end{figure*}

\begin{figure*}
  \includegraphics[trim=0.0cm 0.0cm 0.0cm 0.0cm, clip, width=0.9\linewidth]{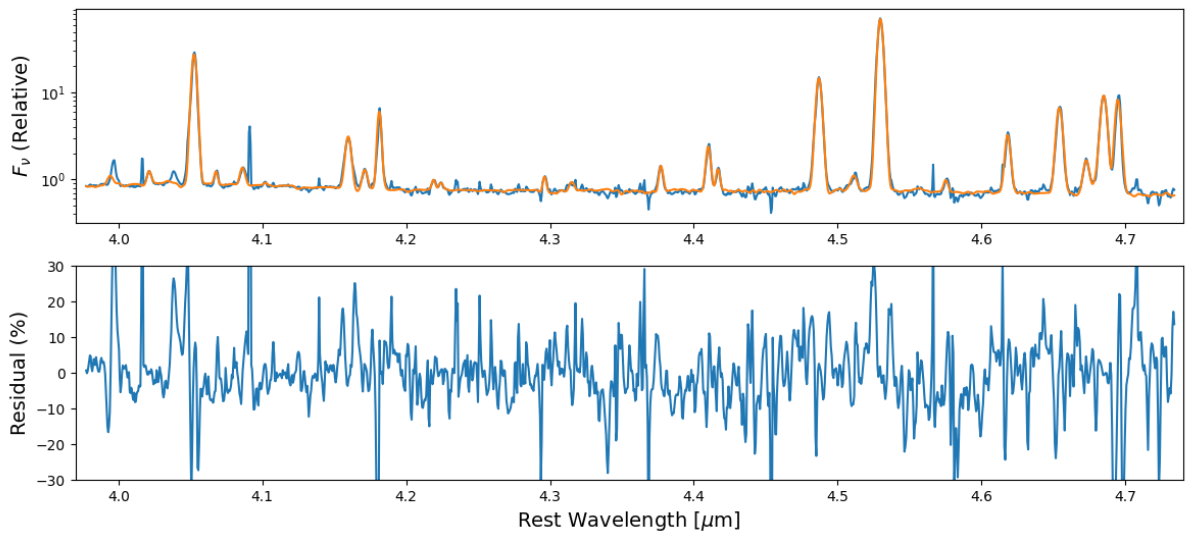}

\caption{pPXF 3-component fits to NIRSpec spectra of NW reference (NWR) region, continued. Notable residuals include poorly fit or unidentified lines at 3.95 $\mu$m, 4.03 $\mu$m, and 4.71 $\mu$m.}
\label{fig:ppxf-fits2}
\end{figure*}

Emission line identifications were made using spectra extracted from the NW reference (NWR) region (Figures~\ref{fig:ppxf-fits} \& \ref{fig:ppxf-fits2}). This region was particularly useful for separating blended lines since it has relatively narrow-line profiles.  We fit the spectrum from each NIRSpec grating-detector pair separately, using {\sc ppxf} with EMILES SPS templates plus a polynomial continuum and 3-component Gaussian.  We similarly fit the spectra of the Fe {\sc ii} Clump and Cyg A Bullet B1, extracted in an $0\farcs3 \times 0\farcs3$ aperture (Fig. 10). The flux from B1 is relatively weak and many of the expected emission lines are blended with unrelated lines, making them difficult to measure.  We present the total flux of each ionic line in Table 1, relative to Pa $\alpha$, with atomic line wavelengths primarily from the compilation in Cloudy\footnote{\url{https://linelist.pa.uky.edu/atomic/}} and H$_2$ line wavelengths from \cite{1987ApJ...322..412B} and \cite{2019A&A...630A..58R}.  Cloudy photoionization model predictions are given in comparison and used to verify line identifications, considering all lines predicted to be $>0.1\%$ of the Pa $\alpha$ flux.  Even though we fit them, we do not tabulate the large number of H$_2$ lines in the NIRSpec spectrum.  We also defer analysis of the spectral lines at $> 5 ~\mu$m, found in the MIRI spectrum.  

 {\it Caveats and Notes}: (1) There are unidentified weak emission lines apparent in the spectra (at 1.77, 1.91, 2.04, 2.12, 2.16, 2.22, 2.43, 2.45, 2.47, 2.49, 2.53, 2.57, 2.62, 2.68, 2.83, 2.84, 3.15, 3.39, 3.46, 3.75, 3.76, 3.82, 4.02, 4.30, $4.38 ~\mu$m) that were included in the line fitting. (2) The [S {\sc xi}] 1.922~$\mu$m, [Si {\sc vi}] 1.963 $\mu$m, and [Si {\sc vii}] 2.483 $\mu$m line wavelengths used in Cloudy deviate significantly from the observed wavelengths, so we utilize  wavelengths from \cite{2018ApJ...852...52D} for those lines instead.  (3) The He {\sc ii} 1.8743 and 1.8757 $\mu$m lines are too close to Pa $\alpha$ to deblend. (4) The He {\sc i} 2.3214 $\mu$m and [Si {\sc ix}] 3.936 $\mu$m lines fall in detector gap regions and therefore were not measured. (5) We identify foreground (F.g.) Galactic Pa $\alpha$ emission in the spectra, likely from diffuse nebulosity at the low Galactic latitude of Cyg A.  (6) We detect but do not fit the 3.3 $\mu$m PAH feature, which is weakly present across the NIRSpec FOV.

\vfil\eject

\startlongtable
\begin{deluxetable*}{lllllllll}
\tablecaption{Cygnus A Ionic Emission Line Flux Ratios to Pa $\alpha$}
\tablehead{
\colhead{Line ID} & \colhead{Rest} & \colhead{$z=0.0557$} & \colhead{NWR} & \colhead{Cloudy} & \colhead{Fe {\sc ii} Clump} &\colhead{Cloudy} &\colhead{Bullet B1$^a$}&\colhead{Cloudy}}
\startdata
Br 14-4         & 1.5880 & 1.6765 & 0.013  & 0.010  & 0.011  & 0.010  & \nodata & 0.010  \\
$[$Fe II$]$     & 1.5995 & 1.6886 & 0.0004 & 0.0038  & 0.0094 & 0.010  & \nodata & \nodata \\
Br 13-4         & 1.6109 & 1.7007 & 0.013  & 0.013  & 0.013  & 0.013  & \nodata & 0.013  \\
Br 12-4         & 1.6407 & 1.7321 & 0.017  & 0.016  & 0.046  & 0.016  & \nodata & 0.016  \\
$[$Fe II$]$     & 1.6436 & 1.7351 & 0.14   & 0.11   & 0.24   & 0.23   & 0.21   & 0.16   \\
$[$Fe II$]$     & 1.6638 & 1.7564 & 0.0017 & 0.0018  & 0.0025 & 0.0052  & \nodata & \nodata \\
$[$Fe II$]$     & 1.6769 & 1.7703 & 0.0057 & 0.0037  & 0.013  & 0.0099  & \nodata & \nodata \\
Br 11-4         & 1.6806 & 1.7743 & 0.025  & 0.021  & 0.022  & 0.021  & \nodata & 0.021  \\
He II           & 1.6918 & 1.7860 & 0.0018 & 0.0007  & 0.0002 & 0.0009  & \nodata & \nodata \\
He I            & 1.7003 & 1.7950 & 0.0074 & 0.0077  & 0.0072 & 0.0077  & \nodata & 0.0076 \\
$[$Fe II$]$     & 1.7111 & 1.8064 & 0.0000 & 0.0008  & 0.0018 & 0.0022  & \nodata & \nodata \\
$[$Ti VI$]$     & 1.7151 & 1.8106 & 0.0011 & 0.015  & 0.0011 & 0.015  & \nodata & 0.015  \\
Br 10-4         & 1.7362 & 1.8329 & 0.032  & 0.027  & 0.029  & 0.027  & \nodata & 0.027  \\
$[$P VIII$]$    & 1.7393 & 1.8361 & 0.0025 & 0.0007  & 0.0032 & 0.0007  & \nodata & \nodata \\
$[$Fe II$]$     & 1.7449 & 1.8421 & 0.0018 & 0.0009  & 0.0062 & 0.0026  & \nodata & \nodata \\
F.g. Pa$\alpha$ & 1.8751 & \nodata & \nodata & \nodata & \nodata & \nodata & \nodata & \nodata \\
$[$Fe II$]$     & 1.7971 & 1.8972 & 0.0019 & 0.0018  & 0.0059 & 0.0051  & \nodata & \nodata \\
$[$Fe II$]$     & 1.8000 & 1.9003 & 0.0014 & 0.0038  & 0.0090 & 0.010  & \nodata & \nodata \\
$[$Fe II$]$     & 1.8094 & 1.9102 & 0.029  & 0.031  & 0.052  & 0.064  & \nodata & 0.046  \\
Br $\epsilon$   & 1.8174 & 1.9186 & 0.044  & 0.038  & 0.038  & 0.038  & \nodata & 0.038  \\
He II           & 1.8637 & 1.9675 & 0.094  & 0.056  & 0.092  & 0.060  & \nodata & 0.060  \\
$[$P X$]$       & 1.8676 & 1.9716 & 0.014  & 0.0003  & 0.013  & 0.0003  & \nodata & \nodata \\
He I doublet    & 1.8691 & 1.9732 & 0.054  & 0.035  & 0.070  & 0.035  & \nodata & 0.034  \\
He II$^b$       & 1.8743 & 1.9787 & \nodata & 0.015  & \nodata & 0.016  & \nodata & 0.016  \\
Pa $\alpha$     & 1.8751 & 1.9795 & 1.00   & 1.00   & 1.00   & 1.00   & \nodata & 1.00   \\
He II$^b$       & 1.8757 & 1.9802 & \nodata & 0.0010  & \nodata & 0.0012  & \nodata & \nodata \\
$[$Fe IIb$]^b$  & 1.8890 & 1.9942 & \nodata & 0.049  & \nodata & 0.10   & \nodata & 0.075  \\
$[$Fe II$]$     & 1.8954 & 2.0010 & 0.0026 & 0.0007  & 0.0015 & 0.0019  & \nodata & \nodata \\
$[$S XI$]$      & 1.9220$^c$ & 2.0291 & 0.049  & 0.0049  & 0.034  & 0.0047  & \nodata & 0.020  \\
$[$Ni II$]$     & 1.9388 & 2.0468 & 0.0052 & 0.0023  & 0.0096 & 0.0067  & \nodata & \nodata \\
Br $\delta$     & 1.9446 & 2.0529 & 0.078  & 0.059  & 0.069  & 0.059  & \nodata & 0.059  \\
$[$Fe II$]$     & 1.9536 & 2.0624 & 0.0025 & 0.0028  & 0.0096 & 0.0075  & \nodata & \nodata \\
He I            & 1.9543 & 2.0632 & 0.0031 & 0.0036  & 0.012  & 0.0035  & \nodata & \nodata \\
$[$Si VI$]$     & 1.9630$^c$ & 2.0723 & 0.25   & 0.43   & 0.25   & 0.42   & 1.40   & 0.42   \\
$[$Al IX$]$     & 2.0444 & 2.1583 & 0.0035 & 0.015  & 0.0054 & 0.015  & \nodata & 0.028  \\
$[$Fe V$]$      & 2.0467 & 2.1607 & 0.0023 & 0.0006  & 0.0023 & 0.0012  & \nodata & \nodata \\
He I            & 2.0581 & 2.1728 & 0.038  & 0.043  & 0.036  & 0.044  & \nodata & 0.042  \\
Br $\gamma$     & 2.1655 & 2.2862 & 0.10   & 0.098  & 0.095  & 0.098  & 0.22   & 0.098  \\
He II           & 2.1885 & 2.3104 & 0.0085 & 0.0062  & 0.0080 & 0.0065  & \nodata & \nodata \\
$[$Fe IIb$]$    & 2.2000 & 2.3225 & 0.0026 & 0.0017  & 0.0020 & 0.0049  & \nodata & \nodata \\
$[$Si VII$]$    & 2.4826$^c$ & 2.6209 & 0.50   & 0.40   & 0.47   & 0.43   & 1.24   & 0.51   \\
$[$Si IX$]$     & 2.5839 & 2.7279 & 0.15   & 0.065  & 0.12   & 0.062  & \nodata & 0.12   \\
Pf 14-5         & 2.6119 & 2.7574 & 0.011  & 0.0086  & 0.013  & 0.0086  & \nodata & \nodata \\
Br $\beta$      & 2.6252 & 2.7714 & 0.30   & 0.19   & 0.26   & 0.19   & \nodata & 0.19   \\
$[$Fe VII$]^b$  & 2.6287 & 2.7751 & \nodata & 0.056  & \nodata & 0.058  & \nodata & 0.061  \\
Pf 13-5         & 2.6744 & 2.8234 & 0.014  & 0.011  & 0.011  & 0.011  & \nodata & 0.011  \\
Pf 12-5         & 2.7575 & 2.9111 & 0.012  & 0.014  & 0.018  & 0.014  & \nodata & 0.014  \\
Pf 11-5         & 2.8722 & 3.0322 & 0.015  & 0.018  & 0.022  & 0.018  & \nodata & 0.018  \\
$[$Al V$]$      & 2.9045 & 3.0663 & 0.010  & 0.048  & 0.018  & 0.045  & \nodata & 0.045  \\
$[$Mg VIII$]$   & 3.0276 & 3.1963 & 0.33   & 0.45   & 0.28   & 0.44   & 0.51   & 0.54   \\
Pf $\epsilon$   & 3.0384 & 3.2076 & 0.029  & 0.026  & 0.025  & 0.026  & \nodata & 0.026  \\
He II           & 3.0908 & 3.2630 & 0.065  & 0.035  & 0.055  & 0.037  & \nodata & 0.037  \\
He II$^b$       & 3.0946 & 3.2670 & \nodata & 0.0011  & \nodata & 0.0013  & \nodata & \nodata \\
$[$K VII$]^b$   & 3.1897 & 3.3673 & \nodata & 0.0065  & \nodata & 0.0081  & \nodata & \nodata \\
$[$Ca IV$]$     & 3.2061 & 3.3847 & 0.042  & 0.63   & 0.051  & 0.57   & \nodata & 0.57   \\
Pf $\delta$     & 3.2961 & 3.4797 & 0.048  & 0.039  & 0.044  & 0.039  & \nodata & 0.039  \\
$[$Al VI$]$     & 3.6593 & 3.8631 & 0.052  & 0.72   & 0.056  & 0.70   & 0.10   & 0.71   \\
$[$Al VIII$]$   & 3.6972 & 3.9031 & 0.0038 & 0.019  & 0.0007 & 0.019  & \nodata & 0.026  \\
Pf $\gamma$     & 3.7395 & 3.9478 & 0.071  & 0.062  & 0.065  & 0.062  & \nodata & 0.062  \\
$[$Si IX $]^d$  & 3.9282 & 4.1470 & \nodata & 0.17   & \nodata & 0.16   & \nodata & 0.29   \\
He I$^b$        & 4.0479 & 4.2734 & \nodata & 0.015  & \nodata & 0.014  & \nodata & 0.014  \\
He II$^b$       & 4.0495 & 4.2750 & \nodata & 0.0021  & \nodata & 0.0026  & \nodata & \nodata \\
Br $\alpha$     & 4.0511 & 4.2768 & 0.58   & 0.48   & 0.58   & 0.48   & 0.48   & 0.48   \\
$[$Ca VII$]$    & 4.0864 & 4.3140 & 0.012  & 0.013  & 0.016  & 0.015  & \nodata & 0.017  \\
$[$Ca V$]$      & 4.1574 & 4.3890 & 0.053  & 0.40   & 0.068  & 0.36   & 0.17   & 0.36   \\
Hu $\gamma$     & 4.1697 & 4.4019 & 0.013  & 0.0097  & 0.013  & 0.0097  & \nodata & \nodata \\
$[$Mg IV$]$     & 4.4871 & 4.7370 & 0.28   & 0.38   & 0.33   & 0.40   & 0.49   & 0.44   \\
$[$Ar VI$]$     & 4.5280 & 4.7802 & 1.28   & 1.27   & 1.31   & 1.32   & 1.71   & 1.32   \\
$[$K III$]$     & 4.6168 & 4.8740 & 0.044  & 0.030  & 0.049  & 0.041  & \nodata & 0.034  \\
Pf $\beta$      & 4.6525 & 4.9117 & 0.12   & 0.11   & 0.13   & 0.11   & \nodata & 0.11   \\
Hu $\epsilon$   & 4.6712 & 4.9314 & 0.020  & 0.017  & 0.020  & 0.017  & \nodata & 0.017  \\
$[$Na VII$]$    & 4.6834 & 4.9443 & 0.17   & 0.070  & 0.16   & 0.072  & \nodata & 0.092  \\
\hline
Hu $\delta$     & 5.1273  & 5.4128  & & & &&\\
$[$Fe {\sc ii}$]$     & 5.3402  & 5.5633  & & & &&\\
$[$Fe VIII$]$   & 5.4451  & 5.677   & & & &&\\
$[$Mg VII$]$    & 5.4927  & 5.7333  & & & &&\\
$[$Mg V$]$      & 5.6070  & 5.9193  & & & &&\\
$[$Ni I$]$      & 5.8933  & 6.1396  & & & &&\\
Hu $\gamma$     & 5.9066  & 6.2356  & & & &&\\
$[$Ar II$]$     & 6.9834  & 7.3723  & & & &&\\
$[$Na III$]$    & 7.3157  & 7.7232  & & & &&\\
Pf $\alpha$     & 7.4578  & 7.8732  & & & &&\\
Hu $\beta$      & 7.5004  & 7.9182  & & & &&\\
$[$Ne VI$]$     & 7.6502  & 8.0763  & & & &&\\
$[$Fe VII$]$    & 7.8104  & 8.2454  & & & &&\\
$[$Ar V $]$     & 7.8997  & 8.3397  & & & &&\\
$[$Na VI$]$     & 8.6083  & 8.9722  & & & &&\\
$[$Ar III$]$    & 8.9890  & 9.4897  & & & &&\\
$[$Mg VII$]$    & 9.0309  & 9.5339  & & & &&\\
$[$Na IV$]$     & 9.0310  & 9.5340  & & & &&\\
$[$Fe VII$]$    & 9.5076  &10.0372  & & & &&\\
$[$S IV$]$      & 10.5076 &11.0929  & & & &&\\
Hu $\alpha$     & 12.3685 &13.0574  & & & &&\\
$[$Ne II$]$     & 12.8101 &13.5236  & & & &&\\
$[$Ar V$]$      & 13.0985 &13.8281  & & & &&\\
$[$Mg V$]$      & 13.5464 &14.3009  & & & &&\\
$[$Ne V$]$      & 14.3178 &15.1153  & & & &&\\
$[$Cl II$]$     & 14.3639 &15.1640  & & & &&\\
$[$Na VI$]$     & 14.3925 &15.1942  & & & &&\\
$[$Ne III$]$    & 15.5509 &16.4171  & & & &&\\
$[$Fe {\sc ii}$]$ & 17.9311 &18.9299  & & & &&\\
$[$S III$]$     & 18.7078 &19.7498  & & & &&\\
$[$Ar III$]$    & 21.8253 &23.0410  & & & &&\\
$[$Ne V$]$      & 24.3109 &25.6650  & & & &&\\
$[$O IV$]$      & 25.8863 &27.3282  & & & &&\\
\enddata
\tablenotetext{a}{Pa $\alpha$ from Cyg A Bullet B1 is blended with other lines and a useful upper limit can not be measured. Bullet line flux ratios are therefore given with respect to the Pa $\alpha$ flux predicted from Br $\gamma$ and the Case B ratio of 12.4. Bullet line fluxes for weak lines below the detection limit were not measured.}
\tablenotetext{b}{Several weak lines predicted by Cloudy that should be detectable are in fact not detected.}
\tablenotetext{c}{We utilize the more accurate [S {\sc xi}], [Si {\sc vi}], and [Si {\sc vii}] coronal line wavelengths from \citep{2018ApJ...852...52D} rather than the wavelengths used by Cloudy.}
\tablenotetext{d}{Line falls in NIRSpec detector gap.}
\label{Table1}
\end{deluxetable*}

\eject

\begin{deluxetable*}{llllll}
\tablecaption{Cloudy Model Parameters}
\tablehead{
\colhead{Model} &\colhead{$R$(kpc)$^a$} &\colhead{$\log n_H$ (cm$^{-3}$)$^b$} & \colhead{$ T_e$ ($10^4$ K)$^c$}& \colhead{$P$ ($10^{-10}$ erg cm$^{-3}$)$^d$} &  \colhead{$M$ ($10^6 M_\odot$)$^e$}}

\startdata
NWR         & 1.6  &0.0, 1.0, 1.5, 2.0, 3.0 & 4.5, 1.9, 1.1, 1.1, 1.0 & 0.062, 0.26, 0.48, 1.5, 14 &19,  6, 15, 0.5, 0.05\\
Fe {\sc ii} Clump & 1.2  &0.0, 1.0, 1.5, 2.0, 3.0 & 6.3, 2.2, 1.2, 1.1, 1.0 & 0.087, 0.30, 0.52, 1.5, 14 &47, 19, 37, 2.1, 0.7\\
Bullet      & 1.3  &0.0, 1.0, 1.5, 2.0, 3.0 & 5.6, 2.1, 1.2, 1.1, 1.0 & 0.077, 0.29, 0.52, 1.5, 15 & 2.3, 0.6, 1.2, 0.06, 0.02\\
\enddata
\tablenotetext{a}{Distance from the galaxy nucleus.}
\tablenotetext{b}{Logarithm of total hydrogen density.}
\tablenotetext{c}{Electron temperature.}
\tablenotetext{d}{Gas pressure.}
\tablenotetext{e}{Gas mass estimated from observed Pa $\alpha$ luminosity and its flux predicted by the model.}
\end{deluxetable*}

\begin{acknowledgments}
This work is based in part on observations made with the NASA/ESA/CSA James Webb Space Telescope. Some of the data presented in this paper were obtained from the Mikulski Archive for Space Telescopes (MAST) at the Space Telescope Science Institute. The specific observations analyzed can be accessed via \dataset[https://doi.org/10.17909/6p74-z566]{https://doi.org/10.17909/6p74-z566}. STScI is operated by the Association of Universities for Research in Astronomy, Inc., under NASA contract NAS5–26555. Support to MAST for these data is provided by the NASA Office of Space Science via grant NAG5–7584 and by other grants and contracts. Some of the data presented herein were obtained at Keck Observatory, which is a private 501(c)3 non-profit organization operated as a scientific partnership among the California Institute of Technology, the University of California, and the National Aeronautics and Space Administration. The Observatory was made possible by the generous financial support of the W. M. Keck Foundation. The authors wish to recognize and acknowledge the very significant cultural role and reverence that the summit of Maunakea has always had within the Native Hawaiian community. We are most fortunate to have the opportunity to conduct observations from this mountain. The Karl G. Jansky Very Large Array (VLA) is a facility of the National Science Foundation operated under cooperative agreement by Associated Universities, Inc.  This research has made use of data obtained from the \textit{Chandra} Data Archive provided by the Chandra X-ray Center (CXC).
P.M.O. and B.S. acknowledge support from STScI grant JWST-GO-04065.
E.L.-R. thanks support by the NASA Astrophysics Decadal Survey Precursor Science (ADSPS) Program (NNH22ZDA001N-ADSPS) with ID 22-ADSPS22-0009 and agreement number 80NSSC23K1585.
A.M.M. acknowledges support from NASA ADAP grant 80NSSC23K0750 and from NSF AAG grant 2009416 and NSF CAREER grant 2239807.
CO and SB acknowledge support from the Natural Sciences and Engineering Research Council (NSERC) of Canada.

\end{acknowledgments}

\vspace{5mm}
\facilities{JWST, Keck Observatory, NRAO VLA, Chandra}

\software{astropy \citep{2013A&A...558A..33A,2018AJ....156..123A},  
          Cloudy \citep{2013RMxAA..49..137F},
          Jdaviz \citep{jdaviz2024}
          }

\bibliography{cyga_nlr.bib}{}

\begin{thebibliography}{}
\expandafter\ifx\csname natexlab\endcsname\relax\def\natexlab#1{#1}\fi
\providecommand{\url}[1]{\href{#1}{#1}}
\providecommand{\dodoi}[1]{doi:~\href{http://doi.org/#1}{\nolinkurl{#1}}}
\providecommand{\doeprint}[1]{\href{http://ascl.net/#1}{\nolinkurl{http://ascl.net/#1}}}
\providecommand{\doarXiv}[1]{\href{https://arxiv.org/abs/#1}{\nolinkurl{https://arxiv.org/abs/#1}}}

\bibitem[{{Antonucci} {et~al.}(1994){Antonucci}, {Hurt}, \& {Kinney}}]{1994Natur.371..313A}
{Antonucci}, R., {Hurt}, T., \& {Kinney}, A. 1994, \nat, 371, 313, \dodoi{10.1038/371313a0}

\bibitem[{{Antonucci} {et~al.}(1993){Antonucci}, {Kinney}, \& {Hurt}}]{1993ApJ...414..506A}
{Antonucci}, R., {Kinney}, A.~L., \& {Hurt}, T. 1993, \apj, 414, 506, \dodoi{10.1086/173098}

\bibitem[{{Argyriou} {et~al.}(2023){Argyriou}, {Glasse}, {Law}, {Labiano}, {{\'A}lvarez-M{\'a}rquez}, {Patapis}, {Kavanagh}, {Gasman}, {Mueller}, {Larson}, {Vandenbussche}, {Glauser}, {Royer}, {Dicken}, {Harkett}, {Sargent}, {Engesser}, {Jones}, {Kendrew}, {Noriega-Crespo}, {Brandl}, {Rieke}, {Wright}, {Lee}, \& {Wells}}]{2023A&A...675A.111A}
{Argyriou}, I., {Glasse}, A., {Law}, D.~R., {et~al.} 2023, \aap, 675, A111, \dodoi{10.1051/0004-6361/202346489}

\bibitem[{{Asmus} {et~al.}(2014){Asmus}, {H{\"o}nig}, {Gandhi}, {Smette}, \& {Duschl}}]{2014MNRAS.439.1648A}
{Asmus}, D., {H{\"o}nig}, S.~F., {Gandhi}, P., {Smette}, A., \& {Duschl}, W.~J. 2014, \mnras, 439, 1648, \dodoi{10.1093/mnras/stu041}

\bibitem[{{Astropy Collaboration} {et~al.}(2013){Astropy Collaboration}, {Robitaille}, {Tollerud}, {Greenfield}, {Droettboom}, {Bray}, {Aldcroft}, {Davis}, {Ginsburg}, {Price-Whelan}, {Kerzendorf}, {Conley}, {Crighton}, {Barbary}, {Muna}, {Ferguson}, {Grollier}, {Parikh}, {Nair}, {Unther}, {Deil}, {Woillez}, {Conseil}, {Kramer}, {Turner}, {Singer}, {Fox}, {Weaver}, {Zabalza}, {Edwards}, {Azalee Bostroem}, {Burke}, {Casey}, {Crawford}, {Dencheva}, {Ely}, {Jenness}, {Labrie}, {Lim}, {Pierfederici}, {Pontzen}, {Ptak}, {Refsdal}, {Servillat}, \& {Streicher}}]{2013A&A...558A..33A}
{Astropy Collaboration}, {Robitaille}, T.~P., {Tollerud}, E.~J., {et~al.} 2013, \aap, 558, A33, \dodoi{10.1051/0004-6361/201322068}

\bibitem[{{Astropy Collaboration} {et~al.}(2018){Astropy Collaboration}, {Price-Whelan}, {Sip{\H{o}}cz}, {G{\"u}nther}, {Lim}, {Crawford}, {Conseil}, {Shupe}, {Craig}, {Dencheva}, {Ginsburg}, {VanderPlas}, {Bradley}, {P{\'e}rez-Su{\'a}rez}, {de Val-Borro}, {Aldcroft}, {Cruz}, {Robitaille}, {Tollerud}, {Ardelean}, {Babej}, {Bach}, {Bachetti}, {Bakanov}, {Bamford}, {Barentsen}, {Barmby}, {Baumbach}, {Berry}, {Biscani}, {Boquien}, {Bostroem}, {Bouma}, {Brammer}, {Bray}, {Breytenbach}, {Buddelmeijer}, {Burke}, {Calderone}, {Cano Rodr{\'\i}guez}, {Cara}, {Cardoso}, {Cheedella}, {Copin}, {Corrales}, {Crichton}, {D'Avella}, {Deil}, {Depagne}, {Dietrich}, {Donath}, {Droettboom}, {Earl}, {Erben}, {Fabbro}, {Ferreira}, {Finethy}, {Fox}, {Garrison}, {Gibbons}, {Goldstein}, {Gommers}, {Greco}, {Greenfield}, {Groener}, {Grollier}, {Hagen}, {Hirst}, {Homeier}, {Horton}, {Hosseinzadeh}, {Hu}, {Hunkeler}, {Ivezi{\'c}}, {Jain}, {Jenness}, {Kanarek}, {Kendrew}, {Kern}, {Kerzendorf}, {Khvalko}, {King}, {Kirkby}, {Kulkarni},
  {Kumar}, {Lee}, {Lenz}, {Littlefair}, {Ma}, {Macleod}, {Mastropietro}, {McCully}, {Montagnac}, {Morris}, {Mueller}, {Mumford}, {Muna}, {Murphy}, {Nelson}, {Nguyen}, {Ninan}, {N{\"o}the}, {Ogaz}, {Oh}, {Parejko}, {Parley}, {Pascual}, {Patil}, {Patil}, {Plunkett}, {Prochaska}, {Rastogi}, {Reddy Janga}, {Sabater}, {Sakurikar}, {Seifert}, {Sherbert}, {Sherwood-Taylor}, {Shih}, {Sick}, {Silbiger}, {Singanamalla}, {Singer}, {Sladen}, {Sooley}, {Sornarajah}, {Streicher}, {Teuben}, {Thomas}, {Tremblay}, {Turner}, {Terr{\'o}n}, {van Kerkwijk}, {de la Vega}, {Watkins}, {Weaver}, {Whitmore}, {Woillez}, {Zabalza}, \& {Astropy Contributors}}]{2018AJ....156..123A}
{Astropy Collaboration}, {Price-Whelan}, A.~M., {Sip{\H{o}}cz}, B.~M., {et~al.} 2018, \aj, 156, 123, \dodoi{10.3847/1538-3881/aabc4f}

\bibitem[{{Baade} \& {Minkowski}(1954)}]{1954ApJ...119..206B}
{Baade}, W., \& {Minkowski}, R. 1954, \apj, 119, 206, \dodoi{10.1086/145812}

\bibitem[{{Bagul} {et~al.}(2024){Bagul}, {Ogle}, {Antonucci}, {Maloney}, \& {Lopez Rodriguez}}]{2024MNRAS.527.2371B}
{Bagul}, A., {Ogle}, P., {Antonucci}, R., {Maloney}, P., \& {Lopez Rodriguez}, E. 2024, \mnras, 527, 2371, \dodoi{10.1093/mnras/stad3328}

\bibitem[{{Bartel} {et~al.}(1995){Bartel}, {Sorathia}, {Bietenholz}, {Carilli}, \& {Diamond}}]{1995PNAS...9211371B}
{Bartel}, N., {Sorathia}, B., {Bietenholz}, M.~F., {Carilli}, C.~L., \& {Diamond}, P. 1995, Proceedings of the National Academy of Science, 92, 11371, \dodoi{10.1073/pnas.92.25.11371}

\bibitem[{{Barthel} {et~al.}(2012){Barthel}, {Haas}, {Leipski}, \& {Wilkes}}]{2012ApJ...757L..26B}
{Barthel}, P., {Haas}, M., {Leipski}, C., \& {Wilkes}, B. 2012, \apjl, 757, L26, \dodoi{10.1088/2041-8205/757/2/L26}

\bibitem[{{Barthel} \& {Arnaud}(1996)}]{1996MNRAS.283L..45B}
{Barthel}, P.~D., \& {Arnaud}, K.~A. 1996, \mnras, 283, L45, \dodoi{10.1093/mnras/283.2.L45}

\bibitem[{{Bianchi} {et~al.}(2019){Bianchi}, {Guainazzi}, {Laor}, {Stern}, \& {Behar}}]{2019MNRAS.485..416B}
{Bianchi}, S., {Guainazzi}, M., {Laor}, A., {Stern}, J., \& {Behar}, E. 2019, \mnras, 485, 416, \dodoi{10.1093/mnras/stz430}

\bibitem[{{Bianchin} {et~al.}(2022){Bianchin}, {Riffel}, {Storchi-Bergmann}, {Riffel}, {Ruschel-Dutra}, {Harrison}, {Dahmer-Hahn}, {Mainieri}, {Sch{\"o}nell}, \& {Dametto}}]{2022MNRAS.510..639B}
{Bianchin}, M., {Riffel}, R.~A., {Storchi-Bergmann}, T., {et~al.} 2022, \mnras, 510, 639, \dodoi{10.1093/mnras/stab3468}

\bibitem[{{Bianchin} {et~al.}(2024){Bianchin}, {U}, {Song}, {Lai}, {Remigio}, {Barcos-Mu{\~n}oz}, {D{\'\i}az-Santos}, {Armus}, {Inami}, {Larson}, {Evans}, {B{\"o}ker}, {Kader}, {Linden}, {Charmandaris}, {Malkan}, {Rich}, {Bohn}, {Medling}, {Stierwalt}, {Mazzarella}, {Law}, {Privon}, {Aalto}, {Appleton}, {Brown}, {Buiten}, {Finnerty}, {Hayward}, {Howell}, {Iwasawa}, {Kemper}, {Marshall}, {McKinney}, {M{\"u}ller-S{\'a}nchez}, {Murphy}, {van der Werf}, {Sanders}, \& {Surace}}]{2024ApJ...965..103B}
{Bianchin}, M., {U}, V., {Song}, Y., {et~al.} 2024, \apj, 965, 103, \dodoi{10.3847/1538-4357/ad2a50}

\bibitem[{{Black} \& {van Dishoeck}(1987)}]{1987ApJ...322..412B}
{Black}, J.~H., \& {van Dishoeck}, E.~F. 1987, \apj, 322, 412, \dodoi{10.1086/165740}

\bibitem[{{Blandford} \& {Payne}(1982)}]{1982MNRAS.199..883B}
{Blandford}, R.~D., \& {Payne}, D.~G. 1982, \mnras, 199, 883, \dodoi{10.1093/mnras/199.4.883}

\bibitem[{{Blietz} {et~al.}(1994){Blietz}, {Cameron}, {Drapatz}, {Genzel}, {Krabbe}, {van der Werf}, {Sternberg}, \& {Ward}}]{1994ApJ...421...92B}
{Blietz}, M., {Cameron}, M., {Drapatz}, S., {et~al.} 1994, \apj, 421, 92, \dodoi{10.1086/173628}

\bibitem[{{Boccardi} {et~al.}(2016{\natexlab{a}}){Boccardi}, {Krichbaum}, {Bach}, {Bremer}, \& {Zensus}}]{2016A&A...588L...9B}
{Boccardi}, B., {Krichbaum}, T.~P., {Bach}, U., {Bremer}, M., \& {Zensus}, J.~A. 2016{\natexlab{a}}, \aap, 588, L9, \dodoi{10.1051/0004-6361/201628412}

\bibitem[{{Boccardi} {et~al.}(2016{\natexlab{b}}){Boccardi}, {Krichbaum}, {Bach}, {Mertens}, {Ros}, {Alef}, \& {Zensus}}]{2016A&A...585A..33B}
{Boccardi}, B., {Krichbaum}, T.~P., {Bach}, U., {et~al.} 2016{\natexlab{b}}, \aap, 585, A33, \dodoi{10.1051/0004-6361/201526985}

\bibitem[{{B{\"o}ker} {et~al.}(2023){B{\"o}ker}, {Beck}, {Birkmann}, {Giardino}, {Keyes}, {Kumari}, {Muzerolle}, {Rawle}, {Zeidler}, {Abul-Huda}, {Alves de Oliveira}, {Arribas}, {Bechtold}, {Bhatawdekar}, {Bonaventura}, {Bunker}, {Cameron}, {Carniani}, {Charlot}, {Curti}, {Espinoza}, {Ferruit}, {Franx}, {Jakobsen}, {Karakla}, {L{\'o}pez-Caniego}, {L{\"u}tzgendorf}, {Maiolino}, {Manjavacas}, {Marston}, {Moseley}, {Ogle}, {Perna}, {Pe{\~n}a-Guerrero}, {Pirzkal}, {Plesha}, {Proffitt}, {Rauscher}, {Rix}, {Rodr{\'\i}guez del Pino}, {Rustamkulov}, {Sabbi}, {Sing}, {Sirianni}, {te Plate}, {{\'U}beda}, {Wahlgren}, {Wislowski}, {Wu}, \& {Willott}}]{2023PASP..135c8001B}
{B{\"o}ker}, T., {Beck}, T.~L., {Birkmann}, S.~M., {et~al.} 2023, \pasp, 135, 038001, \dodoi{10.1088/1538-3873/acb846}

\bibitem[{{Canalizo} {et~al.}(2003){Canalizo}, {Max}, {Whysong}, {Antonucci}, \& {Dahm}}]{2003ApJ...597..823C}
{Canalizo}, G., {Max}, C., {Whysong}, D., {Antonucci}, R., \& {Dahm}, S.~E. 2003, \apj, 597, 823, \dodoi{10.1086/378513}

\bibitem[{{Cappellari}(2017)}]{2017MNRAS.466..798C}
{Cappellari}, M. 2017, \mnras, 466, 798, \dodoi{10.1093/mnras/stw3020}

\bibitem[{{Cappellari}(2023)}]{2023MNRAS.526.3273C}
---. 2023, \mnras, 526, 3273, \dodoi{10.1093/mnras/stad2597}

\bibitem[{{Cappellari} {et~al.}(2011){Cappellari}, {Emsellem}, {Krajnovi{\'c}}, {McDermid}, {Scott}, {Verdoes Kleijn}, {Young}, {Alatalo}, {Bacon}, {Blitz}, {Bois}, {Bournaud}, {Bureau}, {Davies}, {Davis}, {de Zeeuw}, {Duc}, {Khochfar}, {Kuntschner}, {Lablanche}, {Morganti}, {Naab}, {Oosterloo}, {Sarzi}, {Serra}, \& {Weijmans}}]{2011MNRAS.413..813C}
{Cappellari}, M., {Emsellem}, E., {Krajnovi{\'c}}, D., {et~al.} 2011, \mnras, 413, 813, \dodoi{10.1111/j.1365-2966.2010.18174.x}

\bibitem[{{Carilli} {et~al.}(1991){Carilli}, {Bartel}, \& {Linfield}}]{1991AJ....102.1691C}
{Carilli}, C.~L., {Bartel}, N., \& {Linfield}, R.~P. 1991, \aj, 102, 1691, \dodoi{10.1086/115988}

\bibitem[{{Carilli} \& {Barthel}(1996)}]{1996A&ARv...7....1C}
{Carilli}, C.~L., \& {Barthel}, P.~D. 1996, \aapr, 7, 1, \dodoi{10.1007/s001590050001}

\bibitem[{{Carilli} {et~al.}(2019){Carilli}, {Perley}, {Dhawan}, \& {Perley}}]{2019ApJ...874L..32C}
{Carilli}, C.~L., {Perley}, R.~A., {Dhawan}, V., \& {Perley}, D.~A. 2019, \apjl, 874, L32, \dodoi{10.3847/2041-8213/ab1019}

\bibitem[{{Carilli} {et~al.}(2022){Carilli}, {Perley}, {Perley}, {Dhawan}, {Decarli}, {Evans}, \& {Nyland}}]{2022ApJ...937..106C}
{Carilli}, C.~L., {Perley}, R.~A., {Perley}, D.~A., {et~al.} 2022, \apj, 937, 106, \dodoi{10.3847/1538-4357/ac8f45}

\bibitem[{{Cattaneo} {et~al.}(2009){Cattaneo}, {Faber}, {Binney}, {Dekel}, {Kormendy}, {Mushotzky}, {Babul}, {Best}, {Br{\"u}ggen}, {Fabian}, {Frenk}, {Khalatyan}, {Netzer}, {Mahdavi}, {Silk}, {Steinmetz}, \& {Wisotzki}}]{Cattaneo2009}
{Cattaneo}, A., {Faber}, S.~M., {Binney}, J., {et~al.} 2009, \nat, 460, 213, \dodoi{10.1038/nature08135}

\bibitem[{{Costa-Souza} {et~al.}(2024){Costa-Souza}, {Riffel}, {Souza-Oliveira}, {Zakamska}, {Bianchin}, {Storchi-Bergmann}, \& {Riffel}}]{2024ApJ...974..127C}
{Costa-Souza}, J.~H., {Riffel}, R.~A., {Souza-Oliveira}, G.~L., {et~al.} 2024, \apj, 974, 127, \dodoi{10.3847/1538-4357/ad702a}

\bibitem[{{Crain} {et~al.}(2015){Crain}, {Schaye}, {Bower}, {Furlong}, {Schaller}, {Theuns}, {Dalla Vecchia}, {Frenk}, {McCarthy}, {Helly}, {Jenkins}, {Rosas-Guevara}, {White}, \& {Trayford}}]{Crain2015}
{Crain}, R.~A., {Schaye}, J., {Bower}, R.~G., {et~al.} 2015, \mnras, 450, 1937, \dodoi{10.1093/mnras/stv725}

\bibitem[{{Croton} {et~al.}(2006){Croton}, {Springel}, {White}, {De Lucia}, {Frenk}, {Gao}, {Jenkins}, {Kauffmann}, {Navarro}, \& {Yoshida}}]{2006MNRAS.365...11C}
{Croton}, D.~J., {Springel}, V., {White}, S. D.~M., {et~al.} 2006, \mnras, 365, 11, \dodoi{10.1111/j.1365-2966.2005.09675.x}

\bibitem[{{Dasyra} {et~al.}(2024){Dasyra}, {Paraschos}, {Combes}, {Patapis}, {Helou}, {Papachristou}, {Fernandez-Ontiveros}, {Bisbas}, {Spinoglio}, {Armus}, \& {Malkan}}]{2024arXiv240603218D}
{Dasyra}, K.~M., {Paraschos}, G.~F., {Combes}, F., {et~al.} 2024, arXiv e-prints, arXiv:2406.03218, \dodoi{10.48550/arXiv.2406.03218}

\bibitem[{{Del Zanna} \& {DeLuca}(2018)}]{2018ApJ...852...52D}
{Del Zanna}, G., \& {DeLuca}, E.~E. 2018, \apj, 852, 52, \dodoi{10.3847/1538-4357/aa9edf}

\bibitem[{{Djorgovski} {et~al.}(1991){Djorgovski}, {Weir}, {Matthews}, \& {Graham}}]{1991ApJ...372L..67D}
{Djorgovski}, S., {Weir}, N., {Matthews}, K., \& {Graham}, J.~R. 1991, \apjl, 372, L67, \dodoi{10.1086/186025}

\bibitem[{{Dopita} {et~al.}(2002){Dopita}, {Groves}, {Sutherland}, {Binette}, \& {Cecil}}]{2002ApJ...572..753D}
{Dopita}, M.~A., {Groves}, B.~A., {Sutherland}, R.~S., {Binette}, L., \& {Cecil}, G. 2002, \apj, 572, 753, \dodoi{10.1086/340429}

\bibitem[{{Dugan} {et~al.}(2017){Dugan}, {Gaibler}, \& {Silk}}]{Dugan2017}
{Dugan}, Z., {Gaibler}, V., \& {Silk}, J. 2017, \apj, 844, 37, \dodoi{10.3847/1538-4357/aa7566}

\bibitem[{{Fabian}(2012)}]{Fabian2012}
{Fabian}, A.~C. 2012, \araa, 50, 455, \dodoi{10.1146/annurev-astro-081811-125521}

\bibitem[{{Ferland} {et~al.}(2013){Ferland}, {Porter}, {van Hoof}, {Williams}, {Abel}, {Lykins}, {Shaw}, {Henney}, \& {Stancil}}]{2013RMxAA..49..137F}
{Ferland}, G.~J., {Porter}, R.~L., {van Hoof}, P.~A.~M., {et~al.} 2013, \rmxaa, 49, 137.
\newblock \doarXiv{1302.4485}

\bibitem[{{Foltz} {et~al.}(1986){Foltz}, {Weymann}, {Peterson}, {Sun}, {Malkan}, \& {Chaffee}}]{1986ApJ...307..504F}
{Foltz}, C.~B., {Weymann}, R.~J., {Peterson}, B.~M., {et~al.} 1986, \apj, 307, 504, \dodoi{10.1086/164440}

\bibitem[{{F{\"o}rster Schreiber} {et~al.}(2019){F{\"o}rster Schreiber}, {{\"U}bler}, {Davies}, {Genzel}, {Wisnioski}, {Belli}, {Shimizu}, {Lutz}, {Fossati}, {Herrera-Camus}, {Mendel}, {Tacconi}, {Wilman}, {Beifiori}, {Brammer}, {Burkert}, {Carollo}, {Davies}, {Eisenhauer}, {Fabricius}, {Lilly}, {Momcheva}, {Naab}, {Nelson}, {Price}, {Renzini}, {Saglia}, {Sternberg}, {van Dokkum}, \& {Wuyts}}]{Forster2019}
{F{\"o}rster Schreiber}, N.~M., {{\"U}bler}, H., {Davies}, R.~L., {et~al.} 2019, \apj, 875, 21, \dodoi{10.3847/1538-4357/ab0ca2}

\bibitem[{{Gaibler} {et~al.}(2012){Gaibler}, {Khochfar}, {Krause}, \& {Silk}}]{Gaibler2012}
{Gaibler}, V., {Khochfar}, S., {Krause}, M., \& {Silk}, J. 2012, \mnras, 425, 438, \dodoi{10.1111/j.1365-2966.2012.21479.x}

\bibitem[{{Gardner} {et~al.}(2023){Gardner}, {Mather}, {Abbott}, {Abell}, {Abernathy}, {Abney}, {Abraham}, {Abraham}, {Abul-Huda}, {Acton}, {Adams}, {Adams}, {Adler}, {Adriaensen}, {Aguilar}, {Ahmed}, {Ahmed}, {Ahmed}, {Albat}, {Albert}, {Alberts}, {Aldridge}, {Allen}, {Allen}, {Altenburg}, {Altunc}, {Alvarez}, {{\'A}lvarez-M{\'a}rquez}, {Alves de Oliveira}, {Ambrose}, {Anandakrishnan}, {Andersen}, {Anderson}, {Anderson}, {Anderson}, {Anderson}, {Aprea}, {Archer}, {Arenberg}, {Argyriou}, {Arribas}, {Artigau}, {Arvai}, {Atcheson}, {Atkinson}, {Averbukh}, {Aymergen}, {Bacinski}, {Baggett}, {Bagnasco}, {Baker}, {Balzano}, {Banks}, {Baran}, {Barker}, {Barrett}, {Barringer}, {Barto}, {Bast}, {Baudoz}, {Baum}, {Beatty}, {Beaulieu}, {Bechtold}, {Beck}, {Beddard}, {Beichman}, {Bellagama}, {Bely}, {Berger}, {Bergeron}, {Bernier}, {Bertch}, {Beskow}, {Betz}, {Biagetti}, {Birkmann}, {Bjorklund}, {Blackwood}, {Blazek}, {Blossfeld}, {Bluth}, {Boccaletti}, {Boegner}, {Bohlin}, {Boia}, {B{\"o}ker}, {Bonaventura}, {Bond},
  {Bosley}, {Boucarut}, {Bouchet}, {Bouwman}, {Bower}, {Bowers}, {Bowers}, {Boyce}, {Boyer}, {Boyer}, {Boyer}, {Boyer}, {Bradley}, {Brady}, {Brandl}, {Brannen}, {Breda}, {Bremmer}, {Brennan}, {Bresnahan}, {Bright}, {Broiles}, {Bromenschenkel}, {Brooks}, {Brooks}, {Brown}, {Brown}, {Brown}, {Bruce}, {Bryson}, {Bujanda}, {Bullock}, {Bunker}, {Bureo}, {Burt}, {Bush}, {Bushouse}, {Bussman}, {Cabaud}, {Cale}, {Calhoon}, {Calvani}, {Canipe}, {Caputo}, {Cara}, {Carey}, {Case}, {Cesari}, {Cetorelli}, {Chance}, {Chandler}, {Chaney}, {Chapman}, {Charlot}, {Chayer}, {Cheezum}, {Chen}, {Chen}, {Cherinka}, {Chichester}, {Chilton}, {Chittiraibalan}, {Clampin}, {Clark}, {Clark}, {Clark}, {Claybrooks}, {Cleveland}, {Cohen}, {Cohen}, {Col{\'o}n}, {Coleman}, {Colina}, {Comber}, {Comeau}, {Comer}, {Conde Reis}, {Connolly}, {Conroy}, {Contos}, {Contreras}, {Cook}, {Cooper}, {Cooper}, {Correia}, {Correnti}, {Cossou}, {Costanza}, {Coulais}, {Cox}, {Coyle}, {Cracraft}, {Crew}, {Curtis}, {Cusveller}, {Da Costa Maciel}, {Dailey},
  {Daugeron}, {Davidson}, {Davies}, {Davis}, {Davis}, {Day}, {de Chambure}, {de Jong}, {De Marchi}, {Dean}, {Decker}, {Delisa}, {Dell}, {Dellagatta}, {Dembinska}, {Demosthenes}, {Dencheva}, {Deneu}, {DePriest}, {Deschenes}, {Dethienne}, {Detre}, {Diaz}, {Dicken}, {DiFelice}, {Dillman}, {Disharoon}, {Dixon}, {Doggett}, {Dominguez}, {Donaldson}, {Doria-Warner}, {Santos}, {Doty}, {Douglas}, {Doyon}, {Dressler}, {Driggers}, {Driggers}, {Dunn}, {DuPrie}, {Dupuis}, {Durning}, {Dutta}, {Earl}, {Eccleston}, {Ecobichon}, {Egami}, {Ehrenwinkler}, {Eisenhamer}, {Eisenhower}, {Eisenstein}, {El Hamel}, {Elie}, {Elliott}, {Elliott}, {Engesser}, {Espinoza}, {Etienne}, {Etxaluze}, {Evans}, {Fabreguettes}, {Falcolini}, {Falini}, {Fatig}, {Feeney}, {Feinberg}, {Fels}, {Ferdous}, {Ferguson}, {Ferrarese}, {Ferreira}, {Ferruit}, {Ferry}, {Filippazzo}, {Firre}, {Fix}, {Flagey}, {Flanagan}, {Fleming}, {Florian}, {Flynn}, {Foiadelli}, {Fontaine}, {Fontanella}, {Forshay}, {Fortner}, {Fox}, {Framarini}, {Francisco}, {Franck}, {Franx},
  {Franz}, {Friedman}, {Friend}, {Frost}, {Fu}, {Fullerton}, {Gaillard}, {Galkin}, {Gallagher}, {Galyer}, {Garc{\'\i}a Mar{\'\i}n}, {Gardner}, {Garland}, {Garrett}, {Gasman}, {G{\'a}sp{\'a}r}, {Gastaud}, {Gaudreau}, {Gauthier}, {Geers}, {Geithner}, {Gennaro}, {Gerber}, {Gereau}, {Giampaoli}, {Giardino}, {Gibbons}, {Gilbert}, {Gilman}, {Girard}, {Giuliano}, {Gkountis}, {Glasse}, {Glassmire}, {Glauser}, {Glazer}, {Goldberg}, {Golimowski}, {Gonzaga}, {Gordon}, {Gordon}, {Goudfrooij}, {Gough}, {Graham}, {Grau}, {Green}, {Greene}, {Greene}, {Greenfield}, {Greenhouse}, {Greve}, {Greville}, {Grimaldi}, {Groe}, {Groebner}, {Grumm}, {Grundy}, {G{\"u}del}, {Guillard}, {Guldalian}, {Gunn}, {Gurule}, {Gutman}, {Guy}, {Guyot}, {Hack}, {Haderlein}, {Hagan}, {Hagedorn}, {Hainline}, {Haley}, {Hami}, {Hamilton}, {Hammann}, {Hammel}, {Hanley}, {Hansen}, {Hardy}, {Harnisch}, {Harr}, {Harris}, {Hart}, {Hartig}, {Hasan}, {Hashim}, {Hashimoto}, {Haskins}, {Hawkins}, {Hayden}, {Hayden}, {Healy}, {Hecht}, {Heeg}, {Hejal}, {Helm},
  {Hengemihle}, {Henning}, {Henry}, {Henry}, {Henshaw}, {Hernandez}, {Herrington}, {Heske}, {Hesman}, {Hickey}, {Hilbert}, {Hines}, {Hinz}, {Hirsch}, {Hitcho}, {Hodapp}, {Hodge}, {Hoffman}, {Holfeltz}, {Holler}, {Hoppa}, {Horner}, {Howard}, {Howard}, {Huber}, {Hunkeler}, {Hunter}, {Hunter}, {Hurd}, {Hurst}, {Hutchings}, {Hylan}, {Ignat}, {Illingworth}, {Irish}, {Isaacs}, {Jackson}, {Jaffe}, {Jahic}, {Jahromi}, {Jakobsen}, {James}, {James}, {James}, {Jamieson}, {Jandra}, {Jayawardhana}, {Jedrzejewski}, {Jeffers}, {Jensen}, {Joanne}, {Johns}, {Johnson}, {Johnson}, {Johnson}, {Johnson}, {Johnson}, {Johnson}, {Johnstone}, {Jollet}, {Jones}, {Jones}, {Jones}, {Jones}, {Jones}, {Jordan}, {Jordan}, {Jue}, {Jurkowski}, {Justis}, {Justtanont}, {Kaleida}, {Kalirai}, {Kalmanson}, {Kaltenegger}, {Kammerer}, {Kan}, {Kanarek}, {Kao}, {Karakla}, {Karl}, {Kassin}, {Kauffman}, {Kavanagh}, {Kelley}, {Kelly}, {Kendrew}, {Kennedy}, {Kenny}, {Keski-Kuha}, {Keyes}, {Khan}, {Kidwell}, {Kimble}, {King}, {King}, {Kinzel}, {Kirk},
  {Kirkpatrick}, {Klaassen}, {Klingemann}, {Klintworth}, {Knapp}, {Knight}, {Knollenberg}, {Knutsen}, {Koehler}, {Koekemoer}, {Kofler}, {Kontson}, {Kovacs}, {Kozhurina-Platais}, {Krause}, {Kriss}, {Krist}, {Kristoffersen}, {Krogel}, {Krueger}, {Kulp}, {Kumari}, {Kwan}, {Kyprianou}, {Labador}, {Labiano}, {Lafreni{\`e}re}, {Lagage}, {Laidler}, {Laine}, {Laird}, {Lajoie}, {Lallo}, {Lam}, {LaMassa}, {Lambros}, {Lampenfield}, {Lander}, {Langston}, {Larson}, {Larson}, {LaVerghetta}, {Law}, {Lawrence}, {Lee}, {Lee}, {Lee}, {Leisenring}, {Leveille}, {Levenson}, {Levi}, {Levine}, {Lewis}, {Lewis}, {Lewis}, {Libralato}, {Lidon}, {Liebrecht}, {Lightsey}, {Lilly}, {Lim}, {Lim}, {Ling}, {Link}, {Link}, {Lipinski}, {Liu}, {Lo}, {Lobmeyer}, {Logue}, {Long}, {Long}, {Long}, {Long}, {L{\'o}pez-Caniego}, {Lotz}, {Love-Pruitt}, {Lubskiy}, {Luers}, {Luetgens}, {Luevano}, {Lui}, {Lund}, {Lundquist}, {Lunine}, {L{\"u}tzgendorf}, {Lynch}, {MacDonald}, {MacDonald}, {Macias}, {Macklis}, {Maghami}, {Maharaja}, {Maiolino},
  {Makrygiannis}, {Malla}, {Malumuth}, {Manjavacas}, {Marini}, {Marrione}, {Marston}, {Martel}, {Martin}, {Martin}, {Martinez}, {Maschmann}, {Masci}, {Masetti}, {Maszkiewicz}, {Matthews}, {Matuskey}, {McBrayer}, {McCarthy}, {McCaughrean}, {McClare}, {McClare}, {McCloskey}, {McClurg}, {McCoy}, {McElwain}, {McGregor}, {McGuffey}, {McKay}, {McKenzie}, {McLean}, {McMaster}, {McNeil}, {De Meester}, {Mehalick}, {Meixner}, {Mel{\'e}ndez}, {Menzel}, {Menzel}, {Merz}, {Mesterharm}, {Meyer}, {Meyett}, {Meza}, {Midwinter}, {Milam}, {Miller}, {Miller}, {Miskey}, {Misselt}, {Mitchell}, {Mohan}, {Montoya}, {Moran}, {Morishita}, {Moro-Mart{\'\i}n}, {Morrison}, {Morrison}, {Morse}, {Moschos}, {Moseley}, {Mosier}, {Mosner}, {Mountain}, {Muckenthaler}, {Mueller}, {Mueller}, {Muhiem}, {M{\"u}hlmann}, {Mullally}, {Mullen}, {Munger}, {Murphy}, {Murray}, {Muzerolle}, {Mycroft}, {Myers}, {Myers}, {Myers}, {Myers}, {Myrick}, {Nagle}, {Nayak}, {Naylor}, {Neff}, {Nelan}, {Nella}, {Nguyen}, {Nguyen}, {Nickson}, {Nidhiry}, {Niedner},
  {Nieto-Santisteban}, {Nikolov}, {Nishisaka}, {Noriega-Crespo}, {Nota}, {O'Mara}, {Oboryshko}, {O'Brien}, {Ochs}, {Offenberg}, {Ogle}, {Ohl}, {Olmsted}, {Osborne}, {O'Shaughnessy}, {{\"O}stlin}, {O'Sullivan}, {Otor}, {Ottens}, {Ouellette}, {Outlaw}, {Owens}, {Pacifici}, {Page}, {Paranilam}, {Park}, {Parrish}, {Paschal}, {Patapis}, {Patel}, {Patrick}, {Pattishall}, {Paul}, {Paul}, {Pauly}, {Pavlovsky}, {Pe{\~n}a-Guerrero}, {Pedder}, {Peek}, {Pelham}, {Penanen}, {Perriello}, {Perrin}, {Perrine}, {Perrygo}, {Peslier}, {Petach}, {Peterson}, {Pfarr}, {Pierson}, {Pietraszkiewicz}, {Pilchen}, {Pipher}, {Pirzkal}, {Pitman}, {Player}, {Plesha}, {Plitzke}, {Pohner}, {Poletis}, {Pollizzi}, {Polster}, {Pontius}, {Pontoppidan}, {Porges}, {Potter}, {Prescott}, {Proffitt}, {Pueyo}, {Quispe Neira}, {Radich}, {Rager}, {Rameau}, {Ramey}, {Ramos Alarcon}, {Rampini}, {Rapp}, {Rashford}, {Rauscher}, {Ravindranath}, {Rawle}, {Rawlings}, {Ray}, {Regan}, {Rehm}, {Rehm}, {Reid}, {Reis}, {Renk}, {Reoch}, {Ressler}, {Rest},
  {Reynolds}, {Richon}, {Richon}, {Ridgaway}, {Riedel}, {Rieke}, {Rieke}, {Rifelli}, {Rigby}, {Riggs}, {Ringel}, {Ritchie}, {Rix}, {Robberto}, {Robinson}, {Robinson}, {Robinson}, {Rock}, {Rodriguez}, {Rodr{\'\i}guez del Pino}, {Roellig}, {Rohrbach}, {Roman}, {Romelfanger}, {Romo}, {Rosales}, {Rose}, {Roteliuk}, {Roth}, {Rothwell}, {Rouzaud}, {Rowe}, {Rowlands}, {Roy}, {Royer}, {Rui}, {Rumler}, {Rumpl}, {Russ}, {Ryan}, {Ryan}, {Saad}, {Sabata}, {Sabatino}, {Sabbi}, {Sabelhaus}, {Sabia}, {Sahu}, {Saif}, {Salvignol}, {Samara-Ratna}, {Samuelson}, {Sanders}, {Sappington}, {Sargent}, {Sauer}, {Savadkin}, {Sawicki}, {Schappell}, {Scheffer}, {Scheithauer}, {Scherer}, {Schiff}, {Schlawin}, {Schmeitzky}, {Schmitz}, {Schmude}, {Schneider}, {Schreiber}, {Schroeven-Deceuninck}, {Schultz}, {Schwab}, {Schwartz}, {Scoccimarro}, {Scott}, {Scott}, {Seaton}, {Seely}, {Seery}, {Seidleck}, {Sembach}, {Shanahan}, {Shaughnessy}, {Shaw}, {Shay}, {Sheehan}, {Sheth}, {Shih}, {Shivaei}, {Siegel}, {Sienkiewicz}, {Simmons}, {Simon},
  {Sirianni}, {Sivaramakrishnan}, {Slade}, {Sloan}, {Slocum}, {Slowinski}, {Smith}, {Smith}, {Smith}, {Smith}, {Smith}, {Smith}, {Smolik}, {Soderblom}, {Sohn}, {Sokol}, {Sonneborn}, {Sontag}, {Sooy}, {Soummer}, {Southwood}, {Spain}, {Sparmo}, {Speer}, {Spencer}, {Sprofera}, {Stallcup}, {Stanley}, {Stansberry}, {Stark}, {Starr}, {Stassi}, {Steck}, {Steeley}, {Stephens}, {Stephenson}, {Stewart}, {Stiavelli}, {}, {Strada}, {Straughn}, {Streetman}, {Strickland}, {Strobele}, {Stuhlinger}, {Stys}, {Such}, {Sukhatme}, {Sullivan}, {Sullivan}, {Sumner}, {Sun}, {Sunnquist}, {Swade}, {Swam}, {Swenton}, {Swoish}, {Tam Litten}, {Tamas}, {Tao}, {Taylor}, {Taylor}, {te Plate}, {Van Tea}, {Teague}, {Telfer}, {Temim}, {Texter}, {Thatte}, {Thompson}, {Thompson}, {Thomson}, {Thronson}, {Tierney}, {Tikkanen}, {Tinnin}, {Tippet}, {Todd}, {Tran}, {Trauger}, {Trejo}, {Vinh Truong}, {Tsukamoto}, {Tufail}, {Tumlinson}, {Tustain}, {Tyra}, {Ubeda}, {Underwood}, {Uzzo}, {Vaclavik}, {Valenduc}, {Valenti}, {Van Campen}, {van de Wetering},
  {Van Der Marel}, {van Haarlem}, {Vandenbussche}, {van Dishoeck}, {Vanterpool}, {Vernoy}, {Vila Costas}, {Volk}, {Voorzaat}, {Voyton}, {Vydra}, {Waddy}, {Waelkens}, {Wahlgren}, {Walker}, {Wander}, {Warfield}, {Warner}, {Wasiak}, {Wasiak}, {Wehner}, {Weiler}, {Weilert}, {Weiss}, {Wells}, {Welty}, {Wheate}, {Wheeler}, {White}, {Whitehouse}, {Whiteleather}, {Whitman}, {Williams}, {Willmer}, {Willott}, {Willoughby}, {Wilson}, {Wilson}, {Wilson}, {Windhorst}, {Wislowski}, {Wolfe}, {Wolfe}, {Wolff}, {Wondel}, {Woo}, {Woods}, {Worden}, {Workman}, {Wright}, {Wu}, {Wu}, {Wun}, {Wymer}, {Yadetie}, {Yan}, {Yang}, {Yates}, {Yeager}, {Yerger}, {Young}, {Young}, {Yu}, {Yu}, {Zak}, {Zeidler}, {Zepp}, {Zhou}, {Zincke}, {Zonak}, \& {Zondag}}]{2023PASP..135f8001G}
{Gardner}, J.~P., {Mather}, J.~C., {Abbott}, R., {et~al.} 2023, \pasp, 135, 068001, \dodoi{10.1088/1538-3873/acd1b5}

\bibitem[{{Girdhar} {et~al.}(2022){Girdhar}, {Harrison}, {Mainieri}, {Bittner}, {Costa}, {Kharb}, {Mukherjee}, {Arrigoni Battaia}, {Alexander}, {Calistro Rivera}, {Circosta}, {De Breuck}, {Edge}, {Farina}, {Kakkad}, {Lansbury}, {Molyneux}, {Mullaney}, {Silpa}, {Thomson}, \& {Ward}}]{Girdhar2022}
{Girdhar}, A., {Harrison}, C.~M., {Mainieri}, V., {et~al.} 2022, \mnras, 512, 1608, \dodoi{10.1093/mnras/stac073}

\bibitem[{{Gorski} {et~al.}(2024){Gorski}, {Aalto}, {K{\"o}nig}, {Wethers}, {Yang}, {Muller}, {Onishi}, {Sato}, {Falstad}, {Mangum}, {Linden}, {Combes}, {Mart{\'\i}n}, {Imanishi}, {Wada}, {Barcos-Mu{\~n}oz}, {Stanley}, {Garc{\'\i}a-Burillo}, {van der Werf}, {Evans}, {Henkel}, {Viti}, {Harada}, {D{\'\i}az-Santos}, {Gallagher}, \& {Gonz{\'a}lez-Alfonso}}]{2024A&A...684L..11G}
{Gorski}, M.~D., {Aalto}, S., {K{\"o}nig}, S., {et~al.} 2024, \aap, 684, L11, \dodoi{10.1051/0004-6361/202348821}

\bibitem[{{Guillard} {et~al.}(2012){Guillard}, {Ogle}, {Emonts}, {Appleton}, {Morganti}, {Tadhunter}, {Oosterloo}, {Evans}, \& {Evans}}]{Guillard2012}
{Guillard}, P., {Ogle}, P.~M., {Emonts}, B.~H.~C., {et~al.} 2012, \apj, 747, 95, \dodoi{10.1088/0004-637X/747/2/95}

\bibitem[{{Hamann} {et~al.}(2000){Hamann}, {Netzer}, \& {Shields}}]{2000ApJ...536..101H}
{Hamann}, F.~W., {Netzer}, H., \& {Shields}, J.~C. 2000, \apj, 536, 101, \dodoi{10.1086/308936}

\bibitem[{Heckman \& Best(2014)}]{Heckman2014}
Heckman, T.~M., \& Best, P.~N. 2014, Annual Review of Astronomy and Astrophysics, 52, 589, \dodoi{https://doi.org/10.1146/annurev-astro-081913-035722}

\bibitem[{{Heckman} \& {Best}(2023)}]{Heckman2023}
{Heckman}, T.~M., \& {Best}, P.~N. 2023, Galaxies, 11, 21, \dodoi{10.3390/galaxies11010021}

\bibitem[{{Holt} {et~al.}(2008){Holt}, {Tadhunter}, \& {Morganti}}]{Holt2008}
{Holt}, J., {Tadhunter}, C.~N., \& {Morganti}, R. 2008, \mnras, 387, 639, \dodoi{10.1111/j.1365-2966.2008.13089.x}

\bibitem[{{H{\"o}nig}(2019)}]{2019ApJ...884..171H}
{H{\"o}nig}, S.~F. 2019, \apj, 884, 171, \dodoi{10.3847/1538-4357/ab4591}

\bibitem[{{Hummer} \& {Storey}(1987)}]{1987MNRAS.224..801H}
{Hummer}, D.~G., \& {Storey}, P.~J. 1987, \mnras, 224, 801, \dodoi{10.1093/mnras/224.3.801}

\bibitem[{{Jackson} {et~al.}(1998){Jackson}, {Tadhunter}, \& {Sparks}}]{1998MNRAS.301..131J}
{Jackson}, N., {Tadhunter}, C., \& {Sparks}, W.~B. 1998, \mnras, 301, 131, \dodoi{10.1046/j.1365-8711.1998.02008.x}

\bibitem[{{Jakobsen} {et~al.}(2022){Jakobsen}, {Ferruit}, {Alves de Oliveira}, {Arribas}, {Bagnasco}, {Barho}, {Beck}, {Birkmann}, {B{\"o}ker}, {Bunker}, {Charlot}, {de Jong}, {de Marchi}, {Ehrenwinkler}, {Falcolini}, {Fels}, {Franx}, {Franz}, {Funke}, {Giardino}, {Gnata}, {Holota}, {Honnen}, {Jensen}, {Jentsch}, {Johnson}, {Jollet}, {Karl}, {Kling}, {K{\"o}hler}, {Kolm}, {Kumari}, {Lander}, {Lemke}, {L{\'o}pez-Caniego}, {L{\"u}tzgendorf}, {Maiolino}, {Manjavacas}, {Marston}, {Maschmann}, {Maurer}, {Messerschmidt}, {Moseley}, {Mosner}, {Mott}, {Muzerolle}, {Pirzkal}, {Pittet}, {Plitzke}, {Posselt}, {Rapp}, {Rauscher}, {Rawle}, {Rix}, {R{\"o}del}, {Rumler}, {Sabbi}, {Salvignol}, {Schmid}, {Sirianni}, {Smith}, {Strada}, {te Plate}, {Valenti}, {Wettemann}, {Wiehe}, {Wiesmayer}, {Willott}, {Wright}, {Zeidler}, \& {Zincke}}]{2022A&A...661A..80J}
{Jakobsen}, P., {Ferruit}, P., {Alves de Oliveira}, C., {et~al.} 2022, \aap, 661, A80, \dodoi{10.1051/0004-6361/202142663}

\bibitem[{{JDADF Developers} {et~al.}(2024){JDADF Developers}, Averbukh, Bradley, Buikhuizen, Busko, Cherinka, Conroy, Earl, Fox, Geda, Green, Jones, Karatay, Kotler, Lim, Morris, Nguyen, O'Steen, Ogaz, \& Volfman}]{jdaviz2024}
{JDADF Developers}, Averbukh, J., Bradley, L., {et~al.} 2024, Jdaviz,  Zenodo, \dodoi{10.5281/zenodo.5513927}

\bibitem[{{Kondapally} {et~al.}(2023){Kondapally}, {Best}, {Raouf}, {Thomas}, {Dav{\'e}}, {Shabala}, {R{\"o}ttgering}, {Hardcastle}, {Bonato}, {Cochrane}, {Ma{\l}ek}, {Morabito}, {Prandoni}, \& {Smith}}]{Kondapally2023}
{Kondapally}, R., {Best}, P.~N., {Raouf}, M., {et~al.} 2023, \mnras, 523, 5292, \dodoi{10.1093/mnras/stad1813}

\bibitem[{{Konigl} \& {Kartje}(1994)}]{1994ApJ...434..446K}
{Konigl}, A., \& {Kartje}, J.~F. 1994, \apj, 434, 446, \dodoi{10.1086/174746}

\bibitem[{{Krolik} {et~al.}(1981){Krolik}, {McKee}, \& {Tarter}}]{1981ApJ...249..422K}
{Krolik}, J.~H., {McKee}, C.~F., \& {Tarter}, C.~B. 1981, \apj, 249, 422, \dodoi{10.1086/159303}

\bibitem[{{Lanz} {et~al.}(2016){Lanz}, {Ogle}, {Alatalo}, \& {Appleton}}]{Lanz2016}
{Lanz}, L., {Ogle}, P.~M., {Alatalo}, K., \& {Appleton}, P.~N. 2016, \apj, 826, 29, \dodoi{10.3847/0004-637X/826/1/29}

\bibitem[{{Leftley} {et~al.}(2024){Leftley}, {Nesvadba}, {Bicknell}, {Janssen}, {Mukherjee}, {Petrov}, {Shende}, \& {Zovaro}}]{2024A&A...689A.314L}
{Leftley}, J.~H., {Nesvadba}, N.~P.~H., {Bicknell}, G.~V., {et~al.} 2024, \aap, 689, A314, \dodoi{10.1051/0004-6361/202449848}

\bibitem[{{Leipski} {et~al.}(2010){Leipski}, {Haas}, {Willner}, {Ashby}, {Wilkes}, {Fazio}, {Antonucci}, {Barthel}, {Chini}, {Siebenmorgen}, {Ogle}, \& {Heymann}}]{2010ApJ...717..766L}
{Leipski}, C., {Haas}, M., {Willner}, S.~P., {et~al.} 2010, \apj, 717, 766, \dodoi{10.1088/0004-637X/717/2/766}

\bibitem[{{Lopez-Rodriguez} {et~al.}(2018{\natexlab{a}}){Lopez-Rodriguez}, {Antonucci}, {Chary}, \& {Kishimoto}}]{2018ApJ...861L..23L}
{Lopez-Rodriguez}, E., {Antonucci}, R., {Chary}, R.-R., \& {Kishimoto}, M. 2018{\natexlab{a}}, \apjl, 861, L23, \dodoi{10.3847/2041-8213/aacff5}

\bibitem[{{Lopez-Rodriguez} {et~al.}(2023){Lopez-Rodriguez}, {Kishimoto}, {Antonucci}, {Begelman}, {Globus}, \& {Blandford}}]{Lopez-Rodriguez2023}
{Lopez-Rodriguez}, E., {Kishimoto}, M., {Antonucci}, R., {et~al.} 2023, \apj, 951, 31, \dodoi{10.3847/1538-4357/accb96}

\bibitem[{{Lopez-Rodriguez} {et~al.}(2018{\natexlab{b}}){Lopez-Rodriguez}, {Alonso-Herrero}, {Diaz-Santos}, {Gonzalez-Martin}, {Ichikawa}, {Levenson}, {Martinez-Paredes}, {Nikutta}, {Packham}, {Perlman}, {Ramos Almeida}, {Rodriguez-Espinosa}, \& {Telesco}}]{2018MNRAS.478.2350L}
{Lopez-Rodriguez}, E., {Alonso-Herrero}, A., {Diaz-Santos}, T., {et~al.} 2018{\natexlab{b}}, \mnras, 478, 2350, \dodoi{10.1093/mnras/sty1197}

\bibitem[{{Mandal} {et~al.}(2024){Mandal}, {Mukherjee}, {Federrath}, {Bicknell}, {Nesvadba}, \& {Mignone}}]{Mandal2024}
{Mandal}, A., {Mukherjee}, D., {Federrath}, C., {et~al.} 2024, \mnras, 531, 2079, \dodoi{10.1093/mnras/stae1295}

\bibitem[{{Mandal} {et~al.}(2021){Mandal}, {Mukherjee}, {Federrath}, {Nesvadba}, {Bicknell}, {Wagner}, \& {Meenakshi}}]{Mandal2021}
---. 2021, \mnras, 508, 4738, \dodoi{10.1093/mnras/stab2822}

\bibitem[{{Mathews} \& {Ferland}(1987)}]{1987ApJ...323..456M}
{Mathews}, W.~G., \& {Ferland}, G.~J. 1987, \apj, 323, 456, \dodoi{10.1086/165843}

\bibitem[{{McNamara} \& {Nulsen}(2012)}]{McNamara2012}
{McNamara}, B.~R., \& {Nulsen}, P.~E.~J. 2012, New Journal of Physics, 14, 055023, \dodoi{10.1088/1367-2630/14/5/055023}

\bibitem[{{Meena} {et~al.}(2023){Meena}, {Crenshaw}, {Schmitt}, {Revalski}, {Chapman}, {Fischer}, {Kraemer}, {Robinson}, {Falcone}, \& {Polack}}]{2023ApJ...943...98M}
{Meena}, B., {Crenshaw}, D.~M., {Schmitt}, H.~R., {et~al.} 2023, \apj, 943, 98, \dodoi{10.3847/1538-4357/aca75f}

\bibitem[{{Meenakshi} {et~al.}(2022){Meenakshi}, {Mukherjee}, {Wagner}, {Nesvadba}, {Bicknell}, {Morganti}, {Janssen}, {Sutherland}, \& {Mandal}}]{Meenakshi2022}
{Meenakshi}, M., {Mukherjee}, D., {Wagner}, A.~Y., {et~al.} 2022, \mnras, 516, 766, \dodoi{10.1093/mnras/stac2251}

\bibitem[{{Morganti} {et~al.}(2005){Morganti}, {Tadhunter}, \& {Oosterloo}}]{Morganti2005}
{Morganti}, R., {Tadhunter}, C.~N., \& {Oosterloo}, T.~A. 2005, \aap, 444, L9, \dodoi{10.1051/0004-6361:200500197}

\bibitem[{{Morrissey} {et~al.}(2018){Morrissey}, {Matuszewski}, {Martin}, {Neill}, {Epps}, {Fucik}, {Weber}, {Darvish}, {Adkins}, {Allen}, {Bartos}, {Belicki}, {Cabak}, {Callahan}, {Cowley}, {Crabill}, {Deich}, {Delecroix}, {Doppman}, {Hilyard}, {James}, {Kaye}, {Kokorowski}, {Kwok}, {Lanclos}, {Milner}, {Moore}, {O'Sullivan}, {Parihar}, {Park}, {Phillips}, {Rizzi}, {Rockosi}, {Rodriguez}, {Salaun}, {Seaman}, {Sheikh}, {Weiss}, \& {Zarzaca}}]{2018ApJ...864...93M}
{Morrissey}, P., {Matuszewski}, M., {Martin}, D.~C., {et~al.} 2018, \apj, 864, 93, \dodoi{10.3847/1538-4357/aad597}

\bibitem[{{Mouri} {et~al.}(2000){Mouri}, {Kawara}, \& {Taniguchi}}]{Mouri2000}
{Mouri}, H., {Kawara}, K., \& {Taniguchi}, Y. 2000, \apj, 528, 186, \dodoi{10.1086/308142}

\bibitem[{{Mukherjee} {et~al.}(2018){Mukherjee}, {Bicknell}, {Wagner}, {Sutherland}, \& {Silk}}]{2018MNRAS.479.5544M}
{Mukherjee}, D., {Bicknell}, G.~V., {Wagner}, A.~Y., {Sutherland}, R.~S., \& {Silk}, J. 2018, \mnras, 479, 5544, \dodoi{10.1093/mnras/sty1776}

\bibitem[{{Neill} {et~al.}(2023){Neill}, {Matuszewski}, \& {Martin}}]{Neill2023}
{Neill}, D., {Matuszewski}, M., \& {Martin}, C. 2023, {kderp: Keck Cosmic Web Imager Data Extraction and Reduction Pipeline in IDL}, Astrophysics Source Code Library, record ascl:2301.018

\bibitem[{{Nenkova} {et~al.}(2008){Nenkova}, {Sirocky}, {Ivezi{\'c}}, \& {Elitzur}}]{2008ApJ...685..147N}
{Nenkova}, M., {Sirocky}, M.~M., {Ivezi{\'c}}, {\v{Z}}., \& {Elitzur}, M. 2008, \apj, 685, 147, \dodoi{10.1086/590482}

\bibitem[{{Nesvadba} {et~al.}(2021){Nesvadba}, {Wagner}, {Mukherjee}, {Mandal}, {Janssen}, {Zovaro}, {Neumayer}, {Bagchi}, \& {Bicknell}}]{Nesvadba2021}
{Nesvadba}, N.~P.~H., {Wagner}, A.~Y., {Mukherjee}, D., {et~al.} 2021, \aap, 654, A8, \dodoi{10.1051/0004-6361/202140544}

\bibitem[{{Ogle} {et~al.}(1997){Ogle}, {Cohen}, {Miller}, {Tran}, {Fosbury}, \& {Goodrich}}]{1997ApJ...482L..37O}
{Ogle}, P.~M., {Cohen}, M.~H., {Miller}, J.~S., {et~al.} 1997, \apjl, 482, L37, \dodoi{10.1086/310675}

\bibitem[{{Osterbrock} \& {Miller}(1975)}]{1975ApJ...197..535O}
{Osterbrock}, D.~E., \& {Miller}, J.~S. 1975, \apj, 197, 535, \dodoi{10.1086/153541}

\bibitem[{{Pereira-Santaella} {et~al.}(2022){Pereira-Santaella}, {{\'A}lvarez-M{\'a}rquez}, {Garc{\'\i}a-Bernete}, {Labiano}, {Colina}, {Alonso-Herrero}, {Bellocchi}, {Garc{\'\i}a-Burillo}, {H{\"o}nig}, {Ramos Almeida}, \& {Rosario}}]{2022A&A...665L..11P}
{Pereira-Santaella}, M., {{\'A}lvarez-M{\'a}rquez}, J., {Garc{\'\i}a-Bernete}, I., {et~al.} 2022, \aap, 665, L11, \dodoi{10.1051/0004-6361/202244725}

\bibitem[{{Perley} {et~al.}(2017){Perley}, {Perley}, {Dhawan}, \& {Carilli}}]{2017ApJ...841..117P}
{Perley}, D.~A., {Perley}, R.~A., {Dhawan}, V., \& {Carilli}, C.~L. 2017, \apj, 841, 117, \dodoi{10.3847/1538-4357/aa725b}

\bibitem[{{Perley} {et~al.}(2011){Perley}, {Chandler}, {Butler}, \& {Wrobel}}]{Perley2011}
{Perley}, R.~A., {Chandler}, C.~J., {Butler}, B.~J., \& {Wrobel}, J.~M. 2011, \apjl, 739, L1, \dodoi{10.1088/2041-8205/739/1/L1}

\bibitem[{{Perley} {et~al.}(1984){Perley}, {Dreher}, \& {Cowan}}]{1984ApJ...285L..35P}
{Perley}, R.~A., {Dreher}, J.~W., \& {Cowan}, J.~J. 1984, \apjl, 285, L35, \dodoi{10.1086/184360}

\bibitem[{{Privon} {et~al.}(2012){Privon}, {Baum}, {O'Dea}, {Gallimore}, {Noel-Storr}, {Axon}, \& {Robinson}}]{2012ApJ...747...46P}
{Privon}, G.~C., {Baum}, S.~A., {O'Dea}, C.~P., {et~al.} 2012, \apj, 747, 46, \dodoi{10.1088/0004-637X/747/1/46}

\bibitem[{{Revalski} {et~al.}(2022){Revalski}, {Crenshaw}, {Rafelski}, {Kraemer}, {Polack}, {Falc{\~a}o}, {Fischer}, {Meena}, {Martinez}, {Schmitt}, {Collins}, \& {Falcone}}]{2022ApJ...930...14R}
{Revalski}, M., {Crenshaw}, D.~M., {Rafelski}, M., {et~al.} 2022, \apj, 930, 14, \dodoi{10.3847/1538-4357/ac5f3d}

\bibitem[{{Reynolds} {et~al.}(2015){Reynolds}, {Lohfink}, {Ogle}, {Harrison}, {Madsen}, {Fabian}, {Wik}, {Madejski}, {Ballantyne}, {Boggs}, {Christensen}, {Craig}, {Fuerst}, {Hailey}, {Lanz}, {Miller}, {Saez}, {Stern}, {Walton}, \& {Zhang}}]{2015ApJ...808..154R}
{Reynolds}, C.~S., {Lohfink}, A.~M., {Ogle}, P.~M., {et~al.} 2015, \apj, 808, 154, \dodoi{10.1088/0004-637X/808/2/154}

\bibitem[{{Riffel}(2021)}]{2021MNRAS.506.2950R}
{Riffel}, R.~A. 2021, \mnras, 506, 2950, \dodoi{10.1093/mnras/stab1877}

\bibitem[{{Rodr{\'\i}guez-Ardila} {et~al.}(2004){Rodr{\'\i}guez-Ardila}, {Pastoriza}, {Viegas}, {Sigut}, \& {Pradhan}}]{Rodriguez-ardila2004}
{Rodr{\'\i}guez-Ardila}, A., {Pastoriza}, M.~G., {Viegas}, S., {Sigut}, T.~A.~A., \& {Pradhan}, A.~K. 2004, \aap, 425, 457, \dodoi{10.1051/0004-6361:20034285}

\bibitem[{{Rosen} {et~al.}(1999){Rosen}, {Hardee}, {Clarke}, \& {Johnson}}]{Rosen1999}
{Rosen}, A., {Hardee}, P.~E., {Clarke}, D.~A., \& {Johnson}, A. 1999, \apj, 510, 136, \dodoi{10.1086/306565}

\bibitem[{{Rossi} {et~al.}(2020){Rossi}, {Bodo}, {Massaglia}, \& {Capetti}}]{Rossi2020}
{Rossi}, P., {Bodo}, G., {Massaglia}, S., \& {Capetti}, A. 2020, \aap, 642, A69, \dodoi{10.1051/0004-6361/202038725}

\bibitem[{{Roueff} {et~al.}(2019){Roueff}, {Abgrall}, {Czachorowski}, {Pachucki}, {Puchalski}, \& {Komasa}}]{2019A&A...630A..58R}
{Roueff}, E., {Abgrall}, H., {Czachorowski}, P., {et~al.} 2019, \aap, 630, A58, \dodoi{10.1051/0004-6361/201936249}

\bibitem[{{Roy} {et~al.}(2024){Roy}, {Heckman}, {Overzier}, {Saxena}, {Duncan}, {Miley}, {Villar Mart{\'\i}n}, {Gab{\'a}nyi}, {Aydar}, {Bosman}, {Rottgering}, {Pentericci}, {Onoue}, \& {Reynaldi}}]{2024arXiv240111612R}
{Roy}, N., {Heckman}, T., {Overzier}, R., {et~al.} 2024, arXiv e-prints, arXiv:2401.11612, \dodoi{10.48550/arXiv.2401.11612}

\bibitem[{{Rupke}(2014)}]{2014ascl.soft09004R}
{Rupke}, D. S.~N. 2014, {IFSRED: Data Reduction for Integral Field Spectrographs}, Astrophysics Source Code Library, record ascl:1409.004

\bibitem[{{Schlafly} \& {Finkbeiner}(2011)}]{2011ApJ...737..103S}
{Schlafly}, E.~F., \& {Finkbeiner}, D.~P. 2011, \apj, 737, 103, \dodoi{10.1088/0004-637X/737/2/103}

\bibitem[{{Schlafly} {et~al.}(2016){Schlafly}, {Meisner}, {Stutz}, {Kainulainen}, {Peek}, {Tchernyshyov}, {Rix}, {Finkbeiner}, {Covey}, {Green}, {Bell}, {Burgett}, {Chambers}, {Draper}, {Flewelling}, {Hodapp}, {Kaiser}, {Magnier}, {Martin}, {Metcalfe}, {Wainscoat}, \& {Waters}}]{2016ApJ...821...78S}
{Schlafly}, E.~F., {Meisner}, A.~M., {Stutz}, A.~M., {et~al.} 2016, \apj, 821, 78, \dodoi{10.3847/0004-637X/821/2/78}

\bibitem[{{Sebokolodi} {et~al.}(2020){Sebokolodi}, {Perley}, {Eilek}, {Carilli}, {Smirnov}, {Laing}, {Greisen}, \& {Wise}}]{2020ApJ...903...36S}
{Sebokolodi}, M. L.~L., {Perley}, R., {Eilek}, J., {et~al.} 2020, \apj, 903, 36, \dodoi{10.3847/1538-4357/abb80e}

\bibitem[{{Snios} {et~al.}(2018){Snios}, {Nulsen}, {Wise}, {de Vries}, {Birkinshaw}, {Worrall}, {Duffy}, {Kraft}, {McNamara}, {Carilli}, {Croston}, {Edge}, {Godfrey}, {Hardcastle}, {Harris}, {Laing}, {Mathews}, {McKean}, {Perley}, {Rafferty}, \& {Young}}]{2018ApJ...855...71S}
{Snios}, B., {Nulsen}, P. E.~J., {Wise}, M.~W., {et~al.} 2018, \apj, 855, 71, \dodoi{10.3847/1538-4357/aaaf1a}

\bibitem[{{Somerville} \& {Dav{\'e}}(2015)}]{Somerville2015}
{Somerville}, R.~S., \& {Dav{\'e}}, R. 2015, \araa, 53, 51, \dodoi{10.1146/annurev-astro-082812-140951}

\bibitem[{{Steffen} {et~al.}(1997){Steffen}, {G{\'o}mez}, {Raga}, \& {Williams}}]{1997ApJ...491L..73S}
{Steffen}, W., {G{\'o}mez}, J.~L., {Raga}, A.~C., \& {Williams}, R.~J.~R. 1997, \apjl, 491, L73, \dodoi{10.1086/311066}

\bibitem[{{Stockton} {et~al.}(1994){Stockton}, {Ridgway}, \& {Lilly}}]{1994AJ....108..414S}
{Stockton}, A., {Ridgway}, S.~E., \& {Lilly}, S.~J. 1994, \aj, 108, 414, \dodoi{10.1086/117080}

\bibitem[{{Tadhunter}(2007)}]{2007NewAR..51..153T}
{Tadhunter}, C. 2007, \nar, 51, 153, \dodoi{10.1016/j.newar.2006.11.020}

\bibitem[{{Tadhunter} {et~al.}(2003){Tadhunter}, {Marconi}, {Axon}, {Wills}, {Robinson}, \& {Jackson}}]{2003MNRAS.342..861T}
{Tadhunter}, C., {Marconi}, A., {Axon}, D., {et~al.} 2003, \mnras, 342, 861, \dodoi{10.1046/j.1365-8711.2003.06588.x}

\bibitem[{{Tadhunter}(1991)}]{1991MNRAS.251P..46T}
{Tadhunter}, C.~N. 1991, \mnras, 251, 46P, \dodoi{10.1093/mnras/251.1.46P}

\bibitem[{{Tadhunter} {et~al.}(1994){Tadhunter}, {Metz}, \& {Robinson}}]{1994MNRAS.268..989T}
{Tadhunter}, C.~N., {Metz}, S., \& {Robinson}, A. 1994, \mnras, 268, 989, \dodoi{10.1093/mnras/268.4.989}

\bibitem[{{Tadhunter} {et~al.}(1999){Tadhunter}, {Packham}, {Axon}, {Jackson}, {Hough}, {Robinson}, {Young}, \& {Sparks}}]{1999ApJ...512L..91T}
{Tadhunter}, C.~N., {Packham}, C., {Axon}, D.~J., {et~al.} 1999, \apjl, 512, L91, \dodoi{10.1086/311882}

\bibitem[{{Tadhunter} {et~al.}(2000){Tadhunter}, {Sparks}, {Axon}, {Bergeron}, {Jackson}, {Packham}, {Hough}, {Robinson}, \& {Young}}]{Tadhunter2000}
{Tadhunter}, C.~N., {Sparks}, W., {Axon}, D.~J., {et~al.} 2000, \mnras, 313, L52, \dodoi{10.1046/j.1365-8711.2000.03442.x}

\bibitem[{{Ueno} {et~al.}(1994){Ueno}, {Koyama}, {Nishida}, {Yamauchi}, \& {Ward}}]{1994ApJ...431L...1U}
{Ueno}, S., {Koyama}, K., {Nishida}, M., {Yamauchi}, S., \& {Ward}, M.~J. 1994, \apjl, 431, L1, \dodoi{10.1086/187458}

\bibitem[{{Vayner} {et~al.}(2024){Vayner}, {Zakamska}, {Ishikawa}, {Sankar}, {Wylezalek}, {Rupke}, {Veilleux}, {Bertemes}, {Barrera-Ballesteros}, {Chen}, {Diachenko}, {Goulding}, {Greene}, {Hainline}, {Hamann}, {Heckman}, {Johnson}, {Grace Lim}, {Liu}, {Lutz}, {L{\"u}tzgendorf}, {Mainieri}, {McCrory}, {Murphree}, {Nesvadba}, {Ogle}, {Sturm}, \& {Whitesell}}]{2024ApJ...960..126V}
{Vayner}, A., {Zakamska}, N.~L., {Ishikawa}, Y., {et~al.} 2024, \apj, 960, 126, \dodoi{10.3847/1538-4357/ad0be9}

\bibitem[{{Vazdekis} {et~al.}(2016){Vazdekis}, {Koleva}, {Ricciardelli}, {R{\"o}ck}, \& {Falc{\'o}n-Barroso}}]{2016MNRAS.463.3409V}
{Vazdekis}, A., {Koleva}, M., {Ricciardelli}, E., {R{\"o}ck}, B., \& {Falc{\'o}n-Barroso}, J. 2016, \mnras, 463, 3409, \dodoi{10.1093/mnras/stw2231}

\bibitem[{{Wang} {et~al.}(2024){Wang}, {Wylezalek}, {De Breuck}, {Vernet}, {Rupke}, {Zakamska}, {Vayner}, {Lehnert}, {Nesvadba}, \& {Stern}}]{2024A&A...683A.169W}
{Wang}, W., {Wylezalek}, D., {De Breuck}, C., {et~al.} 2024, \aap, 683, A169, \dodoi{10.1051/0004-6361/202348531}

\bibitem[{{Wang} {et~al.}(2011){Wang}, {Knigge}, {Croston}, \& {Pavlovski}}]{Wang2011}
{Wang}, Y., {Knigge}, C., {Croston}, J.~H., \& {Pavlovski}, G. 2011, \mnras, 418, 1138, \dodoi{10.1111/j.1365-2966.2011.19563.x}

\bibitem[{{Wells} {et~al.}(2015){Wells}, {Pel}, {Glasse}, {Wright}, {Aitink-Kroes}, {Azzollini}, {Beard}, {Brandl}, {Gallie}, {Geers}, {Glauser}, {Hastings}, {Henning}, {Jager}, {Justtanont}, {Kruizinga}, {Lahuis}, {Lee}, {Martinez-Delgado}, {Mart{\'\i}nez-Galarza}, {Meijers}, {Morrison}, {M{\"u}ller}, {Nakos}, {O'Sullivan}, {Oudenhuysen}, {Parr-Burman}, {Pauwels}, {Rohloff}, {Schmalzl}, {Sykes}, {Thelen}, {van Dishoeck}, {Vandenbussche}, {Venema}, {Visser}, {Waters}, \& {Wright}}]{2015PASP..127..646W}
{Wells}, M., {Pel}, J.~W., {Glasse}, A., {et~al.} 2015, \pasp, 127, 646, \dodoi{10.1086/682281}

\bibitem[{{Wilman} {et~al.}(2000){Wilman}, {Edge}, {Johnstone}, {Crawford}, \& {Fabian}}]{2000MNRAS.318.1232W}
{Wilman}, R.~J., {Edge}, A.~C., {Johnstone}, R.~M., {Crawford}, C.~S., \& {Fabian}, A.~C. 2000, \mnras, 318, 1232, \dodoi{10.1046/j.1365-8711.2000.03868.x}

\bibitem[{{Young} {et~al.}(2002){Young}, {Wilson}, {Terashima}, {Arnaud}, \& {Smith}}]{2002ApJ...564..176Y}
{Young}, A.~J., {Wilson}, A.~S., {Terashima}, Y., {Arnaud}, K.~A., \& {Smith}, D.~A. 2002, \apj, 564, 176, \dodoi{10.1086/324200}

\bibitem[{{Zhang} {et~al.}(2024){Zhang}, {Packham}, {Hicks}, {Davies}, {Shimizu}, {Alonso-Herrero}, {Hermosa Mu{\~n}oz}, {Garc{\'\i}a-Bernete}, {Pereira-Santaella}, {Audibert}, {L{\'o}pez-Rodr{\'\i}guez}, {Bellocch}, {Bunker}, {Combes}, {D{\'\i}az-Santos}, {Gandhi}, {Garc{\'\i}a-Burillo}, {Garc{\'\i}a-Lorenzo}, {Gonz{\'a}lez-Mart{\'\i}n}, {Imanishi}, {Labiano}, {Leist}, {Levenson}, {Ramos Almeida}, {Ricci}, {Rigopoulou}, {Rosario}, {Stalevski}, {Ward}, {Esparza-Arredondo}, {Delaney}, {Fuller}, {Haidar}, {H{\"o}nig}, {Izumi}, \& {Rouan}}]{2024arXiv240909771Z}
{Zhang}, L., {Packham}, C., {Hicks}, E. K.~S., {et~al.} 2024, arXiv e-prints, arXiv:2409.09771, \dodoi{10.48550/arXiv.2409.09771}

\end{thebibliography}
\bibliographystyle{aasjournal}

\end{document}